\documentstyle[12pt,epsf,rotate]{article}

\newcommand{\tr}{\mbox{tr}}
\newcommand{\eqn}[1]{(\ref{#1})}

\newcommand{\nn}{\nonumber}

\newcommand{\real}{{\bb R}} 
\newcommand{\zeds}{{\bbs Z}} 

\def\(#1{ ^{(#1)} }

\font\mybb=msbm10 at 12pt
\def\bb#1{\hbox{\mybb#1}}
\font\mybbs=msbm10 at 9pt
\def\bbs#1{\hbox{\mybbs#1}}

\def\e{{\rm e}}

\def\beq{\begin{equation}}
\def\eeq{\end{equation}}
\def\be{\begin{equation}}
\def\ee{\end{equation}}
\def\bea{\begin{eqnarray}}
\def\eea{\end{eqnarray}}
\def\bd{\begin{displaymath}}
\def\ed{\end{displaymath}}

\renewcommand{\theequation}{\thesection.\arabic{equation}}

\makeatletter
\newdimen\normalarrayskip              
\newdimen\minarrayskip                 
\normalarrayskip\baselineskip
\minarrayskip\jot
\newif\ifold             \oldtrue            
\def\arraymode{\ifold\relax\else\displaystyle\fi} 
\def\@arrayskip{\ifold\baselineskip\z@\lineskip\z@
     \else
     \baselineskip\minarrayskip\lineskip2\minarrayskip\fi}
\def\@arrayclassz{\ifcase \@lastchclass \@acolampacol \or
\@ampacol \or \or \or \@addamp \or
   \@acolampacol \or \@firstampfalse \@acol \fi
\edef\@preamble{\@preamble
  \ifcase \@chnum
     \hfil$\relax\arraymode\@sharp$\hfil
     \or $\relax\arraymode\@sharp$\hfil
     \or \hfil$\relax\arraymode\@sharp$\fi}}
\def\@array[#1]#2{\setbox\@arstrutbox=\hbox{\vrule
     height\arraystretch \ht\strutbox
     depth\arraystretch \dp\strutbox
     width\z@}\@mkpream{#2}\edef\@preamble{\halign \noexpand\@halignto
\bgroup \tabskip\z@ \@arstrut \@preamble \tabskip\z@ \cr}%
\let\@startpbox\@@startpbox \let\@endpbox\@@endpbox
  \if #1t\vtop \else \if#1b\vbox \else \vcenter \fi\fi
  \bgroup \let\par\relax
  \let\@sharp##\let\protect\relax
  \@arrayskip\@preamble}
\makeatother

\newcommand{\newsection}[1]
{\vspace{5mm}
\pagebreak[3]
\addtocounter{section}{1}
\setcounter{equation}{0}
\setcounter{subsection}{0}
\begin{flushleft}
{\large\bf \thesection. #1}
\end{flushleft}
\nopagebreak
\medskip
\nopagebreak}

\newcommand{\newsubsection}[1]{
 \vspace{5mm}
\pagebreak[3]
\addtocounter{subsection}{1}
\noindent{ \bf \thesubsection. #1}
\nopagebreak
\vspace{2mm}
\nopagebreak}

\newlength{\extraspace}
\newlength{\extraspaces}
\begin{document}

\renewcommand{\footnotesize}{\small}

\addtolength{\baselineskip}{.8mm}

\thispagestyle{empty}

\begin{flushright}
\baselineskip=12pt
{\sc OUTP}-98-54P\\
hep-th/9808124\\
\hfill{  }\\ August 1998
\end{flushright}
\vspace{.2cm}

\begin{center}

\baselineskip=24pt

{\Large\bf{Matrix D-brane Dynamics, Logarithmic Operators and Quantization of
Noncommutative Spacetime}}\\[10mm]

\baselineskip=12pt

{\bf Nick E. Mavromatos}\footnote{PPARC Advanced Fellow (U.K.).\\ E-mail: {\tt
n.mavromatos1@physics.oxford.ac.uk}} and {\bf Richard J.\ Szabo}\footnote{Work
supported in part by PPARC (U.K.). Address
after September 1, 1998: The Niels Bohr Institute, University of Copenhagen,
Blegdamsvej 17, DK-2100 Copenhagen \O, Denmark.\\ E-mail: {\tt
r.szabo1@physics.oxford.ac.uk}}
\\[5mm]
{\it Department of Physics -- Theoretical Physics\\ University of Oxford\\ 1
Keble Road, Oxford OX1 3NP, U.K.}

\vskip 0.5 in

{\sc Abstract}

\begin{center}
\begin{minipage}{16cm}

We describe the structure of the moduli space of $\sigma$-model couplings for
the worldsheet description of a system of $N$ D-particles, in the case that the
couplings are represented by a pair of logarithmic recoil operators. We
derive expressions for the canonical momenta conjugate to the D-particle
couplings and the Zamolodchikov metric to the first few orders in
$\sigma$-model perturbation theory. We show, using only very general properties
of the operator product expansion in logarithmic conformal field theories, that
the canonical dynamics on moduli space agree with the predictions of the
non-abelian generalization of the Born-Infeld effective action for D-particles
with a symmetrized trace structure. We demonstrate that the Zamolodchikov
metric naturally encodes the short-distance structure of spacetime, and from
this we derive uncertainty relations for the D-particle coordinates directly
from the quantum string theory. We show that the moduli space geometry
naturally
leads to new forms of spacetime indeterminancies involving only spatial
coordinates of target space and illustrate the manner in which the open string
interactions between D-particles lead to a spacetime quantization. We also
derive appropriate non-abelian generalizations of the string-modified
Heisenberg uncertainty relations and the space--time uncertainty principle. The
non-abelian uncertainties exhibit decoherence effects suggesting the interplay
of quantum gravity in multiple D-particle dynamics.

\end{minipage}
\end{center}

\end{center}

\noindent

\vfill
\newpage
\pagestyle{plain}
\setcounter{page}{1}
\stepcounter{subsection}

\newsection{Introduction}

Dirichlet-branes are solitonic backgrounds of superstring theory whose
discovery \cite{polchinski} has drastically changed the understanding of the
non-perturbative and target space structures of string theory. Their dynamics
can be simply described by open strings whose worldsheets are discs with
Dirichlet boundary conditions for the collective coordinates of the soliton
\cite{dbranes}, and they are related to ordinary closed string backgrounds by
duality transformations \cite{polchinski}. In this paper we shall study the
dynamics of a many-body system of D-particles.

The effective field theory for a system of $N$ parallel D-branes, with
infinitesimal separation between them, is a good probe of the short-distance
structure of the spacetime implied by string theory \cite{dbraneshort}. The
main characteristic behind this property of D-brane dynamics is the observation
\cite{bound} that the low energy effective field theory for a system of $N$
D-branes is ten-dimensional maximally supersymmetric $U(N)$ Yang-Mills theory
dimensionally reduced to the world-volume of the D-branes. For the case of
D-particles the Yang-Mills potential is
\beq
V_{D0}[Y]=\frac{{\cal T}^2}{4g_s}\,\sum_{i,j=1}^9\tr\left[Y^i,Y^j\right]^2
\label{D0potential}\eeq
where ${\cal T}=1/2\pi\alpha'$ is the elementary string tension, with $\alpha'$
the string Regge slope whose square root is the intrinsic string length
$\ell_s$, and $g_s$ is the (dimensionless) string coupling constant. The fields
$Y^i(t)$ are $N\times N$ Hermitian matrices in the adjoint representation and
the trace is taken in the fundamental representation of the gauge group $U(N)$.
In the free string limit $g_s\to0$, the field theory involving the potential
\eqn{D0potential} localizes onto those matrix configurations satisfying
\beq
\left[Y^i,Y^j\right]=0~~~~~~,~~~~~~i,j=1,\dots,9
\label{classmatrix}\eeq
and so the D-brane coordinate fields can be simultaneously diagonalized by a
gauge transformation. The corresponding eigenvalues $y_a^i$, $a=1,\dots,N$, of
$Y^i$ then represent the collective transverse coordinates of the $N$ D-branes.
In this limit the parallel D-branes are very far apart from each other and
massless vector states emerge only when fundamental strings start and end on
the same D-particle (fig. 1). The gauge group is then $U(1)^N$. Since the
energy of a string which stretches between two D-branes is
\beq
M\propto{\cal T}\,|y_a-y_b|
\label{stringenergy}\eeq
more massless vector states emerge when the branes are practically on top of
each other. The collection of all massless states corresponding to an
elementary string starting and ending on either the same or different D-brane
forms a $U(N)$ multiplet (fig. 1). The off-diagonal components of the $Y^i$ and
the remnant gauge fields describe the dynamics of the short open strings
interacting with the branes through the Dirichlet condition.

\begin{figure}[htb]
\epsfxsize=1.5in
\bigskip
\centerline{\epsffile{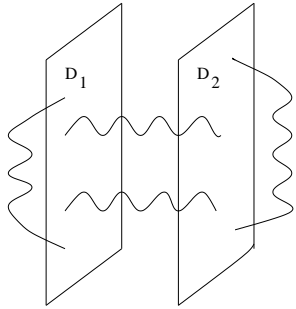}}
\caption{\it\baselineskip=12pt Emergence of the enhanced $U(N)$ gauge symmetry
for bound states of $N=2$ parallel D-branes (planes). An oriented
fundamental string (wavy lines) can start and end either at the same or
different D-brane, giving four massless vector states in the limit of
coinciding branes. These states form a representation of $U(2)$.}
\bigskip
\label{dbrane}\end{figure}

Thus when the D-branes are very far apart the classical vacuum solution of the
field theory has unbroken supersymmetry (or zero energy) and the spacetime
coordinates are represented by a set of {\it commutative} $Y^i$. When the
branes are very close to each other, the full quantum $U(N)$ gauge theory must
be taken into account, whose spectrum consists of D-brane bound states with
broken supersymmetry ($[Y^i,Y^j]\neq0$ for $i\neq j$) and at very short
distances the spacetime is described by {\it noncommutative} structures. The
gauge symmetry is interpreted as a symmetry generalizing the statistics
symmetry for identical particles in quantum mechanics and the D-brane
coordinates are viewed as adjoint Higgs fields in this description. D-brane
field theory therefore explicitly realizes the old ideas of string theory that
at very short distance scales (smaller than the target space Planck length or
the finite size of the string) the classical concepts of spacetime geometry
break down. The noncommutative structure of the spacetime is controlled by the
strength of the string interactions among the constituent D-branes. This is
precisely the structure inherent in the noncommutative geometry formalisms of
stringy spacetimes \cite{ncg}, in which the target space geometry is
represented by the algebra of observables (such as a vertex operator algebra)
corresponding to the interacting states of the theory.

The dimensionally reduced Yang-Mills theory is the relevant field theory for
the description of matrix theory \cite{banks}, which hypothesizes that the
D-particles of type-IIA superstring theory are the partons and the
supersymmetric gauge theory the exact quantum field theory in the infinite
momentum frame of 11-dimensional spacetime. However, this is not the case in
other regimes, for instance in the weak-coupling limit where the relevant
effective action is the disc generating functional. In this paper we shall be
interested in the description of $N$ D-particle dynamics from an elementary
point of view of the bosonic part of a worldsheet $\sigma$-model for the string
interactions. In this formalism, the D-brane coordinate fields appear as
coupling constants associated with boundary deformation vertex operators on the
worldsheet of a free $\sigma$-model. Already at tree-level in the string
coupling $g_s$ (the disc diagram) and in flat target space, the effective
action for $N$ D-branes is a highly non-local object that is not known in
closed form. This complexity is due to the fact that, even at tree-level,
correlation functions on the disc receive contributions from the massive string
states which yield a non-local functional of the massless modes.

The low-energy effective field theory for the $\sigma$-model couplings $y^i(t)$
in the case of a single D-particle is well-known to be described by the
Born-Infeld action \cite{BI}
\beq
\Gamma_{\rm BI}[y]=\frac1{g_s\ell_s}\int dt~\sqrt{1-(\dot y^i)^2}
\label{BIaction}\eeq
which is just the relativistic free particle action for the D0-brane. The
appropriate generalization of \eqn{BIaction} to the case of non-abelian
(Chan-Paton) $\sigma$-model couplings appropriate to the description of multi
D-brane dynamics has been a point of ambiguity in recent literature. Although
it is established that the appropriate global gauge invariant structure in the
action is a trace in the fundamental representation of $U(N)$, the ambiguity
arises in choosing a particular matrix ordering prescription for the action.
The original proposal in \cite{tseytlin}, which employs a symmetrized
matrix ordering, has been argued to hold {\it only} when one incorporates
worldsheet supersymmetry \cite{brecher}, or alternatively when one imposes
certain energy-minimizing BPS-type conditions on the form of the action
\cite{brecher1}. The $U(N)$ Yang-Mills theory should appear as a
``non-relativistic" approximation to the non-abelian Born-Infeld action. An
interesting closed-form expression for the symmetrized action in the case $N=2$
has been obtained recently in \cite{taming}.

In the following we shall show how an appropriate worldsheet formalism yields
the symmetrized form of the effective bosonic action functional for multi
D-brane dynamics, without the need of resorting to supersymmetry arguments. A
crucial feature of the D-brane couplings we shall use is that, not only do
they define a (non-marginal) perturbation about a truly marginal deformation,
but the deformed worldsheet field theory has logarithmic scaling violations,
coming from logarithmic divergences on the worldsheet, and defines not a
conventional two-dimensional conformal field theory, but rather a {\it
logarithmic} conformal field theory \cite{gurarie}. Logarithmic conformal field
theories lie on the border between conformal field theories and generic
two-dimensional renormalizable field theories, and they correspond to the
emergence of hidden continuous symmetries \cite{tsvelik}. It has been suggested
\cite{kogmav} that the appropriate worldsheet description of the collective
coordinates (zero modes) of a soliton in string theory is given by logarithmic
operators. The normalizable target space zero modes for D-branes arise from
translations and rotations (in both spacetime and isospin space) of the
background, and there is a family of backgrounds connected by the symmetries
which act on the moduli space of $\sigma$-model couplings characterizing the
background. These modes are an important ingredient for the proper
incorporation of recoil effects during the scattering of closed string states
off the D-brane background when the soliton state changes during the process of
scattering \cite{paban}--\cite{diffusion}. These effects are important aspects
of the quantization of the collective coordinates of D-branes.

Logarithmic operators have conformal dimensions which are degenerate with those
of the usual primary fields, and as a result of this degeneracy one can no
longer completely diagonalize the usual Virasoro operator $L_0$. Together with
the standard operators they form the basis of a Jordan cell for $L_0$. For a
logarithmic pair $(C,D)$ of conformal dimension $\Delta$, the operator product
expansion of the worldsheet stress-energy tensor $T$ with these fields is
non-trivial and involves a mixing \cite{gurarie}
\bea
T(z)C(w)&\sim&\frac{\Delta}{(z-w)^2}\,C(w)+\frac1{z-w}\,\partial
C(w)+\dots\nn\\T(z)D(w)&\sim&\frac{\Delta}{(z-w)^2}\,D(w)+\frac1{(z-w)^2}
\,C(w)+\frac1{z-w}\,\partial D(w)+\dots
\label{Tope}\eea
where an appropriate normalization of the $D$ operator has been chosen.
Defining the Virasoro operator $L_0$ through the Laurent series expansion
$T(z)=\sum_{n\in\zeds}L_n\,z^{-n-2}$, it follows that the corresponding states
$|C\rangle=C(0)|0\rangle$ and $|D\rangle=D(0)|0\rangle$ generate a $2\times2$
Jordan block for $L_0$,
\beq
L_0|C\rangle=\Delta|C\rangle~~~~~~,
{}~~~~~~L_0|D\rangle=\Delta|D\rangle+|C\rangle
\label{Jordancell}\eeq
This mixing is a consequence of the behaviour of the conformal blocks of the
underlying worldsheet theory which exhibit logarithmic scaling violations. It
is the characteristic non-trivial property of theories involving logarithmic
operators.

In this paper we shall study the disc amplitude in a worldsheet boundary
auxilliary field formalism \cite{wilson}--\cite{dornlast} which can be thought
of as an  ``abelianization" of the $U(N)$ theory. In this framework, before the
auxilliary fields are integrated out, the only difference from the abelian
situation is an extra explicit dependence on the variables parametrizing the
boundary of the string worldsheet. This representation of the Wilson loop
operator enables one to carry out $\sigma$-model perturbation theory in much
the same way as in the abelian (single D-brane) case. Within this formalism, we
will construct the moduli space of the $\sigma$-model couplings which
represents
the effective spacetime of the D-particles and whose geometry is determined by
the Zamolodchikov metric \cite{zam}. The dynamics on moduli space is determined
by the Zamolodchikov $C$-theorem and a set of conditions which ensure the
possibility of canonical quantization \cite{emn}. The crucial observation is
that, because of the logarithmic nature of the D-brane couplings, the
worldsheet deformations become slightly relevant, which in the recoil problem
is precisely the property that leads to a change of state of the 0-brane
background. To restore marginality, we dress the worldsheet theory with
two-dimensional quantum gravity, i.e. Liouville theory \cite{ddk}. We
demonstrate explicitly that the canonical form of the moduli space dynamics
coincides with that of the symmetrized non-abelian Born-Infeld theory.
Physically, the dynamical theory describes the non-relativistic motion of open
strings in the background of a `fat brane', as described in \cite{richfed}.
Although in this framework the explicit form of the D-brane couplings is
associated with those relevant to the recoil problem, we shall see that the
derivation of our results are based only on very general properties of the
operator product expansion in generic logarithmic conformal field theories. The
derivation of the appropriate non-abelian Born-Infeld dynamics in the
kinematical region of interest thereby represents a highly non-trivial
application of the theory of logarithmic operators.

Quantization of the moduli space is then achieved by summing over worldsheet
topologies, in the pinched approximation which gives the dominant terms
\cite{kogmav,paban,emnd,emn}. In the case of a single D-particle, it was shown
in \cite{lm} that, to leading order in the $\sigma$-model coupling constant
expansion, one recovers the usual canonical quantum phase space with position
and momentum having a constant commutator and ``Planck constant" given in terms
of the string coupling $g_s$. Incorporating stringy effects reproduces the
generalized string uncertainty principle \cite{ven,kmm}
\beq
\Delta y^i\,\Delta p_j\geq\mbox{$\frac\hbar2$}\,\delta_j^i\left(1+{\cal
O}(\alpha_s')\,(\Delta p_i)^2+\dots\right)
\label{stringuncert}\eeq
which corresponds to adding corrections to the Heisenberg commutation relations
of the form \\
\beq
\left[\!\left[\widehat{y}^i,\widehat{p}_j\right]\!\right]=i\hbar\,\delta_i^j
\left(1+{\cal O}(\alpha_s')\,\widehat{p}_i^2+\dots\right)
\label{modHeisen}\eeq
where $\alpha_s'=g_s^2\alpha'$ is the 0-brane scale. The result
\eqn{stringuncert} can also be derived from a Heisenberg microscope
approach to the uncertainty principle for D-branes \cite{kmw}. Minimizing the
modified uncertainty relation \eqn{stringuncert} leads to a minimal measurable
length $\Delta y^i\geq{\cal O}(\sqrt{\alpha'_s})$. Note that this length scale
vanishes in the weak-coupling regime $g_s\to0$, in which case there is no lower
bound on the measurability of distances in the spacetime and free D-particles
can probe distances smaller than the string length.

In the multi D-particle case we shall find that the fluctuating worldsheet
topologies yield the appropriate non-abelian generalization of the result
\eqn{stringuncert}, and in addition lead to a
proper quantization of the noncommutative spacetime implied by the D-brane
field theory. As we will see, this leads to new forms of uncertainty relations
involving only the coordinates of spacetime, in the spirit of
\cite{zumi,doplicher}, which are superior to the phase space uncertainty
relation \eqn{stringuncert}. The simplest such relation has the form
\cite{yoneya}
\beq
\Delta y^i\,\Delta t\geq\ell_s^2=\alpha'
\label{tyuncert}\eeq
The space--time uncertainty principle \eqn{tyuncert} follows from the
energy-time uncertainty relation of quantum mechanics applied to strings, and
it can be derived from very basic worldsheet conformal symmetry arguments. The
same relation can be derived within the framework of the effective field theory
for D-instantons \cite{yoneya2b} and it is also naturally encoded in the
effective supersymmetric Yang-Mills theory for D-particles \cite{liyoneya}. It
can be shown \cite{liyoneya} that, for the nonrelativistic scattering of two
D-particles of BPS mass $1/\sqrt{\alpha_s'}$ with impact parameter of order
$\Delta y^i$, the space--time uncertainty relation \eqn{tyuncert} gives the
minimal spatial and temporal lengths
\beq
\Delta y^i\geq g_s^{1/3}\,\ell_s\equiv\ell_{\rm
P}^{(11)}~~~~~~,~~~~~~\Delta t\geq g_s^{-1/3}\,\ell_s
\label{minyt}\eeq
where $\ell_{\rm P}^{(11)}$ is the 11-dimensional Planck length which is the
characteristic distance scale of M-theory \cite{banks}. The space--time
uncertainty principle therefore implies that, for each state of a D-particle,
no information can be stored below the Planck distance in the transverse space.

The following results represent the first examples of such relations within the
framework of a flat space worldsheet D-brane field theory. In this
$\sigma$-model formalism we shall find the appearence of quantum smearing of
multi D-particle coordinates arising from the string interactions between
constituent branes. The appearence of minimal measurable spacetime lengths in
this way is reminescent of the lower bounds which arise from the existence of
internal (ultraviolet regularization) symmetries of the target space
\cite{intern}. The internal symmetry group is the enhanced $U(N)$ gauge
symmetry which comes from the string interactions. For each constituent
D-particle we shall obtain phase space and space--time uncertainty relations of
the form of \eqn{stringuncert} and \eqn{tyuncert} when string interactions are
turned on. There is no noncommutativity between different directions on a given
brane and one obtains the standard stringy smearings of the coordinates.
However, among the matrix off-diagonal components, representing the fundamental
string degrees of freedom, there are uncertainties between different directions
of the fundamental string, in addition to the usual smearing, which leads to a
proper quantum noncommutativity among the D-brane fields. The open string
interactions are in this way responsible for non-trivial quantum mechanical
correlations between different spatial coordinate directions of the
D-particles. As discussed in \cite{nmprl}, these noncommutative uncertainty
relations are determined entirely by the geometry of moduli space. The
Zamolodchikov metric on this space involves the various non-trivial kinematical
quantities characterizing the multi D-brane dynamics, and it naturally encodes
the small-scale structure of spacetime. The noncommutative structures of
spacetime are determined by the transformations which diagonalize the
Zamolodchikov metric. These noncommutative smearings arise from an expansion of
the moduli space around the background of a (Lie algebraic) commutative
spacetime determined as in \eqn{classmatrix} which has the effect of encoding
the noncommutative string interactions into a gauge transformation. The gauge
field interactions are then ultimately responsible for the occurence of the
quantum noncommutativity. This is reminscent of the matrix string framework for
nonperturbative string theory \cite{Mgeom,newmatrix}, which encodes the
geometry of the genus expansion through singular gauge transformations of
commutative spacetime coordinates and naturally yields the characteristic
spatial scale in \eqn{minyt} \cite{newmatrix}. The following results therefore
yield a geometric picture of the string interactions among D-branes and hence
of the short-distance noncommutativity of spacetime.

The present worldsheet framework thus gives an explicit realization of the
spacetime noncommutativity described in \cite{doplicher}, where, based on very
general requirements arising from the Heisenberg uncertainty principle and
classical general relativity, uncertainty relations among different coordinate
directions are postulated in the form
\beq
\sum_{i<j}\Delta y^i\,\Delta y^j\geq\ell_{\rm P}^2
\label{dopluncert}\eeq
However, there are several crucial differences in the present approach. The
first one is that all of our uncertainties are {\it derived} from statistical
distribution functions that are induced from the worldsheet genus expansion,
without the need of postulating auxilliary relations. In particular, we shall
find uncertainties of the sort \eqn{dopluncert} as implied by a stronger
smearing of the coordinates involving a statistical connected correlation
function of the matrix fields. The present approach therefore distinguishes the
quantum noncommutativity of spacetime from the algebraic one, in contrast to
the approaches of \cite{zumi,doplicher, yoneya2b} where the two structures are
identified. Secondly, the noncommutative smearings that we find depend on the
energy content of the system and suggest the emergence of quantum decoherence
in multi D-brane dynamics. In particular, we shall derive a triple space--time
uncertainty relation which implies that the scattering of D-particles at high
energies can probe very small distances through their open string interactions.
The emergence of decoherence effects is characteristic of certain approaches to
spacetime quantum gravity, so that the present formulation of matrix D-brane
dynamics seems to naturally encode the effects of quantum gravity.

The structure of the remainder of this paper is as follows. In section 2 we
briefly describe the formalism of coupling constant quantization in Liouville
string theory. In section 3 we describe the relevant brane configurations that
we shall study, introducing their low-energy effective field theory (the
non-abelian Born-Infeld action) and the associated logarithmic recoil
operators. In section 4 we carry out a detailed perturbative calculation, up to
third order in the $\sigma$-model coupling constants, of the canonical momentum
of the multi D-brane system and show that the result coincides with the
predictions of the symmetrized form of the non-abelian Born-Infeld action. In
section 5 we show that the resulting moduli space dynamics takes the canonical
form of that in Liouville string theory. With this correspondence established,
in section 6 we carry out the sum over worldsheet topologies in the pinched
approximation which leads to a quantization of the D-particle couplings. Then
we derive the spacetime uncertainty relations and discuss their physical
significances. Section 7 contains some concluding remarks and possible physical
tests of the noncommutativity of spacetime. At the end of the paper there are
four appendices containing some of the more technical calculations. In appendix
A we describe the structure of generic correlation functions of the logarithmic
operators, and in appendix B we describe the technical details of the
computation of the canonical momentum of section 4, including a description of
a particular renormalization scheme that must be used for the auxilliary field
representation of the Wilson loop operator. Appendix C summarizes the
complicated boundary integrations used in the paper, and finally in appendix D
we show how to cancel the leading modular divergences in the genus expansion of
section 6 by imposing momentum conservation in the scattering of string states
off the multiple D-brane background.

\newsection{Helmholtz Conditions and Coupling Constant Quantization for
Two-dimensional $\sigma$-models}

In this section we will briefly review the formalism of coupling constant
quantization for two-dimensional $\sigma$-models. Consider a worldsheet
$\sigma$-model that is given by a deformed conformal field theory on a compact
Riemann surface $\Sigma$ with metric $\gamma_{\alpha\beta}$. The deformation is
described by a set of coupling constants $g^i$ associated with vertex operators
$V_i(z,\bar z)$ that have conformal dimensions $(\Delta_i,\bar\Delta_i)$ and
operator product expansion coefficients $c_{jk}^i$. The action is of the form
\beq
  S_\sigma[x;g] = S_0[x] + \int _{\Sigma}d^2z~\sqrt\gamma~g^i\,V_i
\label{smodel}\eeq
where $S_0[x]$ is the action of the unperturbed conformal field theory and an
implicit sum over repeated indices is always understood. The vertex operators
$V_i$ are constructed from the fields of $S_0[x]$. As we will discuss, because
of special properties of the Zamolodchikov renormalization group flow
\cite{zam}, the summation over worldsheet genera leads to a canonical
quantization of the system of moduli space variables $\{g^i\}$ in a non-trivial
way \cite{emnd,emn}. In this picture the ultraviolet worldsheet renormalization
group scale $\log\Lambda$ plays the role of time for the quantum mechanical
system of variables $\{g^i\}$.\footnote{Strictly speaking, $\Lambda$ is the
ratio of the infrared to ultraviolet scales on the worldsheet. In what follows,
however, we shall set the size of the surface $\Sigma$ to unity.}

When the vertex operators $V_i$ describe a relevant deformation (i.e.
$\Delta_i+\bar\Delta_i<2$), the running coupling constants $g^i(\Lambda)$
acquire non-trivial flow under the renormalization group which is described by
the flat worldsheet $\beta$-function
\beq
\beta^i[g]\equiv\frac{dg^i}{d\log\Lambda}=\left(\Delta_i+\bar\Delta_i
-2\right)g^i-\pi c^i_{jk}\,g^jg^k
\label{betgi}\eeq
The flows in the space of running coupling constants interpolate between
various two-dimensional renormalizable quantum field theories. Conformally
invariant theories are infrared or ultraviolet fixed points of these flows.
Studying the global aspects of this moduli space leads to a geometrical
understanding of certain equivalences between various conformal field theories
and their associated target spaces.

One can restore conformal invariance at the quantum level by including
worldsheet gravitational effects and dressing the action \eqn{smodel} by
Liouville theory. This amounts to dressing the vertex operators in \eqn{smodel}
as $V_i\to[V_i]_\varphi$, where $\varphi$ is the Liouville field which scales
the worldsheet metric as
\beq
\gamma_{\alpha\beta}=\e^{(2/\sqrt{\alpha'}\,Q)\varphi}~
\widehat{\gamma}_{\alpha\beta}
\label{Liouvilledef}\eeq
with $\widehat{\gamma}_{\alpha\beta}$ a fixed fiducial metric on $\Sigma$ and
$Q$ is a constant related to the central charge $c$ of the two-dimensional
quantum gravity. In the Liouville framework, $\log\Lambda$ is therefore
identified with the worldsheet zero mode of the Liouville
field~\cite{emnliouv}. This dressing can be viewed as a renormalization of the
corresponding coupling constants in \eqn{smodel} as
\beq
g^i(\varphi)=g^i\,\e^{\alpha_i\varphi/\sqrt{\alpha'}}+\frac{\pi/\sqrt{\alpha'}}
{Q+2\alpha_i}\,c^i_{jk}\,g^jg^k\varphi~\e^{\alpha_i\varphi/\sqrt{\alpha'}}
+\dots
\label{giphi}\eeq
The dressed deformation $[V_i]_\varphi$ is then truly marginal provided that
\beq
\mbox{$\frac12$}\alpha_i\left(\alpha_i+Q\right)=\Delta_i+\bar\Delta_i-2
\label{marginalcond}\eeq
The gravitationally dressed version of \eqn{smodel} is
$S_0[x]+S_{\rm L}[x;\varphi]$, where
\bea
S_{\rm L}[x;\varphi]&=&\frac1{4\pi\alpha'}\int_\Sigma
d^2z~\sqrt{\widehat{\gamma}}~\left[\widehat{\gamma}^{\alpha\beta}
\partial_\alpha\varphi\partial_\beta\varphi-Q\varphi
R^{(2)}(\widehat{\gamma})\right]-\frac Q{4\pi\alpha'}
\oint_{\partial\Sigma}d\widehat{s}~\varphi K(\widehat{\gamma})\nn\\&
&~~~~~~~~~~+\int_\Sigma
d^2z~\sqrt{\widehat{\gamma}}~g^i(\varphi)\,V_i
\label{Liouvilleaction}\eea
is the Liouville action coupled to the ``matter" part of \eqn{smodel}
\cite{ddk}. Here $R^{(2)}$ is the scalar curvature of the worldsheet $\Sigma$
and $K$ is the extrinsic curvature at the worldsheet boundary $\partial\Sigma$.

The most general renormalization group flow for a $\sigma$-model coupling
$g^i$, corresponding to a vertex operator $V_i$, in Liouville string theory has
the form of a friction equation of motion \cite{emn,polkle,schm}
\be
\alpha'\,\ddot g^i (\phi ) + \sqrt{\alpha'}\,Q\,\dot g^i (\phi )=-\beta ^i[g] =
G^{ij}\frac{\partial}{\partial g^j}C[g;\phi]
\label{eikosidyo}\ee
where the dots denote differentiation with respect to the Liouville zero mode
\beq
\phi=-\sqrt{\alpha'}\,Q\log\Lambda
\label{0modedef}\eeq
and
\beq
Q=\left(\mbox{$\frac13$}\,|c_*-C[g;\phi]|+\mbox{$\frac14$}\,\beta^iG_{ij}
\beta^j\right)^{1/2}
\label{Qdef}\eeq
is the central charge deficit with $c_*$ the central charge at the critical
dimension. The quantity $C[g;\phi]$ is the Zamolodchikov $C$-function
\cite{zam}. It interpolates in moduli space among two-dimensional field
theories on $\Sigma$ according to the $C$-theorem, which for flat worldsheets
reads
\be
{\partial C\over\partial\log\Lambda}=-\beta^iG_{ij}\beta^j \label{cteo}
\ee
where
\beq
G_{ij}=\Lambda^4\left\langle V_i(z,\bar z)\,V_j(z,\bar z)\right\rangle_{\rm L}
\label{Zmetric}\eeq
is the Zamolodchikov metric on moduli space. Here $\langle\cdot\rangle_{\rm L}$
denotes the average in the non-critical $\sigma$-model \eqn{smodel} dressed
with the Liouville action \eqn{Liouvilleaction}, and $G^{ij}$ denotes the
matrix inverse of \eqn{Zmetric}.

In \eqn{eikosidyo} we took into account the gradient flow property of the
$\beta$-functions
\be
    \frac \partial{\partial g^i}C = G_{ij}\beta ^j
\label{gradient}\ee
which is an off-shell corollary of the flat worldsheet $C$-theorem~\cite{zam}.
When the $C$-function is regarded as an effective action in moduli space, the
corresponding classical equations of motion therefore coincide with the
renormalization group equations $\beta^i[g]=0$. The Zamolodchikov metric
\eqn{Zmetric} can also be used to determine the short distance behaviour of
3-point correlation functions of the $\sigma$-model. For a scale-invariant
field theory, the short-distance operator product expansion is
\beq
V_i(z_1,\bar z_1)V_j(z_2,\bar z_2)\sim
c_{ij}^k~z_{12}^{\Delta_i+\Delta_j-\Delta_k}\,\bar
z_{12}^{\bar\Delta_i+\bar\Delta_j-\bar\Delta_k}~V_k
\left(\mbox{$\frac12(z_1+z_2),\frac12(\bar z_1+\bar z_2)$}\right)
\label{OPE}\eeq
for $|z_1|\sim|z_2|$, where
\beq
z_{ij}=z_i-z_j
\label{diffvar}\eeq
Then the three-point function of the deformation operators
\bea
\left\langle V_i(z_1,\bar z_1)V_j(z_2,\bar z_2)V_k(z_3,\bar
z_3)\right\rangle_{\rm L}&=&C_{ijk}~z_{12}^{\Delta_i+\Delta_j-\Delta_k}\,\bar
z_{12}^{\bar\Delta_i+\bar\Delta_j-\bar\Delta_k}\,z_{23}^{\Delta_j+\Delta_k-
\Delta_i}\,\bar z_{23}^{\bar\Delta_j+\bar\Delta_k-\bar\Delta_i}\nonumber\\&
&~~~~~~~~~~\times\,z_{13}^{\Delta_i+\Delta_k-\Delta_j}\,\bar
z_{13}^{\bar\Delta_i+\bar\Delta_k-\bar\Delta_j}
\label{3ptfnsgen}\eea
can be determined as
\beq
C_{ijk}=c^l_{ij}\,G_{lk}
\label{3ptcoeffsgen}\eeq
in the limit $|z_{23}|\sim|z_{12}|\gg|z_{13}|$. The coefficients $C_{ijk}$ are
completely symmetric in their indices. From \eqn{3ptcoeffsgen} it follows that
the asymptotic behaviours of the first three sets of correlation functions of
the vertex operators can be related as
\beq
\langle V_iV_j\rangle_{\rm L}\sim c_{ij}^k\,\langle V_k\rangle_{\rm L}\sim
G^{kl}\,\langle V_iV_jV_k\rangle_{\rm L}\,\langle V_l\rangle_{\rm L}
\label{123rel}\eeq

It is well-known that higher-genus effects will quantize the effective coupling
constants $g^i(\phi)$ \cite{emnd,emn}. For a full quantum description, we must
ensure that the equations (\ref{eikosidyo}), which are characteristic of
frictional motion in a potential $C[g;\phi]$, are consistent with the canonical
quantization conditions. We therefore need an action formalism for the
renormalization group flow. In general such equations of motion cannot be cast
in a Lagrangian form, but in the case of non-critical strings this is possible
due to the non-trivial metric $G_{ij}$. In this framework, the Liouville zero
mode \eqn{0modedef} is identified as the physical time coordinate
\cite{emn,time}, observed in standard units.

The conditions for the existence of an underlying Lagrangian $L$ whose
equations of motion are equivalent (but not necessarily identical) to
(\ref{eikosidyo}) are determined by the existence of a non-singular matrix
$\omega_{ij}$ with
\be
\omega_{ij}\left(\alpha'\,\ddot g^j+\sqrt{\alpha'}\,Q\,\dot
g^j+\beta^j\right)=\frac d{d\phi}\left(\frac{\partial L}{\partial {\dot
g}^i}\right)-\frac{\partial L}{\partial g^i}
\label{triantadyo}\ee
which obeys the Helmholtz conditions \cite{hojman}
\bea
\omega_{ij}&=&\omega_{ji}\label{helm1}\\\frac{\partial\omega_{ij}}{\partial
{\dot g}^k}
&=& \frac{\partial \omega_{ik}}{\partial {\dot g}^j}\label{helm2}\\
\frac{1}{2}\frac{D}{D \phi }\left(\omega_{ik}\frac{\partial f^k}{\partial {\dot
g}^j}-\omega_{jk}\frac{\partial f^k}{\partial{\dot g}^i}\right)&=&\omega_{ik}
\frac{\partial f^k}{\partial g^j} -\omega_{jk} \frac{\partial f^k}{\partial
g^i}
\label{helm3}\\ \frac{D}{D \phi}\omega_{ij} &=&
-\frac{1}{2\alpha'}\left(\omega_{ik}\frac{\partial f^k}{\partial {\dot
g}^j}+\omega_{jk}\frac{\partial f^k}{\partial{\dot g}^i}\right)
\label{helm4}\eea
where
\be
f^i \equiv -\sqrt{\alpha'}\,Q\,{\dot g }^i - \beta ^i[g] \qquad , \qquad
\frac{D}{D \phi } \equiv\frac\partial{\partial\phi} + {\dot g^i}
\frac\partial{\partial g^i}+\frac{f^i}{\alpha'}\frac{\partial}{\partial\dot
g^i}
\label{34}\ee
If the conditions (\ref{helm1})--(\ref{helm4}) are met, then
\be
\alpha'\,\omega_{ij}=\frac{\partial ^2 L } {\partial {\dot g ^i}\partial {\dot
g^j}}
\label{35}\ee
and the Lagrangian in (\ref{35}) can be determined up to total derivatives
according to \cite{hojman}
\bea
{\cal S}  ~\equiv~ \int d\phi~L &=&- \int d\phi~\int _{0}^1 d\kappa~g ^i
E_i (\phi, \kappa g, \kappa {\dot g},\kappa {\ddot g} )\nn\\ E_i (\phi,g,{\dot
g},\ddot g)&\equiv&\omega_{ij}\left(\alpha'\,\ddot g^j+\sqrt{\alpha'}\,Q\,\dot
g^j+\beta ^j\right)
\label{36b}\eea

In the case of non-critical strings one can identify \cite{emn}
\beq
\omega_{ij}=-\mbox{$\frac1{\sqrt{\alpha'}}$}\,G_{ij}
\label{wGrel}\eeq
Near a fixed point in moduli space, where the variation of $Q$ is small,
the action \eqn{36b} then becomes \cite{emnd,emn}
\be
{\cal S}= \int d\phi~\left(-\mbox{$\frac{\sqrt{\alpha'}}{2}$}\,\dot
g^i\,G_{ij}[g;\phi]\,\dot
g^j-\mbox{$\frac1{\sqrt{\alpha'}}$}\,C[g;\phi]+\dots\right)
\label{36c}\ee
where the dots denote terms that can be removed by a change of renormalization
scheme. Within a critical string (on-shell) approach, the action (\ref{36b},
\ref{36c}) can be considered as an effective action generating the string
scattering amplitudes. Here it should be considered as a target space
`off-shell' action for non-critical strings~\cite{emn}. From \eqn{36c} it
follows that the canonical momenta $p_i$ conjugate to the couplings $g^i$ are
given by
\be
   p_i = \sqrt{\alpha'}\,G_{ij}\,{\dot g}^j
\label{momenta2}\ee

Let us briefly sketch the validity of the conditions
(\ref{helm1})--(\ref{helm4}) for the choice \eqn{wGrel}. Since $G_{ij}$
is symmetric, the first Helmholtz condition (\ref{helm1}) is satisfied. The
conditions \eqn{helm2} and \eqn{helm3} hold automatically because of the
gradient flow property (\ref{gradient}) of the $\beta$-function, and the fact
that $G_{ij}$ and $C[g;\phi]$ are functions of the coordinates $g^i$ and not of
the conjugate momenta. Finally, the fourth Helmholtz condition \eqn{helm4}
yields the equation
\be
  \frac{D}{D \phi } G_{ij} = \frac Q{\sqrt{\alpha'}}\, G_{ij}
\label{37}\ee
which implies an expanding scale factor for the metric in moduli space
\be
G_{ij}[\phi;g(\phi)]=\e^{Q\phi/\sqrt{\alpha'}}\,{\widehat G}_{ij}[\phi;g(\phi)]
\label{38}\ee
where ${\widehat G}_{ij}$ is a Liouville renormalization group invariant
function, i.e. a fixed fiducial metric on moduli space. This is exactly the
form of the Zamolodchikov metric for Liouville strings \cite{emnd,emnliouv}.
Thus there is an underlying Lagrangian dynamics in the non-critical string
problem.

The action \eqn{36c} allows canonical quantization, which as we have mentioned
is induced by including higher genus effects in the string theory
\cite{emnd,emn}. In the canonical quantization scheme the couplings $g^i$ and
their canonical momenta \eqn{momenta2} are replaced by quantum mechanical
operators (in target space) ${\widehat g}^i$ and $\widehat{p}_i$ obeying
\beq
\left[\!\left[\widehat{g}^i,\widehat{p}_j\right]\!\right]=i\hbar_{\cal
M}\,\delta_j^i
\label{triantaena2}\eeq
where the quantum commutator $[[\cdot\,,\,\cdot]]$ is defined on the moduli
space $\cal M$ of deformed conformal field theories of the form (\ref{smodel}),
and $\hbar_{\cal M}$ is an appropriate ``Planck constant". We can use the
Schr\"odinger representation in which the canonical momentum operators obey
\cite{emn}
\be
\Bigl\langle\widehat{p}_i\Bigr\rangle_{\rm
L}=\left\langle\mbox{$-i\frac{\delta}{\delta g^i}$}\right\rangle_{\rm
L}=\Bigl\langle V_i\Bigr\rangle_{\rm L}
\label{momenta}\ee
Thus the canonical commutation relation \eqn{triantaena2} in general yields, on
account of (\ref{momenta}), a non-trivial commutator between the
couplings $g^i$ and the associated vertex operators of the (genera resummed)
$\sigma$-models.

\newsection{Matrix $\sigma$-models and Fat Brane Dynamics}

To describe the moduli space dynamics of a multi D-brane system, we shall use
the construction described in \cite{richfed} which for the present purposes
lends the best physical interpretation. In this picture, the assembly of
D-branes, including all elementary string interactions, is regarded as a
composite `fat brane' which couples to a single fundamental string with a
matrix-valued coupling. In a T-dual (Neumann boundary conditions)
framework\footnote{For subtleties in applying the T-dual picture see
\cite{brecher,dornlast}. In this paper, as in
\cite{richfed}, we assume that the Neumann picture is the fundamental
picture to describe the propagation of strings in fat brane backgrounds. The
Dirichlet picture is then {\it derived} by applying T-duality as a canonical
functional integral transformation.}, the resulting effective theory is
described by a $\sigma$-model on an `effective' topology of a disc, propagating
in the background of a non-abelian $U(N)$ Chan-Paton gauge field.

Consider the $U(N)$-invariant matrix
$\sigma$-model action
\bea
S_N[X;A]&=&\frac1{4\pi\alpha'}\int_{\Sigma\{z_{ab}\}}d^2z~\tr~\eta_{\mu\nu}
\partial X^\mu\bar\partial
X^\nu-\frac1{2\pi\alpha'}\oint_{\partial\Sigma\{z_{ab}\}}
\tr~Y_i\left(x^0(s)\right)~dX^i(s)\nn\\&
&~~~~~~~~~~~~~~~~~~~~+\oint_{\partial\Sigma\{z_{ab}\}}\tr~
A^0\left(x^0(s)\right)~dX^0(s)
\label{Matrixaction}\eea
where $\eta_{\mu\nu}$ is a (critical) flat 9+1-dimensional spacetime metric.
The worldsheet fields $X$, $Y$ and $A$ are $N\times N$ Hermitian matrices which
transform in the adjoint representation of $U(N)$.\footnote{In this paper we
shall consider only the case of oriented open strings. For unoriented open
strings, the global symmetry group $U(N)$ is replaced with $O(N)$ everywhere.}
The traces in \eqn{Matrixaction} are taken in the fundamental
representation.\footnote{Repeated upper and lower spacetime indices, which are
raised and lowered with the flat metric $\eta_{\mu\nu}$, are always assumed to
be summed over. We also normalize the generators $T^a$ of $U(N)$ as
$\tr\,T^aT^b=\delta^{ab}$ and hence use the flat metric $\delta^{ab}$ to raise
and lower colour indices.} The surface $\Sigma\{z_{ab}\}$ is a sphere with a
set of marked points $z_{ab}$, $1\leq a,b\leq N$, on it. For each $a=b$ it has
the topology of a disc $\Sigma$, while for each pair $a\neq b$ it has the
topology of an annulus. The variable $s\in[0,1]$ parametrizes the circle
$\partial\Sigma$. In \cite{richfed} it was shown that the action
\eqn{Matrixaction} describes an assembly of $N$ parallel D-particles with
fundamental oriented open strings stretching between each pair of them. The
diagonal component $Y_{aa}$ of the matrix field $Y$ parametrizes the Dirichlet
boundary condition on D-particle $a$, while the off-diagonal component
$Y_{ab}=Y_{ba}^*$ represents the Dirichlet boundary condition for the
fundamental oriented open string whose endpoints attach to D-particles $a$ and
$b$. The matrix field $A^0$ parametrizes the usual Neumann boundary conditions
in the temporal direction of the target space. The action \eqn{Matrixaction} is
written in terms of Neumann boundary conditions on the configuration fields,
which is the correct description of the dynamics of the D-branes in
this way, but it is straightforward to apply a functional T-duality
transformation on the fields of \eqn{Matrixaction} to express it in the usual,
equivalent Dirichlet parametrization \cite{richfed}. The configuration
\beq
A^\mu=\left(A^0,-\mbox{$\frac1{2\pi\alpha'}$}\,Y^i\right)
\label{gaugefield}\eeq
can be interpreted as a ten-dimensional $U(N)$ isospin gauge field
dimensionally reduced to the worldline of the D-particle \cite{polchinski,kss}.

However, the action \eqn{Matrixaction} on its own does not properly take into
account the interactions between the D-particles and the fundamental strings.
To do so we must transform it in two ways \cite{richfed}. First, we must
include the sum over all worldsheet topologies, incorporating the Liouville
dressing discussed in the previous section. Due to the induced quantum
fluctuations of the couplings $Y_i^{ab}$, this provides an infinitesimal
separation between the $N$ constituent D-particles proportional to the string
coupling $g_s$ \footnote{Strictly speaking, it is a renormalized coupling
constant $g_s^{\rm ren}$ that appears -- see \cite{richfed} for details.} and
also allows the endpoints of the fundamental strings to fluctuate in spacetime.
We must then integrate out all the fluctuations among the fat brane
constituents, i.e. over all of the marked points of $\Sigma\{z_{ab}\}$. This
necessarily makes the action non-local. By $U(N)$-invariance, the resulting
$\sigma$-model partition function then becomes the expectation value, in a free
(scalar) $\sigma$-model, of the path-ordered $U(N)$ Wilson loop operator
$W[\partial\Sigma;A]$ along the boundary of the worldsheet disc $\Sigma$,
\bea
Z_N[A]&\equiv&\sum_{\rm
genera}~\int[dX]~\int_\Sigma\prod_{a,b=1}^Nd^2z_{ab}~\e^{-S_N[X;A]}
\nonumber\\&\simeq&\left\langle W[\partial\Sigma;A]\right\rangle_0~\equiv~\int
Dx~\e^{-N^2S_0[x]}~\tr~P\exp\left(ig_s\oint_{\partial\Sigma}A_\mu(x^0(s))
{}~dx^\mu(s)\right)\nn\\& &~~~~
\label{partfatbrane}\eea
where $dX$ is the normalized invariant Haar measure for integration on the Lie
algebra of $N\times N$ Hermitian matrices and
\beq
S_0[x]=\frac1{4\pi\alpha'}\int_\Sigma d^2z~\eta_{\mu\nu}\partial
x^\mu\bar\partial
x^\nu
\label{freesigma}\eeq
is the free $\sigma$-model action for the fundamental string. The path integral
measure $Dx$ is normalized so that $\langle1\rangle_0=1$. The partition
function \eqn{partfatbrane} describes the dynamics of a fat brane, which is
depicted in fig. 2.

\begin{figure}[htb]
\epsfxsize=1.5in
\bigskip
\centerline{\epsffile{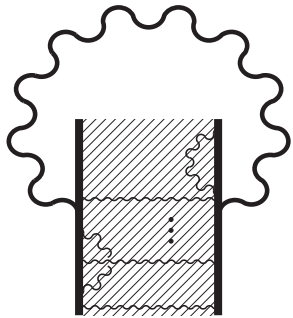}}
\caption{\it\baselineskip=12pt Schematic representation of a fat brane. The
bold strips denote the assembly of $N$ parallel D-branes and the thin wavy
lines represent the fundamental strings which start and end on them. The
shading represents the integration over all of these string interactions, as
well as the sum over worldsheet genera. The matrix $\sigma$-model describes the
interaction of the fat brane with a single fundamental string, represented by
the thick wavy line, which starts and ends on the fat brane with a
matrix-valued coupling constant $Y$.}
\bigskip
\label{fatbrane}\end{figure}

The low-energy effective action for the D-brane configurations is now obtained
by integrating out the fundamental string configurations $x$ in
\eqn{partfatbrane}. To lowest order in the gauge-covariant derivative
expansion, the result is $Z_N[A]\simeq\e^{-N^2\Gamma_{\rm NBI}[A]}$, where
\beq
\Gamma_{\rm NBI}[A]=\frac{c_0}{\sqrt{2\pi\alpha'}\,g_s}\int dt~\tr\left({\rm
Sym}+i\zeta\,{\rm
Asym}\right)\left(\det_{\mu,\nu}\left[\eta_{\mu\nu}I_N+2\pi\alpha'
g_s^2F_{\mu\nu}\right]\right)^{1/2}
\label{nbiaction}\eeq
is the non-abelian Born-Infeld action for the dimensionally-reduced gauge field
$A_\mu$. Here $c_0$ is a numerical constant and $t=x^0(s=0)$ is the worldsheet
zero-mode of the temporal embedding field. $I_N$ is the $N\times N$ identity
matrix, Sym denotes the symmetrized matrix product
\beq
{\rm Sym}(M_1,\dots,M_n)=\frac1{n!}\sum_{\pi\in S_n}M_{\pi_1}\cdots M_{\pi_n}
\label{Symdef}\eeq
and Asym is the antisymmetrized matrix product
\beq
{\rm Asym}(M_1,\dots,M_n)=\frac1{n!}\sum_{\pi\in S_n}({\rm
sgn}\,\pi)M_{\pi_1}\cdots M_{\pi_n}
\label{Asymdef}\eeq
The symmetric product (and similarly for the Asym operation) on functions
$f(M_1,\dots,M_n)$ of $n$ matrices $M_k$ is defined by first formally expanding
$f$ as a Taylor series and then applying the Sym operation to each monomial,
\beq
{\rm
Sym}\,f(M_1,\dots,M_n)=\sum_{k_1,\dots,k_n\geq0}\frac{f^{(k_1,\dots,k_n)}(0,
\dots,0)}{k_1!\cdots k_n!}\,~{\rm Sym}\left(M_1^{k_1},\dots,M_n^{k_n}\right)
\label{Symfdef}\eeq
The symmetrization and antisymmetrization operations have the effect of
removing the ambiguity in the definition of the spacetime determinant in
\eqn{nbiaction} for matrices with noncommuting entries.

The components of the field strength tensor in \eqn{nbiaction} are given by
\beq
2\pi\alpha'F_{0i}=\mbox{$\frac
d{dt}$}\,Y_i-ig_s[A_0,Y_i]~~~~~~,~~~~~~(2\pi\alpha')^2F_{ij}=g_s[Y_i,Y_j]
\label{fieldstrength}\eeq
and the constant $\zeta\in\real$ is left arbitrary so that it interpolates
among the proposals for the true trace structure inherent in the
non-abelian generalization of the Born-Infeld action. The case $\zeta=0$
corresponds to the original proposal in \cite{tseytlin} while the trace
structure with $\zeta=1$ was suggested (in a different context) in
\cite{argnappi}. In \cite{brecher} the two-loop worldsheet $\beta$-function for
the model \eqn{partfatbrane} was calculated to be
\beq
\beta_i^{ab}\equiv\frac{\partial
Y_i^{ab}}{\partial\log\Lambda}=-(2\pi\alpha'g_s)^2\left(D^\mu F_{\mu
i}\right)^{ab}+2(2\pi\alpha'g_s)^3\left(D^\mu\left[F_{\mu\nu},F^\nu_{~i}
\right]\right)^{ab}+{\cal O}\left((\alpha'g_s)^4\right)
\label{betamatrix2loop}\eeq
where
\beq
D_0=\mbox{$\frac d{dt}$}-ig_s\left[A_0,\cdot\,\right]~~~~~~,~~~~~~
D_i=\mbox{$\frac{ig_s}{2\pi\alpha'}$}\left[Y_i,\cdot\,\right]
\label{gcderivsred}\eeq
are the components of the dimensionally reduced gauge-covariant derivative. It
is readily seen that \eqn{betamatrix2loop} coincides with the variation of the
action \eqn{nbiaction} with $\zeta=1$ up to the order indicated in
\eqn{betamatrix2loop}, so that the worldsheet renormalization group equations
$\beta_i^{ab}=0$ coincide with the equations of motion of the D-branes. The
first term in \eqn{betamatrix2loop} yields the (reduced) Yang-Mills equations
of motion, while the second term represents the first order stringy correction
to the Yang-Mills dynamics. We shall return to this issue in the next section.

In this paper we will study the target space quantum dynamics from the
worldsheet $\sigma$-model point of view, which will provide dynamical
worldsheet origins for the noncommutativity of spacetime and matrix D-brane
dynamics in general. We shall study the simplest background of a
Galilean-boosted D-brane,
\be
Y_i(x^0)^{ab}=Y_i^{ab}+U_i^{ab}\,x^0
\label{boost}\ee
corresponding to the case of non-relativistic heavy D-particles. The velocity
matrix $U_i$ describes the velocities of the constituent D-branes in the fat
brane. Alternatively, the choice of couplings \eqn{boost} can be thought of as
parametrizing the action of the spacetime Euclidean group on the fat brane.
However, this background is trivial from the point of view of the dynamics of
the D-branes. In the Neumann picture the D-brane configurations are essentially
gauge fields, so the only part of \eqn{boost} which contributes to the action
\eqn{Matrixaction} is the velocity operator. But we can also Galilean transform
to the rest frame where $U_i=0$. We shall see explicitly in the next section
that the quantum dynamics determined by the configuration \eqn{boost} are
trivial.

The problem is resolved by considering again the genus expansion of the matrix
$\sigma$-model \eqn{Matrixaction}. An analysis of the annulus amplitude reveals
that there are logarithmic divergences arising from modular parameter
integrations of the form $\int dq/q$ \cite{kogmav}--\cite{recoil}. These
divergences can be removed by replacing the velocity operator in \eqn{boost} by
$\lim_{\epsilon\to0^+}U_i^{ab}D(x^0;\epsilon)$, where
\beq
D(x^0;\epsilon)=x^0\,\Theta(x^0;\epsilon)
\label{Depsilonop}\eeq
and
\beq
\Theta(s;\epsilon)=\frac1{2\pi
i}\int_{-\infty}^\infty\frac{dq}{q-i\epsilon}~\e^{iqs}
\label{Thetafnreg}\eeq
is the regulated step function, with
$\Theta(s)\equiv\lim_{\epsilon\to0^+}\Theta(s;\epsilon)=0$ for $s<0$ and
$\Theta(s)=1$ for $s>0$. The infinitesimal parameter $\epsilon$ regulates the
ambiguous value of $\Theta(s)$ at $s=0$, and the integral representation
\eqn{Thetafnreg} is used since $x^0$ will eventually be a quantum operator.
When this velocity term is inserted into the boundary integral of the
$\sigma$-model action, the $\epsilon\to0^+$ divergences arising from the
regulated step function can be used to cancel the logarithmic divergences of
the annulus amplitudes \cite{kogmav}--\cite{recoil}. This relates the target
space regularization parameter $\epsilon$ to the worldsheet ultraviolet scale
$\Lambda$ by~\cite{kmw}
\beq
\epsilon^{-2}=-2\alpha'\log\Lambda
\label{scalerel}\eeq
We shall describe these cancellations explicitly in section 6.

This new velocity operator is called the impulse operator \cite{recoil} and it
has non-zero matrix elements between different states of the fat brane. It
describes recoil effects from the emission or scattering of closed string
states off the fat brane, and in an impulse approximation, it ensures that
(classically) the fat brane starts moving only at time $x^0=0$. But this is not
all that is required. The operator \eqn{Depsilonop} on its own does not lead to
a closed conformal algebra. Computing its operator product expansion with the
stress-energy tensor shows \cite{kmw} that it is only the {\it pair} of
operators $D(x^0;\epsilon),C(x^0;\epsilon)$, where
\beq
C(x^0;\epsilon)=\epsilon\,\Theta(x^0;\epsilon)
\label{Cepsilonop}\eeq
that define a closed algebra under the action of the worldsheet stress-energy
tensor. They form a pair of logarithmic operators of the conformal field theory
\cite{gurarie}. Thus, in order to maintain conformal invariance of the
worldsheet theory, one cannot just work with the operator \eqn{Depsilonop},
because \eqn{Cepsilonop} will be induced by conformal transformations. If we
rescale the worldsheet cutoff
\beq
\Lambda\to\Lambda'=\Lambda\,\e^{-t/\sqrt{\alpha'}}
\label{Lambdarescale}\eeq
by a linear renormalization group scale $t$, then \eqn{scalerel} induces a
transformation
\beq
\epsilon^2\to\epsilon'^2=\frac{\epsilon^2}{1-4\sqrt{\alpha'}\,\epsilon^2\,t}
\label{epsilonrescale}\eeq
and we find
\beq
D(x^0;\epsilon')=D(x^0;\epsilon)+t\,\sqrt{\alpha'}\,C(x^0;\epsilon)~~~~~~,~~~~~~
C(x^0;\epsilon')=C(x^0;\epsilon)
\label{DCscaletransfs}\eeq
If we now modify the initial position of the fat brane to
$\lim_{\epsilon\to0^+}\sqrt{\alpha'}\,Y_i^{ab}C(x^0;\epsilon)$, then this scale
transformation
will induce, by conformal invariance, a transformation of the velocities and
positions as
\beq
U_i\to U_i~~~~~~,~~~~~~Y_i\to Y_i+U_i\,t
\label{Galileanboost}\eeq
i.e. a Galilean evolution of the fat brane in target space.

To properly incorporate non-trivial dynamics of the fat brane, one must
therefore consider instead of \eqn{boost} the recoil operator
\beq
Y_i(x^0)^{ab}=\lim_{\epsilon\to0^+}\left(\sqrt{\alpha'}\,
Y_i^{ab}C(x^0;\epsilon)+U_i^{ab}D(x^0;\epsilon)\right)
\label{recoilop}\eeq
The conformal algebra reveals that the operators \eqn{Depsilonop} and
\eqn{Cepsilonop} have the same conformal dimension \cite{kmw}
\beq
\Delta_\epsilon=-\alpha'|\epsilon|^2/2
\label{Deltaep}\eeq
which vanishes as $\epsilon\to0^+$. For finite $\epsilon$ the operator
\eqn{recoilop}, when inserted into the action \eqn{Matrixaction}, yields a
deformation operator of conformal dimension $1-\alpha'|\epsilon|^2/2$ which
therefore describes a relevant deformation of the $\sigma$-model and the
resulting string theory is non-critical. From \eqn{Galileanboost} it follows
that the corresponding matrix-valued $\beta$-functions are
\beq
\beta_{Y_i}=\Delta_\epsilon
Y_i+\sqrt{\alpha'}\,U_i~~~~~~,~~~~~~\beta_{U_i}=\Delta_\epsilon U_i
\label{betaepsilon}\eeq
As the ``dressing" by the operators $C$ and $D$ is determined entirely by the
temporal coordinate $x^0$, we identify this field as the Liouville field
$\varphi$. Marginality of the deformation is then restored by taking the limit
$\epsilon\to0^+$. In this sense, the gravitational dressing is provided by the
temporal embedding fields of the string, giving a natural interpretation to the
Liouville zero mode as the time coordinate $t=x^0(s=0)$ that appears in
\eqn{nbiaction}. The relation with the worldsheet renormalization scale is
then set by \eqn{scalerel}. Thus, if we consider the initial velocity matrix
$U_i$ of the fat brane as an unrenormalized coupling, then \eqn{recoilop} is
interpreted as the Liouville-dressed renormalized coupling contants \eqn{giphi}
of the matrix $\sigma$-model. We shall make this correspondence somewhat more
precise in section 6. Some properties of the correlators of the logarithmic
pair $C,D$, which will be required in the following, are described in appendix
A.

\newsection{Canonical Momentum of Collective D-brane Configurations}

In this section we shall compute, as prescribed in section 2, the canonical
momenta conjugate to the matrix-valued couplings $Y_i$ in the $\sigma$-model of
the previous section. The necessity to carry out this complicated calculation
is many-fold. For instance, we shall see that the perturbative theory requires
a renormalization of the D-brane couplings, which is unambiguously fixed by the
momentum. This will be important in the following sections where we shall map
the fat brane problem onto the Liouville string problem of section 2.
Furthermore, this quantity enables the most direct comparison with the
non-abelian Born-Infeld theory and illustrates the usage of the generic
features of logarithmic conformal field theory in the calculation of matrix
D-brane dynamical quantities.

\newsubsection{Perturbation Expansion}

We shall need a proper path integral representation of the
$U(N)$ Wilson loop operator, representing the pertinent vertex operator for the
description of a system of $N$ D-branes in the $\sigma$-model framework. We
introduce one-dimensional complex auxilliary fields $\bar\xi_a(s),\xi_a(s)$ on
the boundary $\partial\Sigma$ of the worldsheet. They transform in the
fundamental representation of the $U(N)$ gauge group, and their propagator is
\bea
\left\langle\!\left\langle\bar\xi_a(s_1)\xi_b(s_2)\right\rangle\!
\right\rangle&\equiv&\lim_{\epsilon\to0^+}\int
D\bar\xi~D\xi~\bar\xi_a(s_1)\xi_b(s_2)\exp\left(-\sum_{c=1}^N
\int_0^1ds~\bar\xi_c(s-\epsilon)\frac
d{ds}\xi_c(s)\right)\nn\\&=&\delta_{ab}\,\Theta(s_2-s_1)
\label{auxprop}\eea
where again $\epsilon$ regulates the ambiguous value of $\Theta(s)$ at $s=0$.

Using the propagator \eqn{auxprop} and Wick's theorem we can undo the path
ordering in the Wilson loop operator in \eqn{partfatbrane} by writing it as
\cite{wilson}--\cite{dornlast}
\bea
& &W[\partial\Sigma;A]=\frac1{N\,{\cal W}_{U(1)}[\partial\Sigma;A]}\nn\\&
&\times\lim_{\epsilon\to0^+}\left\langle\!\!\left\langle\sum_{c=1}^N\bar
\xi_c(0)\exp\left(ig_s\sum_{a,b=1}^N\int_0^1ds~\bar\xi_a(s-\epsilon)
A_\mu^{ab}(x^0(s))\xi_b(s)\,
\frac d{ds}x^\mu(s)\right)\xi_c(1)\right\rangle\!\!\right\rangle
\nn\\& &~~~~\label{wilsonloop}\eea
This representation of the Wilson loop operator also requires a renormalization
scheme for the auxilliary quantum field theory which we describe in appendix B.
It puts the partition function \eqn{partfatbrane} into the form of a functional
integral over a local action. Note that it corresponds to the partition
function for the boundary fields $\bar\xi,\xi$ minimally coupled to the gauge
field $A_\mu$. The additional factor
\beq
{\cal W}_{U(1)}[\partial\Sigma;A]=\lim_{\epsilon\to0^+}
\left\langle\!\!\left\langle\exp\left(ig_s\sum_{a,b=1}^N\int_0^1ds~
\bar\xi_a(s-\epsilon)A_\mu^{ab}(x^0(s))\xi_b(s)\,\frac
d{ds}x^\mu(s)\right)\right\rangle\!\!\right\rangle
\label{U(1)factor}\eeq
is induced by the vacuum graphs of the auxilliary quantum field theory. With a
periodic definition of the step function $\Theta(s)$ on the circle
$\partial\Sigma$ (for instance with a discretized version of \eqn{Thetafnreg}),
the auxilliary fields induce loop contractions of the colour indices of the
gauge field $A_\mu$ leading to the $U(1)$ subgroup projection \eqn{U(1)factor}
of the Wilson loop operator.

This formalism gives a one-parameter family of Dirichlet boundary conditions
for the fundamental string fields, labelled by $s\in[0,1]$, in the
corresponding T-dual formalism \cite{dornotto,dorn}, i.e. the dual
configuration fields are
\beq
\widetilde{Y}_i(x^0;s)=\lim_{\epsilon\to0^+}\sum_{a,b=1}^N
\bar\xi_a(s-\epsilon)Y_i^{ab}(x^0(s))\xi_b(s)
\label{dualfields}\eeq
Now, instead of being forced to sit on a unique hypersurface as in the abelian
D-brane case, there are an infinite set of hypersurfaces on which the string
endpoints are situated. Alternatively, we obtain a one-parameter family of bare
matrix-valued vertex operators
\beq
V^i_{ab}(x;s)=\frac{ig_s}{2\pi\alpha'}\,\frac
d{ds}x^i(s)\lim_{\epsilon\to0^+}\bar\xi_a(s-\epsilon)\xi_b(s)
\label{barevertexops}\eeq
and renormalized matrix couplings \eqn{recoilop}. Thus the trade-off for
removing the non-locality of the effective theory \eqn{partfatbrane} is the
extra explicit boundary dependence of operators involved.

We will use the representation \eqn{wilsonloop} to compute the classical
canonical momentum $\Pi^j_{ab}(s)$ in the moduli space of the collective
D-brane configurations $Y_j^{ab}(s)$. According to \eqn{momenta}, the momentum
can be computed as the one-point function of the deformation vertex operators
\eqn{barevertexops} in the statistical ensemble \eqn{partfatbrane},
\bea
\Pi^j_{ab}(s)&\equiv&N\,{\cal
W}_{U(1)}[\partial\Sigma;A]\left(-\frac{\delta}{\delta
Y_j^{ab}(x^0(s))}Z_N[A]\right)\nn\\&=&\widetilde{\Pi}_{ab}^j(s)
-N\left\langle W[\partial\Sigma;A]\left(-\frac\delta
{\delta Y_j^{ab}(x^0(s))}{\cal
W}_{U(1)}[\partial\Sigma;A]\right)\right\rangle_0
\label{Pitot}\eea
where
\bea
\widetilde{\Pi}_{ab}^j(s)&=&\lim_{\epsilon\to0^+}
\sum_{c=1}^N~\Biggm\langle~\Biggm\langle\!\!\Biggm\langle
\bar\xi_c(0)\,V^j_{ab}(x;s)\nn\\& &~~~~\times\exp\left(ig_s
\sum_{d,e=1}^N\int_0^1ds'~\bar\xi_d(s'-\epsilon)
A_\mu^{de}(x^0(s'))\xi_e(s')\,\frac
d{ds'}x^\mu(s')\right)\xi_c(1)\Biggm\rangle\!\!\Biggm\rangle~\Biggm\rangle_0
\nn\\& &~~~~\label{Pidef}\eea
is the contribution from the $SU(N)$ part of the gauge group. The second term
in \eqn{Pitot} involves traces of the gauge field $A_\mu$ which we identify
as the center of mass coordinates of the fat brane, i.e. $Y_j^{\rm
cm}\equiv\frac1N\,\tr\,Y_j$. The expression \eqn{Pitot} thus shows that the
momenta of the collective center of mass motion of the fat brane and of the
constituent D-branes comprising the fat brane completely decouple. In this
paper we shall be interested in only the former contribution, since the latter
one essentially represents the dynamics of a single D-brane (i.e. gauge group
$U(1)$) and here we are interested in the non-abelian modification determined
by the constituent D-particles. In effect we restrict attention to unimodular
Wilson loops (i.e. gauge group $SU(N)$). For these terms the statistics of the
auxilliary boundary fields $\bar\xi,\xi$ are irrelevant.

{}From now on we shall work in the static gauge $A_0=0$ for the dimensionally
reduced gauge field. Then the canonical momentum \eqn{Pidef} can be expanded as
the power series
\beq
\widetilde{\Pi}_{ab}^j(s)=-\sum_{n=1}^\infty\frac{(-i)^{n+1}}
{n!}\left(\frac{g_s}{2\pi\alpha'}\right)^{n+1}~{\cal P}_{ab}^{(n)j}[Y;s]
\label{Pipowerseries}\eeq
where the ${\cal O}(Y(x^0)^n)$ contribution is
\bea
& &{\cal P}_{ab}^{(n)j}[Y;s]=\lim_{\epsilon\to0^+}
\sum_{c=1}^N~\sum_{{\buildrel{a_1,\dots,a_n}\over
{b_1,\dots,b_n}}}~\int_0^1\prod_{k=1}^nds_k~\Biggl\langle\!\!\Biggl\langle
\bar\xi_c(0)\bar\xi_a(s-\epsilon)
\xi_b(s)\biggr.\biggr.\nn\\& &~~~~~~\Biggl.\Biggl.\times
\left(\prod_{k=1}^n\bar\xi_{a_k}(s_k-\epsilon)
\xi_{b_k}(s_k)\right)\xi_c(1)\Biggr\rangle\!\!\Biggr\rangle\left\langle\frac
d{ds}x^j(s)\prod_{k=1}^nY_{i_k}^{a_kb_k}\left(x^0(s_k)\right)\frac
d{ds_k}x^{i_k}(s_k)\right\rangle_0
\nn\\& &~~~~\label{Piordern}\eea
The correlation functions appearing in \eqn{Piordern} can be evaluated using
Wick's theorem and the propagator \eqn{auxprop} to write the auxilliary field
averages as a sum over permutations,
\beq
\left\langle\!\!\left\langle\prod_{k=1}^m\bar\xi_{a_k}(s_k)\xi_{b_k}(s'_k)
\right\rangle\!\!\right\rangle=\sum_{P\in
S_m}~\prod_{k=1}^m\delta_{a_k,b_{P(k)}}\,\Theta\left(s'_{P(k)}-s_k\right)
\label{etaavgs}\eeq

The evaluation of the momentum contribution \eqn{Piordern} is rather
technically involved and is presented in appendix B. It is also shown there
that one must further specify a renormalization of the auxilliary quantum field
theory in order to remove step function ambiguities which come from the
correlation functions \eqn{etaavgs}. The resulting renormalized expression is
finite at $s=0$. This point defines the (renormalized) target space coordinates
$\widetilde{\Pi}_{ab}^j\equiv\widetilde{\Pi}_{ab}^j(s=0)^{\rm ren}$ as the zero
modes of the worldsheet fields. As shown in appendix B, the order $n$
contribution is given by
\beq
{\cal P}_{ab}^{(n)j}[Y;0]^{\rm
ren}=\int_0^1\prod_{k=1}^nds_k~\left\langle\frac{dx^j(s)}{ds}
\biggm|_{s=0}~{\rm Sym}\left[\prod_{k=1}^nY_{i_k}\left(x^0(s_k)\right)\,\frac
d{ds_k}x^{i_k}(s_k)\right]_{ba}\right\rangle_0
\label{Pirenordern}\eeq
This is to be compared with the corresponding expression in the abelian case
(corresponding to a single D-particle, $N=1$) for which there is no
matrix-ordering problem and the expansion of the abelian Wilson loop operator
proceeds directly without the need of an auxilliary field representation. We
see that the properly renormalized momentum \eqn{Pirenordern} is a natural
non-abelian generalization of the corresponding single D-particle quantity, to
which it reduces in the limit $N=1$. Physically, the symmetrization of the
amplitude occurs because the correlators involve bosonic fields.

In the following we will, for simplicity, normalize $\eta^{00}=1$ and assume
that the target space temporal and spatial embedding fields are uncorrelated,
i.e. $\eta^{0i}=\eta^{i0}=0$. The resulting time-space factorization of
correlators implies that \eqn{Pirenordern} is non-vanishing only when $n$ is
odd. Note that for the configurations \eqn{boost} we have
$\widetilde{\Pi}^i\equiv0$, since all the periodic boundary integrations in
\eqn{Pirenordern} then vanish. When the fat brane configuration is given by the
non-trivial recoil operator \eqn{recoilop}, the correlation functions of the
logarithmic operators $C$ and $D$ can be evaluated using the results of
appendix A. In particular, the correlation functions involving only
$C(x^0;\epsilon)$
operators vanish as $\epsilon\to0^+$. This means that the canonical momentum
vanishes at zero velocities, as expected from physical considerations. It is a
nontrivial function which mixes the velocities and positions of the fat brane.
This explicit vanishing of the correlators of the $C$ operator is required to
consistently yield the correct abelian limit in which the momentum depends only
on the velocity.

\newsubsection{Velocity Renormalization}

In this subsection we will consider the lowest non-trivial contribution
\eqn{Pirenordern}, which using \eqn{derivprop} and \eqn{1ptfns} can be written
as
\bea
{\cal P}_{ab}^{(1)j}(Y,U;0)^{\rm
ren}&=&\lim_{\epsilon\to0^+}[U_i]_{ba}\int_0^1ds'~\left\langle
D(s';\epsilon)\right\rangle_0\left\langle\frac d{ds}x^j(s)\biggm|_{s=0}\,\frac
d{ds'}x^i(s')\right\rangle_0\nn\\&=&\lim_{\epsilon\to0^+}
\frac{4\pi^2a\alpha'}\epsilon~U^j_{ba}~\frac1{\pi\tan\pi s_\Lambda}
\label{order1def}\eea
where $a$ is an arbitrary constant and $s_\Lambda$ is the short-distance cutoff
\eqn{bdrycutoff} on $\partial\Sigma$. The divergent term as $\Lambda\to0$ is
the boundary version of the bulk logarithmic divergence $\log\Lambda$. Naively,
the boundary integral in \eqn{order1def} vanishes since its integrand is a
total
derivative. However, the one-point function of the logarithmic $D$ operator is
divergent as $\epsilon\to0^+$ and one must therefore carefully regularize the
boundary integration. The boundary regulator $s_\Lambda$ is also correlated
with the target space regularization parameter $\epsilon$ as in the bulk
equation \eqn{scalerel}. Although $s_\Lambda$ is given explicitly by
\eqn{bdrycutoff}, we shall assume that the bulk and boundary cutoffs are
independent and take $\tan\pi s_\Lambda\sim\epsilon^2$. This usage of the
logarithmic correlation functions will be the key feature in the determination
of the matrix D-brane dynamics.

The resulting expression \eqn{order1def} diverges as $\epsilon\to0^+$. Part of
this divergence can be removed by renormalizing the velocity matrix of the
D-branes as
\beq
U_i=\sqrt{\alpha'}\,\epsilon\,\bar U_i
\label{renvelocity}\eeq
{}From \eqn{scalerel} and \eqn{betaepsilon} we see that this renormalized
coupling constant is truly marginal,
\beq
\frac{d\bar U_i}{dt}=0
\label{Utrmarg}\eeq
where $t=-\sqrt{\alpha'}\log\Lambda$, and it therefore plays the role of a
uniform velocity for the fat brane dynamics. From \eqn{Pipowerseries} we see
that the remaining $\epsilon^{-2}$ divergence can be absorbed into a
renormalization of the string coupling constant as
\beq
g_s=\sqrt{\alpha'}\,\epsilon\,\bar g_s
\label{gsren}\eeq
As we will see in section 6, $\bar g_s$ is also a truly marginal coupling. Thus
we see that, after a suitable renormalization of the logarithmic deformation,
the leading order contribution to the canonical momentum \eqn{Pipowerseries} is
just the constant velocity of the Galilean boosted fat brane, which coincides
with the corresponding result for a single non-relativistic heavy D-particle
\cite{lm}.

\newsubsection{Logarithmic Algebra}

We now examine the leading order corrections to the velocity of the fat brane,
which are given by
\bea
& &{\cal P}_{ab}^{(3)j}(Y,U;0)^{\rm
ren}=\lim_{\epsilon\to0^+}\int_0^1ds_1~ds_2~ds_3~\left\langle \mbox{$\frac
d{ds}$}\,x^j(s)\Bigm|_{s=0}\,\mbox{$\frac d{ds_1}$}\,x^{i_1}(s_1)\,\mbox{$\frac
d{ds_2}$}\,x^{i_2}(s_2)\,\mbox{$\frac
d{ds_3}$}\,x^{i_3}(s_3)\right\rangle_0\nn\\& &~~~~~~~~~~\times~{\rm
Sym}\Bigl[\alpha'\left\langle
C(s_1;\epsilon)C(s_2;\epsilon)D(s_3;\epsilon)\right\rangle_0\,\left(Y_{i_1}
Y_{i_2}U_{i_3}+Y_{i_1}U_{i_3}Y_{i_2}+U_{i_3}Y_{i_2}Y_{i_1}\right)\nn\\&
&~~~~~~~~~~~~~~~~~~~~+\,\sqrt{\alpha'}\left\langle
C(s_1;\epsilon)D(s_2;\epsilon)D(s_3;\epsilon)\right\rangle_0\,\left(Y_{i_1}
U_{i_2}U_{i_3}+U_{i_2}Y_{i_1}U_{i_3}+U_{i_3}U_{i_2}Y_{i_1}\right)\nn\\&
&~~~~~~~~~~~~~~~~~~~~+\left\langle
D(s_1;\epsilon)D(s_2;\epsilon)D(s_3;\epsilon)\right\rangle_0
\,U_{i_1}U_{i_2}U_{i_3}\Bigr]_{ba}
\label{order3def}\eea
Using Wick's theorem, the propagator \eqn{derivprop} and the
three-point functions \eqn{3ptCCD}--\eqn{3ptDDD} of the logarithmic operators,
after some lengthy tedious algebra we can write \eqn{order3def} as
\bea
& &{\cal P}_{ab}^{(3)j}(Y,U;0)^{\rm
ren}=\lim_{\epsilon\to0^+}\left(4\pi^2\alpha'\right)^2\left[\alpha'
\left(d\epsilon+c\epsilon^3
\alpha'\log\Lambda\right)I_0\left(6\,Y^jY_iU^i+3\,Y_iY^iU^j\right.\right.\nn\\&
&~~~~~~+\,3\,Y^j\left[U_i,Y^i\right]+Y_i\left[U^j,Y^i\right]+\left[Y_iU^i,Y^j
\right]+\left[U^j,Y_iY^i\right]+\left[U_i,Y^j\right]Y^i\nn\\&
&~~~~~~\left.+\left[U_iY^i,Y^j\right]+\left[Y_i,Y^j\right]U^i\right)+
\mbox{$\frac
c2$}\,\epsilon^3\alpha'^2I_u^{(1)}\left(3\,Y_iY^iU^j-6\,Y^jY_iU^i-\left[Y_i,Y^j
\right]U^i\right.\nn\\& &~~~~~~+\,
Y_i\left[U^j,Y^i\right]-\left[Y_iU^i,Y^j\right]+\left[U^j,Y_iY^i\right]-
\left[U_i,Y^j\right]Y^i-\left[U_iY^i,Y^j\right]\nn\\&
&~~~~~~\left.-\,3\,Y^j\left[U_i,Y^i\right]\right)-
c\epsilon^3\alpha'^2I_c^{(1)}\left(3\,Y_iY^iU^j+Y_i\left[U^j,Y^i\right]+
\left[U^j,Y_iY^i\right]\right)\nn\\& &~~~~~~+\,\sqrt{\alpha'}\left(\mbox{$\frac
e\epsilon$}\,+2d\epsilon\alpha'\log\Lambda+c\epsilon^3\alpha'^2(\log\Lambda)^2
\right)I_0\left(6\,Y_iU^iU^j+3\,Y^jU_iU^i\right.\nn\\&
&~~~~~~+\,Y_i\left[U^j,U^i\right]+\left[U^j,Y_iU^i\right]+
U_i\left[U^j,Y^i\right]+3\left[U_i,Y^i\right]U^j+\left[U^j,U_iY^i\right]\nn\\&
&~~~~~~\left.+\left[U_iU^i,Y^j\right]+\left[U_i,Y^j\right]U^i
\right)+\sqrt{\alpha'}
\left(c\epsilon^3\alpha'^2I_q^{(1)}-(d\epsilon\alpha'+c\epsilon^3
\alpha'^2\log\Lambda)I_u^{(1)}\right.\nn\\& &~~~~~~\left.+\,\mbox{$\frac
c4$}\,\epsilon^3\alpha'^2I_u^{(2)}\right)\left(3\,Y^jU_iU^i+\left[U_i,Y^j
\right]U^i+\left[U_iU^i,Y^j\right]
\right)+\sqrt{\alpha'}\left(\mbox{$\frac
c2$}\,\epsilon^3\alpha'^2\left(I_q^{(2)}+I_q^{(3)}\right)\right.\nn\\&
&~~~~~~\left.-(d\epsilon\alpha'+c\epsilon^3\alpha'^2\log
\Lambda)I_c^{(1)}+\mbox{$\frac c4$}\,\epsilon^3\alpha'^2I_c^{(2)}
\right)\left(6\,Y_iU^iU^j+Y_i\left[U^j,U^i\right]+
\left[U^j,Y_iU^i\right]\right.\nn\\&
&~~~~~~\left.+\,U_i\left[U^j,Y^i\right]+3\left[U_i,Y^i\right]U^j+
\left[U^j,U_iY^i\right]\right)\nn\\&
&~~~~~~-\left(U_iU^iU^j+U_iU^jU^i+U^jU_iU^i\right)\left\{I_0\left(
\mbox{$\frac f{\epsilon^3}$}\,+\mbox{$\frac{3e}\epsilon$}\,\alpha'\log\Lambda+
3d\epsilon\alpha'^2(\log\Lambda)^2\right.\right.\nn\\&
&~~~~~~\left.+\,c\epsilon^3\alpha'^3(\log\Lambda)^3
\right)-\left(I_u^{(1)}+2I_c^{(1)}\right)\left(\mbox{$\frac
e{2\epsilon}$}+d\epsilon\alpha'^2\log\Lambda+\mbox{$\frac
c2$}\,\epsilon^3\alpha'^3(\log\Lambda)^3\right)\nn\\&
&~~~~~~+\,\mbox{$\frac34$}
\left(2I_u^{(2)}+I_c^{(2)}\right)\left(d\epsilon\alpha'^2-c\epsilon^3\alpha'^3
\log\Lambda\right)+c\epsilon^3\alpha'^3
\left(\left(I_m^{(1)}+\mbox{$\frac12$}\,I_m^{(2)}\right)\log\Lambda\right.
\nn\\& &~~~~~~\left.\left.\left.-\,\mbox{$\frac12$}\,
I_t^{(1)}-\mbox{$\frac34$}\left(I_t^{(2)}+I_t^{(3)}+I_t^{(4)}\right)+
\mbox{$\frac18$}\left(2I_t^{(5)}+I_t^{(6)}\right)\right)\right\}\right]_{ba}
\label{order3long}\eea
The quantities denoted by $I$ in \eqn{order3long} are the various boundary
integrals that arise and are summarized in appendix C. The constants
$c,d,\dots$ come from the correlation functions of the logarithmic operators.
These constants are for the most part arbitrary integration constants, the
remaining ones being fixed by the leading logarithmic terms in the conformal
blocks. We shall eliminate the arbitrary ones by demanding that, in the limit
$N=1$, \eqn{order3long} reproduce the appropriate result anticipated from
abelian Born-Infeld theory, i.e. that only the $U^jU^2$ term in
\eqn{order3long} survives in the abelian reduction. In doing so, we assume a
more general logarithmic deformation structure than that given by the recoil
operators of the previous section, but the qualitative (and most quantitative)
features remain the same.

Let us start with the first set of $Y^2U$ type terms. From the discussion of
the previous subsection and \eqn{scalerel} it follows that the bulk and
boundary ultraviolet cutoff scales are related as
\beq
4\mu\log\Lambda=\frac1{\tan\pi s_\Lambda}
\label{bulkbdryrel}\eeq
where $\mu$ is a real-valued constant to be determined. Using
\eqn{ic1final}--\eqn{iu1}, it then follows that the $Y^2U$ part of
\eqn{order3long} reduces to
\bea
& &\lim_{\epsilon\to0^+}\frac{4\alpha'}
{\pi^3}\left(4\pi^2\alpha'\right)^2\frac{\log\Lambda}{\tan\pi
s_\Lambda}\left[\left(d\epsilon+c\epsilon^3
\alpha'\log\Lambda\right)\left(6\,Y^jY_iU^i+3\,Y_iY^iU^j\right.\right.\nn\\&
&~~~~~~+\,3\,Y^j\left[U_i,Y^i\right]+Y_i\left[U^j,Y^i\right]+\left[Y_iU^i,Y^j
\right]+\left[U^j,Y_iY^i\right]+\left[U_i,Y^j\right]Y^i\nn\\&
&~~~~~~\left.+\left[U_iY^i,Y^j\right]+\left[Y_i,Y^j\right]U^i\right)
-c\epsilon^3\alpha'\log\Lambda\left((3-6\mu)\,Y_iY^iU^j-6\,Y^jY_iU^i\right.
\nn\\& &~~~~~~-\left[Y_i,Y^j\right]U^i+(1-2\mu)Y_i\left[U^j,Y^i\right]-
\left[Y_iU^i,Y^j\right]+(1-2\mu)\left[U^j,Y_iY^i\right]\nn\\&
&~~~~~~\left.\left.-\left[U_i,Y^j\right]Y^i-\left[U_iY^i,Y^j\right]-3\,Y^j
\left[U_i,Y^i\right]\right)\right]_{ba}
\label{YYUterms}\eea
In the abelian limit $N=1$, all commutators in \eqn{YYUterms} vanish. Requiring
that the coefficients of the $Y^jY_iU^i$ and $Y_iY^iU^j$ terms vanish leads,
respectively, to the equations
\beq
6d\epsilon+12c\epsilon^3\alpha'\log\Lambda=0~~~~~~,~~~~~~3d\epsilon+6\mu
c\epsilon^3\alpha'\log\Lambda=0
\label{YYU0eqs}\eeq
which for finite $\epsilon$ have unique solution
\beq
\mu=1~~~~~~,~~~~~~d=-2c\epsilon^2\alpha'\log\Lambda
\label{YYUsolns}\eeq
Here we have used the fact that the constant $c$ is determined by the leading
logarithmic terms in the conformal blocks of the logarithmic conformal algebra
generated by the $C$ and $D$ operators, and hence that $c\neq0$. We see that
the arbitrariness of certain integration constants which appear from the
logarithmic conformal algebra can be fixed by the appropriate abelian reduction
requirement. Substituting \eqn{YYUsolns} into \eqn{YYUterms} we see that the
set of $Y^2U$ type terms in fact vanishes identically.

Next we examine the second set of $YU^2$ type terms in \eqn{order3long}. Using
\eqn{bulkbdryrel}, \eqn{YYUsolns}, the integrals \eqn{ic1}--\eqn{iq3} and
dropping those terms which vanish as $\epsilon\to0^+$ relative to the rest, we
arrive after some algebra at the expression
\bea
& &\lim_{\epsilon\to0^+}\frac{4\sqrt{\alpha'}}
{\pi^3}\left(4\pi^2\alpha'\right)^2\frac{\log\Lambda}{\tan\pi
s_\Lambda}\left[\left(\mbox{$\frac
e\epsilon$}-\mbox{$\frac53$}\,c\epsilon^3\alpha'^2(\log\Lambda)^2\right)
\left(6\,Y_iU^iU^j+3\,Y^jU_iU^i\right.\right.\nn\\&
&~~~~~~+\,Y_i\left[U^j,U^i\right]+\left[U^j,Y_iU^i\right]+
U_i\left[U^j,Y^i\right]+3\left[U_i,Y^i\right]U^j+\left[U^j,U_iY^i\right]\nn\\&
&~~~~~~\left.\left.+\left[U_iU^i,Y^j\right]+\left[U_i,Y^j\right]U^i
\right)\right]_{ba}
\label{YUUterms}\eea
The reproduction of the correct abelian limit requires the equality
\beq
e=\mbox{$\frac53$}\,c\epsilon^4\alpha'^2(\log\Lambda)^2
\label{YUUsolns}\eeq
of the parameters of the logarithmic conformal algebra. As in \eqn{YYUterms},
this restriction forces the entire contribution \eqn{YUUterms} to vanish
identically for all $N$.

Thus, with the parameters of the logarithmic deformations fixed according to
\eqn{YYUsolns} and \eqn{YUUsolns}, the only contribution to the $n=3$ canonical
momentum is from the cubic velocity terms in \eqn{order3long}, which we
evaluate using \eqn{bulkbdryrel} and the boundary integrals
\eqn{ic1}--\eqn{it6}. Using \eqn{scalerel} and absorbing the remaining
$\epsilon^{-7}$ divergence in the total momentum \eqn{Pipowerseries} using the
renormalizations \eqn{renvelocity} and \eqn{gsren}, we arrive finally at
\beq
{\cal P}_{ab}^{(3)j}(Y,U;0)^{\rm ren}=-128\pi\alpha'^{3/2}
\left(f+\mbox{$\frac{139}8$}\,c\right)\left[\bar U_i\bar U^i\bar
U^j+\bar U_i\bar U^j\bar U^i+\bar U^j\bar U_i\bar U^i\right]_{ba}
\label{order3final}\eeq
The sum of \eqn{order1def} and \eqn{order3final} now involve three constants
$a$, $c$ and $f$ determined from the logarithmic conformal algebra. We can fix
another one of them by requiring that, again in the abelian limit, one recovers
the well-known result predicted from abelian Born-Infeld theory. One finds (see
the next subsection) that the relative coefficient between the $\bar U^j$ and
$\bar U^j\bar U_i\bar U^i$ terms in the abelian theory should be $\frac12$,
which imposes the additional constraint
\beq
8f+139c=64\pi^2a
\label{relconstr}\eeq

The results above now yield the total canonical momentum
\eqn{Pipowerseries} up to order 3 as
\beq
\widetilde{\Pi}_{ab}^j(\bar Y,\bar U)=\frac{4a\bar
g_s^2}{\pi\sqrt{\alpha'}}\left[\bar U^j+\frac{\bar g_s^2}6\left(3\,\bar U_i\bar
U^i\bar U^j+\left[\bar
U_i,\left[\bar U^j,\bar U^i\right]\right]\right)\right]_{ba}+{\cal O}\left(\bar
g_s^6\right)
\label{tildePiorder3}\eeq
The expression \eqn{tildePiorder3} involves one parameter $a$ determined by the
one-point function of the logarithmic $D$ operator. The remaining parameters of
the logarithmic conformal algebra that enter into the three-point functions
\eqn{3ptCCC}--\eqn{3ptDDD} are determined by \eqn{YYUsolns}, \eqn{YUUsolns} and
\eqn{relconstr}. In this way the matrix D-brane dynamics fixes most of the
algebraic information about the logarithmic deformation and localizes the
problem to a small region of moduli space. The fact that these
parameters are scale-dependent is a general feature of logarithmic conformal
field theories \cite{gflcft}. Note that they become scale-independent though
with the correlation \eqn{scalerel}.

\newsubsection{Canonical Momentum in Non-abelian Born-Infeld Theory}

Let us now compare the perturbative result \eqn{tildePiorder3} to that which
comes from the non-abelian Born-Infeld action \eqn{nbiaction}. For this, we
expand the spacetime determinant in \eqn{nbiaction} as a series in powers of
$F_{\mu\nu}$ to get
\bea
&
&\frac1{\sqrt{2\pi\alpha'}\,g_s}\left(\det_{\mu,\nu}\left[\eta_{\mu\nu}I_N+2
\pi\alpha'g_s^2F_{\mu\nu}\right]\right)^{1/2}\nn\\&
&~~~~~=~\left(\sqrt{2\pi\alpha'}\,g_s\right)^3\left[(2\pi\alpha'g_s^2)^{-2}I_N+
\frac14F_{\mu\nu}F^{\mu\nu}+\frac1{12}(2\pi\alpha'g_s^2)
\left[F_{\mu\nu},F^{\nu\lambda}\right]F_\lambda^{~\mu}\right.\nn\\&
&~~~~~~~~~~\left.+\frac1{32}(2\pi\alpha'g_s^2)^2\left(\left(F_{\mu\nu}
F^{\mu\nu}\right)^2-4F_{\mu\nu}F^{\nu\lambda}F_{\lambda\rho}F^{\rho\mu}\right)
+{\cal O}\left((2\pi\alpha'g_s^2)^3\right)\right]
\label{detexpansion}\eea
Since $F_{\mu\nu}=-F_{\nu\mu}$, the symmetrization operation picks out the even
powers of the field strength while the antisymmetrized product picks out the
odd ones. Using \eqn{fieldstrength} in the gauge $A_0=0$, after some algebra we
find that the expansion of the action \eqn{nbiaction} to leading orders in the
string coupling constant is
\bea
\Gamma_{\rm NBI}[Y]&=&c_0\left(\sqrt{2\pi\alpha'}\,g_s\right)^3\int
dt~\left(N(2\pi\alpha'g_s^2)^{-2}+\frac12\left(\frac{g_s}{2\pi\alpha'}
\right)^2\tr~\dot Y_i\dot Y^i\right.\nn\\&
&\left.+\frac1{16}\left(\frac{g_s^2}{2\pi\alpha'}\right)^2\left\{\tr\left(
2\,\dot Y_i\dot Y^i\dot Y_j\dot Y^j+\dot Y_i\dot Y_j\dot Y^i\dot
Y^j\right)+\frac{i\zeta}3\frac1{2\pi\alpha'}\,\tr\Bigl[Y_i,Y_j\Bigr]
\left[\dot Y^i,\dot Y^j\right]\right\}\right.\nn\\& &\Biggl.+\,{\cal
O}\left(g_s^6\right)\Biggr)
\label{nbiexpansion}\eea
where $\dot Y_i\equiv \frac d{dt}Y_i$. The perturbative expansion of the
canonical momentum in non-abelian Born-Infeld theory can now be calculated from
\eqn{nbiexpansion} and after some algebra we find
\bea
\Pi_{ab}^j(t)&\equiv&\frac{\delta\left(\mbox{$\frac1{g_s^3}$}\,\Gamma_{\rm
NBI}[Y]\right)}{\delta\dot
Y^{ab}_j(t)}\nn\\&=&\frac{c_0\,g_s^2}{\sqrt{2\pi\alpha'}}\left\{\dot
Y_{ba}^j+\frac{g_s^2}6\left(\left[\dot Y_i\dot Y^i\dot Y^j+\dot Y_i\dot Y^j\dot
Y^i+\dot Y^j\dot Y_i\dot Y^j\right]_{ba}+\frac{i\zeta}{2\pi\alpha'}\left[\dot
Y_i,\left[Y^j,Y^i\right]\right]_{ba}\right)\right\}\nn\\& &+\,{\cal
O}\left(g_s^6\right)
\label{nbimom}\eea

In particular, for the case of the D-particle configurations \eqn{boost}
corresponding to a Galilean-boosted fat brane, we have
\bea
\Pi_{ab}^j(t)&=&\frac{c_0\,g_s^2}{\sqrt{2\pi\alpha'}}\left(U_{ba}^j+
\frac{g_s^2}6\left[U_iU^iU^j+U_iU^jU^i+U^jU_iU^i\right.\right.\nn\\& &~~~~~~+
\frac{i\zeta}{2\pi\alpha'}\left\{\left[U_i,\left[Y^j,Y^i
\right]\right]+\left(\left[U_i,\left[U^j,Y^i\right]\right]+\left[U_i,
\left[Y^j,U^i\right]\right]\right)t\right.\nn\\&
&~~~~~~\left.\left.\left.+\left[U_i,\left[U^j,U^i\right]\right]t^2
\right\}\right]_{ba}\right)+{\cal O}\left(g_s^6\right)
\label{nbiboost}\eea
We see that the canonical momenta \eqn{tildePiorder3} and \eqn{nbiboost} agree,
up to an overall normalization, when
\beq
\zeta=0
\label{lambda0}\eeq
which corresponds to taking only the symmetrized trace in \eqn{nbiaction}. The
possible occurence of the extra antisymmetrized trace structure in
\eqn{nbiaction} was pointed out in \cite{brecher} where the worldsheet
$\beta$-functions \eqn{betamatrix2loop} were computed. As noted there, however,
when one properly takes into account the worldsheet fermionic fields for the
full superstring theory, it is only the symmetrized trace structure that
survives. This feature was elucidated on in \cite{brecher1} where it was shown
that the symmetrized action is the only potential generalization for which BPS
configurations linearize the non-abelian Born-Infeld action and minimize its
energy. Here we have shown that, within the auxilliary field formalism for the
worldsheet matrix $\sigma$-model, there exists a particular regularization of
the auxilliary quantum field theory which agrees with the results predicted by
the symmetrized action, without the need of introducing worldsheet
supersymmetry.

There may of course be other regularizations of the auxilliary quantum field
theory which reproduce the antisymmetrized trace structure in \eqn{nbiaction},
but we have not been able to find any such one. The renormalization described
in appendix B is the most natural scheme that one can impose and the
symmetrized matrix products which occur are natural from the perspective of
representing bosonic string amplitudes. It is also that which naturally leads
to the correct abelian reduction of the theory. The full, unrenormalized
expression for the canonical momentum in the matrix $\sigma$-model is given in
appendix B. To further check the validity of the non-abelian Born-Infeld
action,
one would need to extend the calculation of ${\cal P}_{ab}^{(n)j}(Y,U;0)^{\rm
ren}$ up to $n=5$. This in turn would require explicit knowledge of the
five-point correlation functions of the logarithmic operators, which are
extremely complicated (see appendix A), and the calculations at higher orders
of perturbation theory become overwhelmingly tedious and difficult to manage.

In any case, the results of this section illustrate a non-trivial application
of logarithmic conformal field theory to the study of solitonic states in
string theory. We note that the results derived in this section are invariant
under T-duality transformations of the string theory. In \cite{dornlast} it was
pointed out that an alternative functional integral representation of the
quantum D-particle dynamics is given by a $\sigma$-model action defined
with a non-abelian Wilson loop operator that has normal boundary derivatives
$\partial_\perp x^i$ for the relevant deformation vertex operators. This model
corresponds to the imposition of dynamical Dirichlet boundary conditions,
rather than dynamical Neumann ones as in \eqn{partfatbrane} which are
equivalent (by T-duality) to the imposition of external Dirichlet boundary
conditions. In contrast to the abelian case, these two models are inequivalent
beyond tree-level because of anomalous Jacobian factors in the path integral
measure which arise in the non-abelian case. By careful investigation of the
worldsheet $\beta$-functions it has been argued in \cite{dornlast} (see also
\cite{brecher}) that the model with dynamical Dirichlet boundary conditions
constitutes the appropriate T-dual description of the quantum D-brane dynamics
represented by the open string model with free (Neumann) boundary conditions.
It is straightforward to see that the perturbative expansion of the canonical
momentum in the theory with boundary normal derivatives is equivalent to the
one employed in this section, since the boundary correlation functions involved
are the same. The results described in this section are therefore independent
of which picture one chooses to work in.

\newsection{Dynamics on Moduli Space}

We can learn more about the fat brane dynamics by studying the structure of the
moduli space $\cal M$ determined by the (dressed) matrix D-brane
configurations. Assuming the generic D-brane couplings to admit decompositions
\eqn{recoilop} into pairs of logarithmic operators, this space is the direct
sum
\beq
{\cal M}={\cal M}_C\oplus{\cal M}_D
\label{modulidecomp}\eeq
of two subspaces which each have classical dimension $9N^2$. According to the
results of the previous section, to lowest order in
the string coupling $g_s$ the decomposition \eqn{modulidecomp} can be
interpreted as the splitting of the fat-brane collective coordinates into phase
space degrees of freedom. However, this will not be true at higher-orders and
in general \eqn{modulidecomp} represents a non-trivial mixing between
configuration space and phase space variables. As discussed in \cite{kogmav},
the logarithmic nature of the deformation makes the geometry on the space
\eqn{modulidecomp} well-defined.

The Zamolodchikov metric on $\cal M$ is given by the two-point function of the
deformation vertex operators \eqn{barevertexops},
\bea
G^{ij}_{ab;cd}(s,s')&=&2N\Lambda^2~\e^{-Q\phi/\sqrt{\alpha'}}\left\langle
V_{ab}^i(x;s)\,V_{cd}^j(x;s')\right
\rangle_{\rm L}\nn\\&\equiv&-2N\Lambda^2\,{\cal
W}_{U(1)}[\partial\Sigma;A]\frac\delta{\delta Y^{ab}_i(x^0(s))}\,{\cal
W}_{U(1)}[\partial\Sigma;A]\frac\delta{\delta Y^{cd}_j(x^0(s'))}\,Z_N[A]
\nn\\& &~~~~\label{Zmetrictot}\eea
where we have taken into account the extrinsic curvature term in the Liouville
dressing \eqn{Liouvilleaction} which in the case of the disc has $K=2$. With
this definition, \eqn{Zmetrictot} determines a fiducial metric on moduli space.
The $SU(N)$ part of \eqn{Zmetrictot} relevant for the constituent D-brane
dynamics is given by the perturbative expansion
\beq
\widetilde{G}^{ij}_{ab;cd}(s,s')=2\Lambda^2\sum_{n=0}^\infty
\frac{(-i)^n}{n!}\left(
\frac{g_s}{2\pi\alpha'}\right)^{n+2}~{\cal G}^{(n)ij}_{ab;cd}[Y;s,s']
\label{Zmetricexp}\eeq
where the ${\cal O}(Y(x^0)^n)$ contribution is
\bea
& &{\cal G}_{ab;cd}^{(n)ij}[Y;s,s']=\lim_{\epsilon\to0^+}
\sum_{c=1}^N~\sum_{{\buildrel{a_1,\dots,a_n}\over
{b_1,\dots,b_n}}}~\int_0^1\prod_{k=1}^nds_k~\Biggl\langle\!\!\Biggl\langle
\bar\xi_c(0)\bar\xi_a(s-\epsilon)
\xi_b(s)\bar\xi_c(s'-\epsilon)\xi_d(s')\biggr.\biggr.\nn\\& &~~~~~~
\Biggl.\Biggl.\times\left(\prod_{k=1}^n\bar\xi_{a_k}(s_k-\epsilon)
\xi_{b_k}(s_k)\right)\xi_c(1)\Biggr\rangle\!\!\Biggr\rangle\left\langle\frac
{d x^i(s)}{ds}\frac{d
x^j(s')}{ds'}\prod_{k=1}^nY_{i_k}^{a_kb_k}\left(x^0(s_k)\right)\frac
d{ds_k}x^{i_k}(s_k)\right\rangle_0
\nn\\& &~~~~\label{Zmetricordern}\eea
The expression \eqn{Zmetricordern} can be evaluated as described in appendix B
by writing it as a sum over permutations $P\in S_{n+3}$. However, it is much
simpler to note that \eqn{Zmetricexp} can be obtained from the canonical
momentum \eqn{Pidef} by functional differentiation,
\beq
\widetilde{G}^{ij}_{ab;cd}(s,s')=2\Lambda^2
\left(-i\frac\delta{\delta Y_i^{ab}(x^0(s))}\widetilde{\Pi}^j_{cd}(s')\right)
\label{ZmetricPidef}\eeq
This differentiation preserves the renormalization of the auxilliary quantum
field theory described in appendix B, so that \eqn{ZmetricPidef} also holds for
the corresponding renormalized quantities. Equating coefficients of the ${\cal
O}(Y(x^0)^n)$ terms in the renormalizations of the perturbative expansions
\eqn{Pipowerseries} and \eqn{Zmetricexp} thus gives for the relevant zero mode
contributions
\bea
& &{\cal G}^{(n)ij}_{ab;cd}[Y;0,0]^{\rm
ren}=\frac1{n+1}\sum_{l=1}^{n+1}\int_0^1\prod_{{\buildrel{k=1}\over{k\neq
l}}}^{n+1}ds_k~\Biggm\langle\frac{dx^i(s)}{ds}\biggm|_{s=0}\frac{dx^j(s')}{ds'}
\biggm|_{s'=0}\nn\\& &~~~~\times~{\rm
Sym}\left[\prod_{k=1}^{l-1}Y_{i_k}\left(x^0(s_k)\right)\,\frac
d{ds_k}x^{i_k}(s_k)~\otimes~\prod_{m=l+1}^{n+1}Y_{i_m}\left(x^0(s_m)
\right)\,\frac d{ds_m}x^{i_m}(s_m)\right]_{db;ca}\Biggm\rangle_0\nn\\& &~~~~
\label{Zmetricrenordern}\eea

Let us now compute \eqn{Zmetricrenordern} when the D-brane configuration fields
are given by logarithmic deformation operators. Then the expression
\eqn{Zmetricrenordern} is non-vanishing only when $n$ is even. The leading
$n=0$ contribution is the identity operator on $\cal M$,
\beq
{\cal G}^{(0)ij}_{ab;cd}(Y,U;0,0)^{\rm
ren}=\frac{8\pi^2\alpha'}{\Lambda^2}\,\eta^{ij}\delta_{da}\delta_{bc}
\label{Gorder0}\eeq
while the next contribution is at $n=2$ which gives
\bea
& &{\cal G}^{(2)ij}_{ab;cd}(Y,U;0,0)^{\rm ren}\nn\\&
&=\frac13\lim_{\epsilon\to0^+}\int_0^1ds_1~ds_2~\left\langle\mbox{$\frac
d{ds}$}\,x^i(s)\Bigm|_{s=0}\,\mbox{$\frac
d{ds'}$}\,x^j(s')\Bigm|_{s'=0}\,\mbox{$\frac
d{ds_1}$}\,x^{i_1}(s_1)\,\mbox{$\frac
d{ds_2}$}\,x^{i_2}(s_2)\right\rangle_0\nn\\& &~~~~~~~~~~\times~{\rm
Sym}\Bigl[\sqrt{\alpha'}\left\langle
C(s_1;\epsilon)D(s_2;\epsilon)\right\rangle_0\,\left\{I_N\otimes(Y_{i_1}
U_{i_2}+U_{i_2}Y_{i_1})+Y_{i_1}\otimes U_{i_2}\right.\nn\\&
&~~~~~~~~~~~~~~~~~~~~~~~~~~~~~~~~~~~~~~~~~~~~~~~~~~\left.+\,U_{i_2}\otimes
Y_{i_1}+(Y_{i_1}U_{i_2}+U_{i_2}Y_{i_1})\otimes I_N\right\}\nn\\&
&~~~~~~~~~~~~~~~~~~~~+\left\langle
D(s_1;\epsilon)D(s_2;\epsilon)\right\rangle_0 \,\left\{I_N\otimes
U_{i_1}U_{i_2}+U_{i_1}\otimes U_{i_2}+
U_{i_1}U_{i_2}\otimes I_N\right\}\Bigr]_{db;ca}\nn\\& &~~~~
\label{Gorder2def}\eea
Using Wick's theorem, and substituting the boundary string propagators
\eqn{derivprop} and the two-point correlation functions \eqn{2ptCD},\eqn{2ptDD}
of the logarithmic operators into \eqn{Gorder2def} yields
\bea
& &{\cal G}^{(2)ij}_{ab;cd}(Y,U;0,0)^{\rm
ren}=\lim_{\epsilon\to0^+}\frac{(4\pi^2\alpha')^2b}3\left[\frac{2\sqrt{\alpha'}}
{\Lambda^2}
I_g^{(1)}\,\eta^{ij}\left\{I_N\otimes\left(Y_kU^k+U^kY_k\right)+Y_k\otimes
U^k\right.\right.\nn\\&
&~~~~~~~~~~~~~~~~~~~~~~~~~~~~~~~~~~~~~~~~~~~~~~~~~~\left.+\,U^k\otimes
Y_k+\left(Y_kU^k+U^kY_k\right)\otimes I_N\right\}\nn\\&
&~~~~~~~~~~+\,\sqrt{\alpha'}\,I_g^{(2)}\left\{I_N\otimes
\left(Y^iU^j+Y^jU^i+U^jY^i+U^iY^j
\right)+Y^i\otimes U^j+Y^j\otimes U^i\right.\nn\\&
&~~~~~~~~~~~~~~~~~~~~\left.+\,U^j\otimes Y^i+U^i\otimes
Y^j+\left(Y^iU^j+Y^jU^i+U^jY^i+U^iY^j\right)\otimes I_N\right\}\nn\\&
&~~~~~~~~~~-\frac{2\alpha'}{\Lambda^2}I_g^{(3)}\,\eta^{ij}\left\{I_N\otimes
U_kU^k+U_k\otimes U^k+U_kU^k\otimes I_N\right\}\nn\\& &~~~~~~~~~~
-\,\alpha'I_g^{(4)}\left\{I_N\otimes\left(U^iU^j+U^jU^i\right)+U^i\otimes
U^j\right.\nn\\& &~~~~~~~~~~~~~~~~~~~~~~~~~\biggl.\left.+\,U^j\otimes
U^i+\left(U^iU^j+U^jU^i\right)\otimes I_N\right\}\biggr]_{db;ca}
\label{Gorder2long}\eea
where we have used \eqn{scalerel} and the boundary integrals $I_g^{(l)}$ are
given in \eqn{ig1}--\eqn{ig4}.

We see that in the limit $\epsilon\to0^+$ the most dominant contribution to
\eqn{Gorder2long} comes from the integral $I_g^{(4)}$ which yields the only
non-vanishing contributions with the renormalizations \eqn{renvelocity} and
\eqn{gsren}, and the bulk-boundary scaling relation \eqn{bulkbdryrel}. Then the
total Zamolodchikov metric up to second order in the perturbative expansion is
\bea
\widetilde{G}_{ab;cd}^{ij}(\bar Y,\bar U)&=&\frac{4\bar
g_s^2}{\alpha'}\left[\eta^{ij}\,I_N\otimes I_N-\frac{\bar
g_s^2b}{9\pi^2}\left\{I_N\otimes\left(\bar U^i\bar U^j+\bar U^j\bar
U^i\right)+\bar U^i\otimes\bar U^j\right.\right.\nn\\&
&~~~~~~\biggl.\left.+\,\bar U^j\otimes\bar U^i+\left(\bar U^i\bar U^j+\bar
U^j\bar U^i\right)\otimes I_N\right\}\biggr]_{db;ca}+{\cal O}\left(\bar
g_s^6\right)
\label{Gfinal}\eea
Now the logarithmic conformal algebra comes into play again and implies an
important property. If we renormalize the position of the fat brane as
\beq
Y_i=\sqrt{\alpha'}\,\epsilon\,\bar Y_i
\label{renposition}\eeq
then the $\beta$-function equations \eqn{betaepsilon} imply that
\beq
\frac{d\bar Y_i}{dt}=\bar U_i
\label{velocityeq}\eeq
The pair of renormalization group equations \eqn{Utrmarg} and \eqn{velocityeq}
are just the Galilean evolution equations for (renormalized) velocities. If we
now further adjust the parameters of the logarithmic conformal algebra as
\beq
a=\pi~~~~~~,~~~~~~b=-\mbox{$\frac{\pi^2}4$}
\label{abset}\eeq
then the canonical momentum \eqn{tildePiorder3} can be written as
\beq
\widetilde{\Pi}_{ab}^j(\bar Y,\bar
U)=\sqrt{\alpha'}\sum_{c,d=1}^N\widetilde{G}^{ji}_{ab;cd}(\bar Y,\bar
U)~\dot{\bar Y}^{cd}_{\!i}
\label{Pimodsprel}\eeq
and so we recover the canonical moduli space dynamics (see \eqn{momenta2}).
Note that the fixing of the coefficients \eqn{abset} does not completely
determine all parameters of the logarithmic operator correlators, as there is
still some freedom coming from the relation \eqn{relconstr}.

The corresponding Liouville problem satisfies the Helmholtz conditions of
section 2 and the associated action \eqn{36c} in the limit $\epsilon\to0^+$
coincides at leading orders with the (symmetrized) non-abelian Born-Infeld
action described in subsection 4.4. The Zamolodchikov $C$-function is given by
the $C$-theorem \eqn{cteo} which in the present case can be expressed as
\bea
\lim_{t\to\infty}\frac{\partial C(\bar Y,\bar U;t)}{\partial
t}&=&\sqrt{\alpha'}~\e^{-Q^2t/\sqrt{\alpha'}}\sum_{a,b,c,d}\bar
U_i^{ab}\,\widetilde{G}_{ab;cd}^{ij}(\bar Y,\bar U)\,\bar
U_j^{cd}\nn\\&=&\frac{4\bar g_s^2}{\sqrt{\alpha'}}\,{\cal F}(\bar
U)~\e^{-Q^2t/\sqrt{\alpha'}}+{\cal O}\left(\bar g_s^6\right)
\label{Cfng4}\eea
where we have introduced the velocity dependent invariant function
\beq
{\cal F}(\bar U)\equiv\tr\,\bar U_i\bar U^i+\frac{\bar
g_s^2}{36}\,\tr\left(2\,\bar U_i\bar U^i\bar U_j\bar U^j+\bar
U_i\bar U_j\bar U^i\bar U^j\right)
\label{calFdef}\eeq
Note that the expression \eqn{Gfinal} for the Zamolodchikov metric is
explicitly time independent and, strictly speaking, only valid for $t\to\infty$
because of the scaling property \eqn{scalerel}. Notice also that in \eqn{Cfng4}
we have reintroduced the appropriate scaling factors required for the monotonic
decreasing property of the $C$-function and also the expansion property
\eqn{38} which is crucial to the validity of the Helmholtz conditions.

The above results show that the geometry of the moduli space, determined by the
Zamolodchikov metric \eqn{Gfinal}, is a complicated function of the fat brane
dynamical parameters, which will be the key to its use in examining the
short-distance spacetime structures probed by D-particles. In the next section
we shall examine the genus expansion of the matrix
$\sigma$-model which will lead to a canonical quantization of the moduli space
dynamics described above. In particular, the velocity matrix $\bar U_i$ will
become a quantum operator. The same is true of the central charge deficit $Q$
which, neglecting irrelevant terms that can be removed by a change of
renormalization scheme, is given by
\beq
Q(\bar Y,\bar U;t)\equiv\sqrt{C(\bar Y,\bar U;t)}
\label{Cdeficit}\eeq
The quantity \eqn{Cdeficit} defines the ``physical" target space time in the
Liouville framework via \cite{emn,time}
\beq
T(\bar Y,\bar U;t)\equiv\phi=Q(\bar Y,\bar U;t)~t
\label{phystimedef}\eeq
where $t=-\sqrt{\alpha'}\log\Lambda$ is the rescaling (flat worldsheet) time
variable and $\phi$ is the zero mode of the Liouville field. Then the time
evolution of the Liouville dressed couplings with respect to the target space
time variable are governed by conventional worldsheet $\beta$-functions upon
replacing bare coupling constants with dressed ones. The definition
\eqn{phystimedef} comes from the normalization of the Liouville field kinetic
term $\partial\varphi\bar\partial\varphi$ appropriate to the Robertson-Walker
metric on spacetime \cite{time,aben}. The physical time \eqn{phystimedef}
becomes a quantum operator upon summing over worldsheet genera \cite{a-c}.
In general, the expression \eqn{Cfng4} which determines it as a function of $t$
is a complicated highly non-linear first order differential equation. If we
assume, however, that $C$ varies slowly with time, then \eqn{Cfng4} can be
solved at linear order in the string tension by quadratures to give
\beq
T(\bar Y,\bar U;t)\simeq\frac{2\bar g_st}{\alpha'^{1/4}}\,\sqrt{{\cal F}(\bar
U)}\,\left(\int_0^td\tau~\e^{2(\tau^2-t^2)\bar g_s^2{\cal F}(\bar
U)/\alpha'}\right)^{1/2}
\label{phystimeexpl}\eeq
The limit of slowly varying $C$-function holds near any fixed point in moduli
space. This assumption is consistent with the assumptions of small
$\sigma$-model and string couplings and also of a slowly-moving
(non-relativistic) fat-brane which is the kinematical region of interest here.
We note that, in contrast to the abelian case, the time variable
\eqn{phystimeexpl} is a complicated function of the various fat-brane
velocities because of the trace structure of the invariant function
\eqn{calFdef}.

\newsection{Quantization and Spacetime Uncertainty Relations}

In this section we will apply the formalism of \cite{emn} to sum over
worldsheet genera of the partition function \eqn{partfatbrane}. The pertinent
deformation couplings represented by the logarithmic operators have vanishing
conformal dimension in the limit $\epsilon\to0^+$ (see \eqn{Deltaep}), and as a
result extra logarithmic divergences appear in pinched annulus diagrams. This
will amount to a quantization of the couplings $Y_i(x^0)^{ab}$ from which we
will be able to derive a set of stringy uncertainty relations.

\newsubsection{Resummation of the Genus Expansion}

We consider the partition function \eqn{partfatbrane} defined on a genus $h$
surface $\Sigma_h$. This surface has $h$ `holes' in it and for all $h$ its
boundary has the topology of a circle, so that, in the notation above,
$\Sigma_0\equiv\Sigma$. The genus expansion is
\beq
\sum_{h=0}^\infty Z_N^h[A]=\sum_{h=0}^\infty\left\langle
W[\partial\Sigma_h;A]\right\rangle_0^h
\label{genusexpdef}\eeq
where the average is taken in the free $\sigma$-model \eqn{freesigma} defined
on $\Sigma_h$. Since we assume that $\partial\Sigma_h$ has the topology of a
disjoint union of $h+1$ circles, the sum over genera commutes with the
representation \eqn{wilsonloop} of the Wilson loop operator in terms of
auxilliary fields and we can write
\beq
\sum_{h=0}^\infty Z_N^h[A]=\left\langle\!\!\left\langle\sum_{c=1}^N\bar
\xi_c(0)~\sum_{h=0}^\infty\left\langle\exp\left(\sum_{a,b=1}^N\,\sum_{k=0}^h
\int_0^1ds_k~Y_i^{ab}\left(x^0(s_k)\right)\,V_{ab}^i(x;s_k)\right)\right
\rangle_0^h~\xi_c(1)\right\rangle\!\!\right\rangle
\label{genuscomm}\eeq
where for simplicity we have set $N{\cal W}_{U(1)}[\partial\Sigma_h;A]=1$ and
we work in the temporal gauge $A_0=0$ as usual. The double brackets in
\eqn{genuscomm} denote, as before, the average over the auxilliary fields as in
\eqn{auxprop} and the boundary vertex operators $V_{ab}^i$ are defined in
\eqn{barevertexops}.

For the recoil operators \eqn{recoilop} we can insert a temporal delta-function
$1=\int_0^\infty dt~\delta(t-x^0(s))$ into \eqn{genuscomm} to get
\bea
\sum_{h=0}^\infty
Z_N^h[A]&=&\lim_{\epsilon\to0^+}\left\langle\!\!\left\langle\sum_{c=1}^N\bar
\xi_c(0)~\sum_{h=0}^\infty\left\langle\exp\left(\sum_{a,b=1}^N\int_{-\infty}^
\infty d\omega~\int_0^\infty dt~Y_i^{ab}(t;\epsilon)~\e^{i\omega
t}\right.\right.\right.\right.\nn\\&
&~~~~~~\left.\left.\left.\left.\times\sum_{k=0}^h\int_0^1ds_k~\e^{-i\omega
x^0(s_k)}\,\Theta(x^0(s_k);\epsilon)\,V_{ab}^i(x;s_k)\right)\right\rangle_0^h~
\xi_c(1)\right\rangle\!\!\right\rangle
\label{timedelta}\eea
where $Y_i^{ab}(t;\epsilon)=\sqrt{\alpha'}\,Y_i^{ab}\epsilon+U_i^{ab}t$. If we
introduce the Fourier transform
\beq
\breve{Y}_i^{ab}(\omega)=\lim_{\epsilon\to0^+}\int_0^\infty dt~\e^{i\omega
t}\,Y_i^{ab}(t;\epsilon)
\label{FourierY}\eeq
and the new boundary vertex operators
\beq
\oint_{\partial\Sigma_h}V^i_{ab}(x;\omega)\equiv\lim_{\epsilon\to0^+}\frac
{ig_s}{2\pi\alpha'}\sum_{k=0}^h\int_0^1ds_k~\e^{-i\omega
x^0(s_k)}\,\Theta(x^0(s_k);\epsilon)
\,\bar\xi_a(s_k-\epsilon)\xi_b(s_k)\,\frac d{ds_k}x^i(s_k)
\label{newvertexops}\eeq
then the sum over genera in \eqn{timedelta} takes the usual form of a set of
$\sigma$-model couplings $\breve{Y}_i^{ab}(\omega)$ associated with the
deformation vertex operators \eqn{newvertexops},
\beq
\sum_{h=0}^\infty Z_N^h[A]=\left\langle\!\!\left\langle\sum_{c=1}^N\bar
\xi_c(0)~\sum_{h=0}^\infty\left\langle\exp\left(\sum_{a,b=1}^N\int_{-\infty}^
\infty d\omega~\breve{Y}_i^{ab}(\omega)\,\oint_{\partial\Sigma_h}V^i_{ab}(x;
\omega)\right)\right\rangle_0^h~\xi_c(1)\right\rangle\!\!\right\rangle
\label{usualsum}\eeq
The representation \eqn{usualsum}, along with \eqn{Thetafnreg}, justifies the
identification of the Liouville field $\varphi$ with the fundamental temporal
embedding field $x^0$, in the limit $\epsilon\to0^+$. The latter field appears
in the tachyon operator part of \eqn{newvertexops}, thereby dressing the
boundary theory analogously to that by two-dimensional quantum gravity. Some
further aspects of this correspondence, such as the properties of the induced
target space geometry, are discussed in \cite{diffusion}.

We now focus on the properties of the (abelianized) average over
fundamental string fields in \eqn{usualsum}. As we will show, the resummation
of
\eqn{usualsum} over {\it pinched} genera yield the dominant worldsheet
divergences, thereby spoiling the conformal symmetry. Conformal invariance
requires absorbing such singularities into renormalized quantities at lower
genera, leading to a generalized version of the Fischler-Susskind mechanism
\cite{fs}. Such degenerate Riemann surfaces involve a string propagator over
thin long worldsheet strips of thickness $\delta\to0$ that are attached to a
disc. These strips can be thought of as two-dimensional quantum gravity
wormholes. Consider first the resummation of one-loop worldsheets, i.e. those
with an annular topology, in the pinched approximation (fig. 3). String
propagation on such a worldsheet can be described formally by adding bilocal
worldsheet operators $\cal B$ \cite{polchinski1} which in the present case are
defined by
\beq
{\cal
B}(\omega,\omega')=\sum_{a,b,c,d}\,\oint_{\partial\Sigma}\oint_{\partial
\Sigma'}V_{ab}^i(x;\omega)~\frac{G_{ij}^{ab;cd}(
\omega,\omega')}{L_0-1}~V_{cd}^j(x;\omega')
\label{bilocal}\eeq
where the Zamolodchikov metric in \eqn{bilocal} is the two-point correlation
function of the vertex operators defined in \eqn{newvertexops} and $L_0$
denotes the usual Virasoro generator. The operator insertion $(L_0-1)^{-1}$ in
\eqn{bilocal} represents the string propagator $\triangle_s$ on the thin strip
of the pinched annulus.

\begin{figure}[htb]
\epsfxsize=2.5in
\bigskip
\centerline{\epsffile{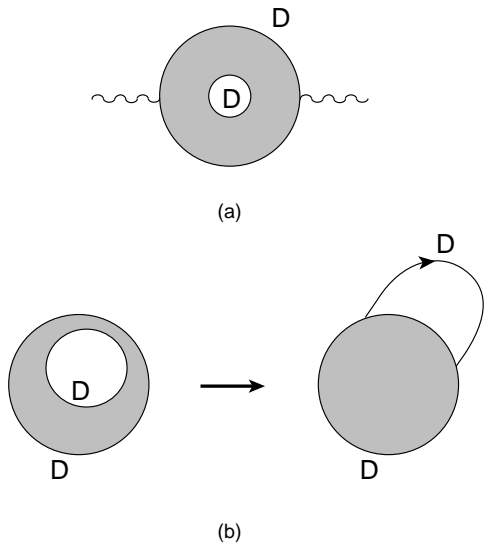}}
\caption{\baselineskip=12pt (a) {\it World-sheet annulus diagram for the
leading
quantum correction to the propagation of a string state $V$ in a D-brane
background, and} (b) {\it the pinched annulus configuration which is the
dominant divergent contribution to the quantum recoil.}}
\bigskip
\label{annulus}\end{figure}

Inserting a complete set of intermediate string states ${\cal E}_I$, we can
rewrite (\ref{bilocal}) as an integral over the variable $q \equiv \e^{2 \pi i
\tau}$, where $\tau$ is the complex modular parameter characterizing the
worldsheet strip. The string propagator over the strip then reads
\be
\triangle_s(z,z')\,=\,\sum_I\int dq~q^{\Delta_I-1}\,\Bigl\{{\cal
E}_I(z)\otimes({\rm ghosts})\otimes{\cal E}_I(z')\Bigr\}_{\Sigma\#\Sigma'}
\label{props}\ee
where $\Delta_I$ are the conformal dimensions of the states ${\cal E}_I$. The
sum in (\ref{props}) is over all states which propagate along the long thin
strip connecting the discs $\Sigma$ and $\Sigma'$ (in the degenerating annulus
handle case of interest here, $\Sigma'=\Sigma$). As indicated in (\ref{props}),
the sum over states must include ghosts, whose central charge cancels that of
the worldsheet matter fields in any critical string model.

In \eqn{props} we have assumed that the Virasoro operator $L_0$ can be
diagonalized in the basis of string states with eigenvalues their conformal
dimensions $\Delta_I$, i.e.
\be
L_0|{\cal E}_I\rangle=\Delta_I|{\cal E}_I\rangle~~~~~~,~~~~~~q^{L_0-1}|{\cal
E}_I\rangle=q^{\Delta_I-1}|{\cal E}_I\rangle
\label{diagonal}\ee
However, this simple diagonalization fails in the presence of the logarithmic
pair of operators \cite{kogmav}, due to the non-trivial mixing between $C$
and $D$ in the Jordan cell of $L_0$. Generally, states with
$\Delta_I=0$ may lead to extra logarithmic divergences in (\ref{props}),
because such states make contributions to the integral of the form $\int
dq/q\sim\log\delta$, in the limit $q \sim \delta \rightarrow 0$
representing a long thin strip of thickness $\delta$. We assume that such
states are discrete in the space of all string states, i.e. that they are
separated from other states by a gap. In that case, there are factorizable
logarithmic divergences in (\ref{props}) which depend on the background
surfaces $\Sigma$ and $\Sigma'$. These are precisely the states corresponding
to the logarithmic recoil operators (\ref{Depsilonop}) and \eqn{Cepsilonop},
with vanishing conformal dimension (\ref{Deltaep}) as $\epsilon \rightarrow
0^+$.

In the case of mixed logarithmic states, the pinched topologies are
characterized by divergences of a double logarithmic type which arise from the
form of the string propagator in (\ref{bilocal}) in the presence of generic
logarithmic operators $C$ and $D$,
\beq
\int dq~q^{\Delta_\epsilon-1}\,\langle C,D|\pmatrix{1&\log
q\cr0&1\cr}|C,D\rangle
\label{CDprop}\eeq
As shown in \cite{kogmav}, the mixing between $C$ and $D$ states along
degenerate handles leads formally to divergent string propagators in physical
amplitudes, whose integrations have leading divergences of the form
\beq
\int\frac{dq}q~\log q\int d^2z~D(z;\epsilon)\int
d^2z'~C(z';\epsilon)\simeq(\log\delta)^2
\int d^2z~D(z;\epsilon)\int d^2z'~C(z';\epsilon)
\label{mixing}\eeq
These $(\log\delta)^2$ divergences can be cancelled by imposing momentum
conservation in the scattering process of the light string states off the
D-brane background \cite{lm}. This cancellation of leading divergences
of the genus expansion in the non-abelian case is demonstrated explicitly in
appendix D. It is shown there that this renormalization requires that the
change in (renormalized) velocity of the fat brane due to the recoil from the
scattering of string states be
\beq
\bar U_i^{ab}=-\frac1{M_s}\,\Bigl(k_1+k_2\Bigr)_i\,\delta^{ab}=\frac{d\bar
Y_i^{ab}}{dt}
\label{recoilvel}\eeq
where $k_{1,2}$ are the initial and final momenta in the scattering process and
$M_s=1/\sqrt{\alpha'}\,\bar g_s$ is the BPS mass of the string soliton
\cite{polchinski}.\footnote{Note that this differs from the mass normalization
of the derived canonical momentum \eqn{tildePiorder3}. In \eqn{recoilvel}, the
$k_{1,2}$ are true physical momenta so that $M_s$ represents the actual BPS
mass of the D-particles.} This means that, to leading order, the constituent
D-branes
move parallel to one another with a common velocity and there are no
interactions among them. Thus the leading recoil effects imply a commutative
structure and the fat brane behaves as a single D-particle. Note that the
relation \eqn{recoilvel} also shows directly that $d\bar g_s/dt=0$.

In addition to this divergence, there are sub-leading $\log\delta$
singularities, corresponding to the diagonal terms $\int d^2z~D(z;\epsilon)\int
d^2z'~D(z';\epsilon)$ and $\int d^2z~C(z;\epsilon)\int d^2z'~C(z';\epsilon)$.
With our choice of basis \eqn{modulidecomp} on the moduli space of D-brane
configurations, these latter terms are the ones we should concentrate upon for
the purposes of deriving the quantum fluctuations of the collective D-particle
coordinates. As we will see, it is these sub-leading divergences in the genus
expansion which lead to interactions between the constituent D-branes and
provide the appropriate noncommutative quantum extension of the leading
dynamics
\eqn{recoilvel}.

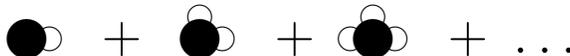
\begin{figure}[htb]
\unitlength=1.00mm
\linethickness{0.4pt}
\bigskip
\begin{center}
\begin{picture}(70.00,10.00)
\LARGE
\put(5.00,2.00){\circle*{100.00}}
\put(8.00,2.00){\circle{3.00}}
\put(15.00,2.00){\makebox(0,0)[l]{$+$}}
\put(28.00,2.00){\circle*{100.00}}
\put(31.00,2.00){\circle{3.00}}
\put(28.00,5.00){\circle{3.00}}
\put(38.00,2.00){\makebox(0,0)[l]{$+$}}
\put(48.00,2.00){\circle{3.00}}
\put(51.00,2.00){\circle*{100.00}}
\put(54.00,2.00){\circle{3.00}}
\put(51.00,5.00){\circle{3.00}}
\put(61.00,2.00){\makebox(0,0)[l]{$+~\dots$}}
\end{picture}
\end{center}
\caption{\it\baselineskip=12pt Resummation of the genus expansion in the
pinched approximation. The solid circles are the worldsheet discs and the thin
lines are  strips attached to them with infinitesimal pinching size $\delta$.
Each strip corresponds to an insertion of a bilocal operator \eqn{bilocal} on
the worldsheet.}
\bigskip
\label{pinchsum}\end{figure}

In the weak-coupling case, we can truncate the genus expansion \eqn{usualsum}
to a sum over pinched annuli (fig. \ref{pinchsum}). This truncation corresponds
to a semi-classical approximation to the full quantum string theory in which we
treat the D-particles as heavy non-relativistic objects in target space. Then
the dominant contributions to the sum are given by the $\log\delta$ modular
divergences described above, and the effects of the dilute gas of wormholes on
the disc are to exponentiate the bilocal operator \eqn{bilocal}. In the pinched
approximation, the genus expansion thus leads to an effective change in the
matrix $\sigma$-model action in \eqn{usualsum} by
\beq
\Delta S\simeq\frac{g_s^2}2\log\delta\sum_{a,b,c,d}\,\int_{-\infty}^\infty
d\omega~d\omega'~\oint_{\partial\Sigma}\oint_{\partial\Sigma'}
V_{ab}^i(x;\omega)~G_{ij}^{ab;cd}(\omega,\omega')~V_{cd}^j(x;\omega')
\label{actionchange}\eeq
The bilocal action \eqn{actionchange} can be cast into the form of a local
worldsheet effective action by using standard tricks of wormhole calculus
\cite{wormholes} and rewriting it as a functional Gaussian integral
\bea
\e^{\Delta
S}&=&\int[d\breve{\rho}]~\exp\left[-\frac12\sum_{a,b,c,d}\,\int_{-\infty}^
\infty d\omega~d\omega'~\breve{\rho}_i^{ab}(\omega)~\oint_{\partial\Sigma}
\oint_{\partial\Sigma'}G^{ij}_{ab;cd}(\omega,\omega')~
\breve{\rho}_j^{cd}(\omega')\right.\nn\\& &\left.~~~~~~~~~~~~~~~~~~~~
+\,g_s\,\sqrt{\log\delta}~\sum_{a,b=1}^N\,\int_{-\infty}^\infty
d\omega~\breve{\rho}_i^{ab}(\omega)\,\oint_{\partial\Sigma}
V_{ab}^i(x;\omega)\right]
\label{Gaussianint}\eea
where $\breve{\rho}_i^{ab}(\omega)$ are quantum coupling constants of the
worldsheet matrix $\sigma$-model. Thus the effect of the resummation over
pinched genera is to induce quantum fluctuations of the collective D-brane
background, leading to a set of effective quantum coordinates
\beq
\breve{Y}_i^{ab}(\omega)~\to~\widehat{\cal
Y}_i^{ab}(\omega)=\breve{Y}_i^{ab}(\omega)+g_s\,\sqrt{\log\delta}~
\breve{\rho}_i^{ab}(\omega)
\label{quantumcoords}\eeq
viewed as position operators in a co-moving target space frame.

Transforming the quantum couplings to the temporal field representation using
the inverse transformations which led to \eqn{usualsum}, we find that the genus
expansion \eqn{genusexpdef} in the pinched approximation is
\beq
\sum_{h^{(p)}}Z_N^{h^{(p)}}[A]~\simeq~\left\langle\int_{\cal
M}[d\rho]~\wp[\rho]
{}~W\!\left[\partial\Sigma;A-\mbox{$\frac1{2\pi\alpha'}$}\,\rho\right]\right
\rangle_0
\label{pinchedpartfn}\eeq
where the sum is over all pinched genera of infinitesimal pinching size, and
\beq
\wp[\rho]=\exp\left[-\frac1{2\Gamma^2}\sum_{a,b,c,d}\,\int_0^1ds~ds'~
\rho_i^{ab}\left(x^0(s)\right)\,G_{ab;cd}^{ij}(s,s')\,\rho_j^{cd}
\left(x^0(s')\right)\right]
\label{Gaussiandistr}\eeq
is a functional Gaussian distribution on moduli space of width
\beq
\Gamma=g_s\,\sqrt{\log\delta}
\label{widthdef}\eeq
In \eqn{pinchedpartfn} we have normalized the functional Haar integration
measure $[d\rho]$ appropriately. We see therefore that the diagonal sub-leading
logarithmic divergences in the modular cutoff scale $\delta$, associated with
degenerate strips in the genus expansion of the matrix $\sigma$-model, can be
treated by absorbing these scaling violations into the width $\Gamma$ of the
probablity distribution characterizing the quantum fluctuations of the
(classical) D-brane configurations $Y_i^{ab}(x^0(s))$. In this way the
interpolation among families of D-brane field theories corresponds to a
quantization of the worldsheet renormalization group flows. Note that the
worldsheet wormhole parameters, being functions on the moduli space
\eqn{modulidecomp}, can be decomposed as
\beq
\rho_i^{ab}(x^0(s))=\lim_{\epsilon\to0^+}\left([\rho_C]_i^{ab}C(x^0;\epsilon)
+[\rho_D]_i^{ab}D(x^0;\epsilon)\right)
\label{wormholedecomp}\eeq
The fields $\rho_{C,D}$ are then renormalized in the same way as the D-brane
couplings, so that the corresponding renormalized wormhole parameters generate
the same type of (Galilean) $\beta$-function equations \eqn{betaepsilon}
\cite{lm}. This will be implicitly assumed in the following.

According to the standard Fischler-Susskind mechanism for cancelling string
loop divergences \cite{fs}, modular infinities should be identified with
worldsheet divergences at lower genera. Thus the strip divergence $\log\delta$
should be associated with a worldsheet ultraviolet cutoff scale which in turn
is identified with the Liouville field as described earlier. We may in effect
take $\delta$ independent from $\Lambda$, in which case we can first let
$\epsilon\to0^+$ in the above and then take the limit $\delta\to0$.
Interpreting $\log\delta$ in this way as a renormalization group time
parameter (interpolating among D-brane field theories), the time dependence of
the renormalized width \eqn{widthdef} expresses the usual properties of the
distribution function describing the time evolution of a wavepacket in moduli
space \cite{schm}. The inducing of a statistical Gaussian spread of the D-brane
couplings is the essence of the quantization procedure.

\newsubsection{String Interactions and Diagonalization of Moduli Space}

The Gaussian distribution functional \eqn{Gaussiandistr} can be used to
determine the quantum fluctuations $\Delta\bar Y_i^{ab}$ in the initial D-brane
positions to leading order in the string coupling constant expansion. For this,
we first need to diagonalize the Zamolodchikov metric \eqn{Gfinal}. As we will
see, the parameters of the diagonalization of the geometry of moduli space
expose the precise nature of the string interactions inherent in the multi
D-brane system. This eigenvalue problem is somewhat intractable in general, but
in the limit $g_s\ll1$ of weakly coupled strings it can be carried through with
some work.

In the free string limit, the interactions between the constituent D-branes
are negligible to lowest order and their position matrices commute. In the
temporal gauge that we are working in, the configuration fields can then be
simultaneously diagonalized by a time independent gauge transformation
\beq
\bar Y^i=\Omega~{\rm
diag}\left(y^i_1,\dots,y^i_N\right)\,\Omega^{-1}~~~~~~,~~~~~~\Omega\in U(N)
\label{Ydiag}\eeq
The eigenvalues $y_a^i\in\real$ represent the positions of the constituent
D-branes themselves which move at velocities $u_a^i=dy_a^i/dt$. The
noncommutativity of spacetime is encoded through the unitary matrix $\Omega$
which represents the string interactions between the D-particles. In this way
we will study the coordinate fluctuations both as a quantum mechanical effect
and geometrically as the perturbations around classical (commutative) spacetime
represented by the diagonal matrix configurations in \eqn{Ydiag}. This limit
corresponds to a configuration of well-separated branes and it
represents a Born-Oppenheimer approximation to the D-particle interactions,
which is valid for small velocities \cite{liyoneya}, whereby the diagonal
D-particle coordinates are separated from the off-diagonal parts of the adjoint
Higgs fields representing the short open string excitations connecting them.

Using \eqn{Ydiag} the Zamolodchikov metric \eqn{Gfinal} can be written as
\beq
\widetilde{G}^{ij}(\bar Y,\bar U)=\frac{4\bar
g_s^2}{\alpha'}\,(\Omega\otimes\Omega)\left(\eta^{ij}\,I_N\otimes
I_N+\frac{\bar g_s^2}{36}\,{\cal U}^{ij}+{\cal O}\left(\bar
g_s^4\right)\right)(\Omega\otimes\Omega)^{-1}
\label{Gdiag}\eeq
where ${\cal U}_{ab;cd}^{ij}={\cal U}_{ab}^{ij}\,\delta_{ad}\delta_{bc}$ is the
$u(N)\otimes u(N)$ diagonal matrix with entries
\beq
{\cal U}_{ab}^{ij}=2u_a^iu_a^j+2u_b^iu_b^j+u_a^iu_b^j+u_a^ju_b^i
\label{Udiagdef}\eeq
We now need to diagonalize the symmetric matrix \eqn{Udiagdef} with respect to
the $9\times9$ spacetime indices $i,j$. For this, we assume that
$\eta_{ij}=\delta_{ij}$ and consider separately the two cases $a=b$ and $a\neq
b$.

Consider first the case $a=b$. Upon examination of the characteristic equation
for the matrix ${\cal U}_{aa}^{ij}=6u_a^iu_a^j$ one easily sees that there are
two eigenvalues $\lambda=6\|u_a\|^2$ and $\lambda=0$, where
$\|u_a\|=\sqrt{\sum_iu_a^iu_a^i}$ is the Euclidean norm of the vector
$u_a\in\real^9$. The dimension of the kernel of ${\cal U}_{aa}^{ij}$ is 8
because there are precisely eight linearly independent vectors in $\real^9$
which are orthogonal to $u_a^i$. Thus the eigenvalues are
\beq
\lambda_{aa}^1=6\,\|u_a\|^2~~~~~~,~~~~~~\lambda_{aa}^2=\dots=
\lambda_{aa}^9=0
\label{aaevs}\eeq
The normalized eigenvector corresponding to $\lambda_{aa}^1$ is just
$u_a/\|u_a\|$ and the remaining ones span the eight-dimensional space
transverse to this line, which we refer to as the ``string frame" because it
represents the coordinate system relative to the fundamental open string
excitations which start and end on the same D-particle $a$. Upon rotation to
the one-dimensional string frame, the $9\times9$ orthogonal matrix $\Xi_{aa}$
which diagonalizes \eqn{Udiagdef} for $a=b$ is just the identity matrix,
\beq
\Xi_{aa}=I_9
\label{Xiaatilde}\eeq

The situation for $a\neq b$ is similar but a bit more technically involved. We
assume that the velocity vectors $u_a$ and $u_b$ are linearly independent.
There are then seven linearly independent vectors which are orthogonal to both
$u_a$
and $u_b$, and therefore there is a zero eigenvalue of mulitplicity 7. The
remaining two eigenvectors are linear combinations of the velocity vectors,
\beq
\psi^{(1,2)i}_{ab}=u_a^i+B^{(1,2)}\,u_b^i
\label{ab12evecs}\eeq
up to an overall normalization. Solving the eigenvalue equations for the
eigenvectors \eqn{ab12evecs} gives after some tedious algebra the two non-zero
eigenvalues,
\bea
\lambda_{ab}^{1,2}&\equiv&\lambda_\pm~=~\|u_a\|^2+\|u_b\|^2+u_a\cdot
u_b~\pm~\Biggl\{\left(\|u_a\|^2+\|u_b\|^2+u_a\cdot u_b\right)^2\Biggr.\nn\\&
&~~~~~~~~~~\left.+\,\frac{\Bigl[(u_a\cdot u_b)^2+\|u_a\|^2\|u_b\|^2+2u_a\cdot
u_b\left(\|u_a\|^2+\|u_b\|^2\right)\Bigr]^2}{\|u_a\|^2\|u_b\|^2-(u_a\cdot
u_b)^2}\right\}^{1/2}\nn\\\lambda_{ab}^3&=&\dots~=~\lambda_{ab}^9~=~0
\label{lambdapm}\eea
and the coefficients
\beq
B^{(1,2)}=\frac{(u_a\cdot u_b)^2+\|u_a\|^2\|u_b\|^2+2u_a\cdot
u_b\left(\|u_a\|^2+\|u_b\|^2\right)-\lambda_{ab}^{1,2}\,u_a\cdot
u_b}{2\|u_a\|^2u_a\cdot u_b+2(u_a\cdot
u_b)^2+2\|u_a\|^4-\lambda_{ab}^{1,2}\,\|u_a\|^2}
\label{abcoeffs}\eeq
where the dot between vectors denotes the usual Euclidean inner product on
$\real^9$. The remaining seven orthonormal eigenvectors are those which span
the space transverse to the plane in $\real^9$ generated by the vectors
\eqn{ab12evecs}, which defines the two-dimensional string frame representing
the fundamental open string which starts on D-brane $a$ and ends on D-brane
$b$. Note that for $a\neq b$ the dimension of this coordinate system increases
by one because of the increase in degrees of freedom of the string which now
stretches between two different branes. Once again the orthogonal
diagonalization transformation matrix $\Xi_{ab}$ is particularly simple in the
string frame. We parametrize the plane spanned by $u_a$ and $u_b$ via
$u_a^i=\|u_a\|\,\delta^{i,1}$ and the angle $\theta_{ab}$ between the two
vectors. Then upon rotation to the two-dimensional string frame we have
\beq
\Xi_{ab}=\pmatrix{{\cal N}\left(\|u_a\|+B^{(1)}\,\|u_b\|\cos\theta_{ab}\right)&
-{\cal N}\,B^{(1)}\,\|u_b\|\sin\theta_{ab}&0&\dots&0\cr
{\cal N}\,B^{(1)}\,\|u_b\|\sin\theta_{ab}&{\cal
N}\left(\|u_a\|+B^{(1)}\,\|u_b\|\cos\theta_{ab}\right)&0&\dots&0\cr
0&0&1&\dots&0\cr\vdots&\vdots&\vdots&\ddots&\vdots\cr0&0&0&\dots&1\cr}
\label{Xiabtilde}\eeq
where we have orthogonalized the $2\times2$ block matrix corresponding to the
string frame and
\beq
{\cal N}=\|u_a+B^{(1)}\,u_b\|^{-1}
\label{normconst}\eeq
is the appropriate normalization constant.

With the above constructions, the Zamolodchikov metric can now be written as a
unitary transformation of a diagonal metric on $\cal M$,
\bea
\widetilde{G}^{ij}_{ab;cd}(\bar Y,\bar U)&=&\frac{4\bar
g_s^2}{\alpha'}\,\eta_{kl}\sum_{e,f=1}^N\Omega_{af}(\bar Y)\,\Omega_{be}(\bar
Y)\,\Xi_{ef}^{ik}(u)\,\Xi_{fe}^{jl}(u)\,\Omega^*_{ce}(\bar
Y)\,\Omega^*_{df}(\bar Y)\nn\\& &~~~~~~~~~~~~~~~~~~~~\times\left(1+\frac{\bar
g_s^2}{36}\,\lambda_{ef}^k(u)+{\cal O}\left(\bar g_s^4\right)\right)
\label{Gdiagexpl}\eea
We see therefore that the diagonalization of the Zamolodchikov metric on moduli
space $\cal M$ naturally encodes within it the geometry of the string
interactions among the D-branes. In particular, we see the enormous complexity
involved in going from the dynamics for a single D-particle ($a=b$) to the
interactions between constituent D-branes ($a\neq b$). These properties will
have important ramifications for the physical consequences of the stringy
spacetime uncertainty relations which we now proceed to derive.

\newsubsection{Quantum Fluctuations of Collective D-brane Configurations}

Given the diagonalization \eqn{Gdiagexpl} of the bilinear form of the Gaussian
distribution functional \eqn{Gaussiandistr}, we can now write down the quantum
fluctuations of the D-brane coordinates. Substituting \eqn{Gdiagexpl} into
\eqn{Gaussiandistr} and redefining the matrix-valued wormhole parameters
$\rho_i^{ab}$ leads to a complex bilinear form in a new set of complex-valued
wormhole parameters. Since the metric of the bilinear form is diagonal, one can
associate a width to each direction $i=1,\dots,9$ and D-brane configuration
$a,b=1,\dots,N$. The coordinate transformation
\be
{\widetilde Y}_{ab}^i=\Xi_{ab}^{ji}\left[\Omega^{*-1}\bar
Y_j\Omega\right]_{ba}\equiv \Xi^{ji}_{ab}\,X_j^{ab}(\bar Y)
\label{coord}\ee
is precisely the one which achieves the desired diagonalization and leads to
the statistical variances
\beq
\left(\Delta\widetilde{Y}^i_{ab}\right)\left(\Delta\widetilde{Y}^i_{ab}
\right)^\dagger=\frac{\alpha'\Gamma^2}{2\bar g_s^2}\left(1-\frac{\bar
g_s^2}{36}\,\lambda_{ab}^i(u)+{\cal O}\left(\bar g_s^4\right)\right)
\label{tildeYuncert}\eeq
Note that, as a result of \eqn{scalerel}, the renormalized string coupling
$\bar g_s$ is imaginary, i.e. $\bar g_s^2<0$, owing to the Minkowskian
signature of the spacetime.

The time dependence in the width \eqn{widthdef} can be absorbed into the usual
renormalization of the string coupling constant by taking the correlation
\beq
\log\delta=2|\bar g_s|^\chi\,\epsilon^{-2}
\label{deltaepcorr}\eeq
between the modular worldsheet and target space scale parameters. The
exponent $\chi\geq0$ is left arbitrary for the moment. Later on we shall fix it
by demanding consistency of certain results with conventional D-particle
mechanics. The variances \eqn{tildeYuncert} are therefore time-independent and
represent not the spread in time of a wavepacket on $\cal M$, but rather the
true quantum fluctuations of the D-brane configurations. The collective
D-particle coordinates $X_i(\bar Y)$ naturally encode the effects of the open
string excitations. Their uncertainties may be computed using the formula
\beq
\left(\Delta\widetilde{Y}^i_{ab}\right)\left(\Delta\widetilde{Y}^i_{ab}
\right)^\dagger\equiv\left(\!\left(\widetilde{Y}^i_{ab}\biggm|\left[
\widetilde{Y}^i_{ab}\right]^\dagger\right)\!\right)-\left|\left(\!\left(
\widetilde{Y}^i_{ab}\right)\!\right)\right|^2=\Xi_{ab}^{ji}\,\Xi_{ab}^{ki}
\left(\!\left(X^{ab}_j\biggm|\left[X^{ab}_k\right]^\dagger\right)\!\right)_{\rm
conn}
\label{YXavgs}\eeq
where the brackets denote statistical averages with respect to the wormhole
probability distribution \eqn{Gaussiandistr} and the average of the $X$ fields
in \eqn{YXavgs} is a connected correlation function. In this subsection we
shall always work in string coordinates, but, by covariance, the qualitative
features are the same in any reference frame.

Let us first consider the relation \eqn{tildeYuncert} in the case $a=b$, which
corresponds to a single D-particle. Using \eqn{aaevs} and \eqn{Xiaatilde} it is
straightforward to see that the variances \eqn{tildeYuncert} and \eqn{YXavgs}
lead to the position uncertainties
\be
\left|\Delta X_i^{aa}\right|=|\bar
g_s|^{\chi/2}\sqrt{\alpha'}\left(1+\frac{|\bar g_s|^2}{12}\,\|u_a
\|^2\,\delta_{i,1}+{\cal O}\left(|\bar g_s|^4\right)\right)\geq|\bar
g_s|^{\chi/2}\sqrt{\alpha'}
\label{minimum}\ee
for the individual D-particle coordinates. For $\chi=0$ the minimal length in
\eqn{minimum} coincides with the standard smearing \cite{ven} due to the finite
size of the string, while for $\chi=\frac23$ it matches the 11-dimensional
Planck length $\ell_{\rm P}^{(11)}$ which arises from the kinematical
properties of D-particles \cite{liyoneya}. A choice of $\chi\neq0$ is more
natural since the modular strip divergences should be small for weakly
interacting strings. Note that the uncertainty \eqn{minimum} is always larger
in the string frame, representing the additional energetic smearing that arises
from the open string excitations on the D-particles. Outside of this frame we
obtain exactly the standard stringy smearings directly from the worldsheet
formalism, without the need of postulating an auxilliary uncertainty relation
as is done in \cite{ven,liyoneya}. With the present normalization of the mass
of the D-particles (see \eqn{tildePiorder3}), we see that the
velocity-dependent shift in \eqn{minimum} is just the kinetic energy of
D-particle $a$.

The coordinate uncertainties for $a\ne b$ are responsible for the emergence of
a true noncommutative quantum spacetime and represent the genuine non-abelian
characteristics of the D-particle dynamics. From \eqn{lambdapm} and
\eqn{Xiabtilde} it follows that, outside the string frame, the uncertainties
$|\Delta X_i^{ab}|$, $i>2$, are given by the same minimal length \eqn{minimum}
as for the individual D-particles. In string coordinates, we may assume, by
symmetry, that $|\Delta X_1^{ab}|\sim|\Delta X_2^{ab}|$. Then
\eqn{tildeYuncert} and \eqn{YXavgs} lead to a system of two linear equations in
two unknowns,
\bea
|\bar g_s|^\chi\alpha'\left(1+\frac{|\bar
g_s|^2}{36}\,\lambda_\pm(u)\right)&=&\left|\Delta X_1^{ab}\right|^2\pm4{\cal
N}^2B^{(1)}\,\|u_b\|\sin\theta_{ab}\left(\|u_a\|+B^{(1)}\|u_b\|\cos\theta_{ab}
\right)\nn\\& &~~~~~~~~~~~~~~~~~~~~\times~{\rm
Re}\left(\!\left(X^{ab}_1\biggm|\left[X^{ab}_2\right]^\dagger\right)\!
\right)_{\rm conn}
\label{abeqns}\eea
which hold up to ${\cal O}(|\bar g_s|^2)$. Adding the two equations
\eqn{abeqns} gives the smearings
\beq
\left|\Delta X_1^{ab}\right|=|\bar
g_s|^{\chi/2}\sqrt{\alpha'}\left[1+\frac{|\bar
g_s|^2}{144}\Bigl(3s_{ab}+t_{ab}\Bigr)+{\cal O}\left(|\bar g_s|^4\right)\right]
\label{absmear}\eeq
where we have introduced the kinematical invariants $s_{ab}=\|u_a+u_b\|^2$ and
$t_{ab}=\|u_a-u_b\|^2$ representing, respectively, the center of mass kinetic
energy and momentum transfer of the scattering of D-particles $a$ and $b$. The
uncertainty in measurement of an open string coordinate thus depends on both
the center of mass and relative energies of the two D-particles to which it is
attached. Its minimum coincides with that of \eqn{minimum}. Note that when
D-particles $a$ and $b$ move orthogonally to one another, i.e. their scattering
angle is $\theta_{ab}=\frac\pi2$, the uncertainty \eqn{absmear} depends only on
the total kinetic energy of the two particles. This is the case that is
discussed in \cite{nmprl}.

Subtracting the two equations \eqn{abeqns} gives the connected correlation
function
\beq
{\rm Re}\left(\!\left(X^{ab}_1\biggm|\left[X^{ab}_2\right]^\dagger\right)\!
\right)_{\rm conn}=\frac{|\bar
g_s|^{2+\chi}\alpha'\,\|u_a+B^{(1)}\,u_b\|^2\,{\cal
X}_{ab}(u)}{144B^{(1)}\,\|u_b\|\sin
\theta_{ab}\left(\|u_a\|+B^{(1)}\,\|u_b\|\cos\theta_{ab}\right)}
{}~~~~~~,~~~~~~a\neq b
\label{x1x2}\eeq
to ${\cal O}(|\bar g_s|^2)$, with
\bea
B^{(1)}&=&\frac{\|u_a\|^2\|u_b\|^2+u_a\cdot u_b\left[\|u_a\|^2+\|u_b\|^2-{\cal
X}_{ab}(u)\right]}{2(u_a\cdot u_b)^2+\|u_a\|^2\left[u_a\cdot
u_b+\|u_a\|^2-\|u_b\|^2-{\cal X}_{ab}(u)\right]}\nn\\& &~~~~\nn\\
{\cal
X}_{ab}(u)&=&\frac14\sqrt{\Bigl(3s_{ab}+t_{ab}\Bigr)^2+\frac{16\,\|u_a\|^2
\|u_b\|^2}{\sin^2\theta_{ab}}\left[1+\cos^2\theta_{ab}+\left(\frac{\|u_a\|}
{\|u_b\|}+\frac{\|u_b\|}{\|u_a\|}\right)\cos\theta_{ab}\right]^2}\nn\\& &~~~~
\label{B1expl}\eea
The result \eqn{x1x2} shows that for the scattering of D-particles, the
position operators of the open strings which mediate the interactions are not
independent random variables and have a non-trivial quantum mechanical
correlation. This is a new form of quantum spacetime uncertainty relations
between different {\it spatial} directions of target space. When $\chi=\frac23$
the right-hand side of \eqn{x1x2} can be written in terms of $(\ell_{\rm
P}^{(11)})^2$ and an additional complicated function of the D-particle kinetic
energies. For the transverse scattering of two D-particles of equal speed, this
function is just the total kinetic energy of the D-particles \cite{nmprl}. In
general though, the right-hand side of \eqn{x1x2} is a horrendously complicated
function of the scattering parameters. It demonstrates the complexity of the
open string interactions between D-branes, in that the smearing of the string
coordinates is a highly non-trivial function of the kinematical invariants of
the D-particles to which they are attached.

The energy dependence of \eqn{minimum}, \eqn{absmear} and \eqn{x1x2} is a
quantum decoherence effect which can be understood from a generalization of the
Heisenberg microscope whereby we scatter a low-energy probe, represented by a
closed string state with definite energy and momentum, off the D-particle
configuration. As the closed string state hits a D-particle, it splits into two
open string states, represented by the recoil of the particle upon impact with
the detector, which absorb energy from the scattering. Formally, such a
splitting is described by means of the conformal field theory formalism
developed in \cite{cardy}. When a closed string state, represented as a bulk
deformation by a closed string matter excitation $O$ on $\Sigma$ of scaling
dimension $\Delta_O$, approaches the boundary $\partial\Sigma$, then one can
infer the operator product expansion \cite{cardy,diehl}
\be
O(z,\bar z;s)\sim\sum_I\,(2s)^{\Delta_I-\Delta_O}\,C_{O,{\cal E}_I}^A\,{\cal
E}_I(s)
\label{openboundary}\ee
provided that the set of boundary conditions $A$ doesn't break the conformal
symmetry. The splitting amplitudes $C_{O,{\cal E}_I}^A$ can be expressed
\cite{cardy} in terms of bulk operator product expansion coefficients
$c_{ij}^k$. In the context of recoiling D-particles, the splitting coefficients
for a closed string state to split into a pair of open string excitations, with
their ends attached to the D-particles, have been shown \cite{diffusion} to be
non-zero by expressing them in terms of the bulk amplitude $c_{O,O}^D$ for an
``in'' closed string state to scatter off the D-brane into an ``out'' string
state, including the recoil operator $D$, the latter being represented as a
worldsheet bulk operator \cite{paban,recoil},
\be
\left(C_{O,D}^A\right)^2\simeq\frac{1}{\sqrt{\log\Lambda}}\,c_{O,O}^D
\label{corelation}\ee
In \eqn{corelation} we have concentrated for simplicity on the leading
divergent
contributions as $\epsilon\to0^+$ which are associated with the $D$ operator.
This allows for closed-to-open string state transitions within the present
framework.

For an isolated D-particle, these open string excitations have their ends
attached to the same point. For two D-particles the ends of the open
string can attach to different points. Since the recoil of the constituent
D-particles causes the fat brane as a whole to recoil as well, the
interactions mediated by the open strings cause a non-trivial correlation
between different coordinate degrees of freedom stretched between the two
particles. Only when there is no recoil ($u_a=u_b=0$) can one measure
independently the positions of the two D-particles. In this way the
uncertainties in length measurements and the position correlations between two
D-branes depend on the energy content of the scattering process and grow with
increasing recoil energies.

Notice that the correlation (\ref{x1x2}) we have derived is not simply a
product of uncertainties $\Delta X_1\Delta X_2 $, as is the usual case in
axiomatic approaches to spacetime quantization based on noncommutative geometry
\cite{doplicher} or as one would have naively expected from the Lie algebraic
noncommutativity of the multiple D-brane matrix coordinates $X_i^{ab}$. The
Schwarz inequality
\beq
\left|\left(\!\left(X_1^{ab}\Bigm|X_2^{ab}\right)\!\right)_{\rm
conn}\right|\leq\Delta X_1^{ab}\,\Delta X_2^{ab}
\label{schwarz}\eeq
leads to a spacetime uncertainty relation in the spirit of \cite{doplicher}.
However, the quantum mechanical correlation \eqn{x1x2} is much stronger than
this uncertainty relation, because two random variables can be independent yet
have non-vanishing variances, and as such it probes much deeper into the short
distance structure of spacetime. The present worldsheet approach associates the
Lie algebraic noncommutativity to a spacetime noncommutativity only rather
subtly through the  relation (\ref{x1x2}). This differs from the approach of
\cite{yoneya2b} which identifies the two types of noncommutative algebras using
the Schild formalism of string theory, in which case the uncertainties in the
D-particle positions are given by
\beq
\Bigl(\Delta
y_i^a\Bigr)^2\equiv\left[\left(Y_i-Y_i^{aa}\right)^2\right]^{aa}=\sum_{b\neq
a}\left|Y_i^{ab}\right|^2
\label{Lieuncert}\eeq
In contrast to our uncertainties, the smearing \eqn{Lieuncert} is a direct
result of the open string interactions between particle $a$ and all of the
other D-branes. The inequalities (\ref{x1x2}, \ref{schwarz}) essentially
summarize the implications of the noncommutative nature of spacetime on the
measurability of lengths. Their energy dependence distinguishes them from the
usual inequalities which arise in axiomatic noncommutative field theories
(which involve only the spacetime Planck length), and moreover the present
uncertainties are derived from Lagrangian dynamics for the system of
D-particles.

\newsubsection{Quantum Phase Space}

The quantum phase space of the multi D-brane system is determined by the
canonical momentum (\ref{tildePiorder3}) which, according to
(\ref{triantaena2}), upon quantization becomes an operator
$\widehat{\Pi}_{ab}^j$ obeying the Heisenberg commutation relations
\beq
\left[\!\left[\widehat{Y}_i^{ab},\widehat{\Pi}_{cd}^j\right]\!\right]=
i\hbar_{\cal M}\,\delta_i^j\delta_c^a\delta_d^b
\label{heisencomms}\eeq
on $\cal M$. The relation \eqn{heisencomms} leads to the moduli space
Heisenberg uncertainty principle
\beq
\Delta\bar Y_i^{ab}\,\Delta\widetilde{\Pi}_{cd}^j\geq\mbox{$\frac12$}\,
\hbar_{\cal M}\,\delta_i^j\delta_c^a\delta_d^b
\label{heisenuncerts}\eeq
The Planck constant $\hbar_{\cal M}$ can be determined by noting that, in the
present context, the partition function \eqn{pinchedpartfn} is identified with
the wavefunction of the multi D-brane system. The lower bound in
\eqn{heisenuncerts} is then saturated if one interprets \eqn{pinchedpartfn} as
a minimum uncertainty wavepacket on moduli space. In the single D-particle
case, such an assumption is consistent with the solution of a generalized
Schr\"odinger equation \cite{diffusion}, stemming from an application of a
worldsheet Wilsonian renormalization group equation, under the identification
of the Liouville field with target time.

Since we have effectively been representing the canonical momentum
$\widetilde{\Pi}^j_{ab}$ as an operator in coupling constant space (see
\eqn{momenta}), the effects of the summation over worldsheet topologies on it
are implicitly already taken into account. This means that the variance
$(\Delta\widetilde{\Pi}_{ab}^j)^2$ can be computed in the worldsheet
$\sigma$-model on a tree-level disc topology. In this way, using the two-point
and one-point functions \eqn{Gfinal} and \eqn{tildePiorder3} we find
\bea
\left(\Delta\widetilde{\Pi}_{ab}^j\right)^2&=&
\widetilde{G}^{jj}_{ab;ab}(\bar Y,\bar U)-\left(\widetilde{\Pi}_{ab}^j(\bar
Y,\bar U)\right)^2\nn\\&=&\frac{4\bar g_s^2}{\alpha'}\,\delta_{ab}+\frac{2\bar
g_s^4}{9\alpha'}\left(2\delta_{ab}\left[(\bar U^j)^2\right]_{ba}-287\left(\bar
U_{ba}^j\right)^2\right)+{\cal O}\left(\bar g_s^6\right)
\label{momentuncert}\eea
to lowest orders in $\bar g_s$. $\hbar_{\cal M}$ can then be found by
performing a Galilean boost to a co-moving target space frame in which the
recoil velocities vanish. For example, setting $a=b=c=d$ in \eqn{heisenuncerts}
with the inequality saturated and substituting in \eqn{minimum} and
\eqn{momentuncert} for $\bar U=0$, we can solve for the moduli space Planck
constant to get
\beq
\hbar_{\cal M}=4|\bar g_s|^{1+\chi/2}
\label{hbarM}\eeq
which we note is time independent. Thus the basic constant $\hbar_{\cal M}$ of
the resulting quantum phase space is proportional to the string coupling $|\bar
g_s|$, which owes to the fact that in the present case quantum mechanics is
induced by string interactions.

The velocity-dependent terms in \eqn{momentuncert} correspond to stringy
corrections. As mentioned at the beginning of section 5, to lowest order in the
string coupling constant expansion, the moduli space coincides with the phase
space of the D-particle system. This means that, with the appropriate mass
normalization, we can identify the canonical momentum with the velocity $\bar
U$, so that to lowest orders the position and velocity operators have a
canonical quantum commutator of the form \eqn{heisencomms}. We can therefore
compute the commutator $[[\widehat{Y}_i^{ab},\widehat{\Pi}^j_{cd}]]$
iteratively, using \eqn{tildePiorder3}, by assuming a position-velocity
commutator of the form \eqn{heisencomms} and identifying the velocity-squared
terms which arise from the commutators involving the $\bar U^3$ terms in
\eqn{tildePiorder3} with squares of the momentum operator $\widehat{\Pi}$.
After some algebra, this leads to the string-modified Heisenberg commutation
relations
\bea
\left[\!\left[\widehat{Y}_i^{ab},\widehat{\Pi}^j_{cd}\right]\!\right]&=&i
\hbar_{\cal M}\left(\delta_i^j\delta_c^a\delta_d^b+\mbox{$\frac1{96}$}\,|\bar
g_s|^2\bar\alpha_s'
\left[\delta_i^j\left(\delta^a_c\,[\widehat{\Pi}_k
\widehat{\Pi}^k]_d^b+\delta_d^b\,[\widehat{\Pi}_k\widehat{\Pi}^k]^a_c+[
\widehat{\Pi}_k]^a_c[\widehat{\Pi}^k]_d^b\right)\right.\right.\nn\\&
&~~~~~~~~~~\left.\left.+\,\delta^a_c\left\{\widehat{\Pi}_i,\widehat{\Pi}^j
\right\}^b_d+\delta_d^b\left\{\widehat{\Pi}_i,\widehat{\Pi}^j\right\}^a_c+
[\widehat{\Pi}_i]^b_d[\widehat{\Pi}^j]^a_c+[\widehat{\Pi}_i]^a_c
[\widehat{\Pi}^j]^b_d\right]+\dots\right)\nn\\& &~~~~
\label{strmodheisencomm}\eea
to leading orders, where $\bar\alpha_s'=\alpha'/|\bar g_s|^4$ is the (time
independent) 0-brane scale with the present mass normalization.

The commutation relation \eqn{strmodheisencomm} represents the appropriate
generalization of the string-modified phase space relations
(\ref{stringuncert}, \ref{modHeisen}) to the multi D-particle case. For
$a=b=c=d$ and $i=j$ it reproduces the standard string-modified phase space
uncertainty principle \cite{ven} for a single recoiling D-particle
\cite{kmw,lm}. However, it also takes into account of the various string
interactions among D-particles (the off-diagonal parts of
\eqn{strmodheisencomm}). Minimizing the off-diagonal components (in both
Lorentz and colour indices) of the uncertainty relations corresponding to
\eqn{strmodheisencomm} leads to non-trivial kinetic energy dependent
uncertainties among the various open string excitations, and also along
different spatial directions. The relation \eqn{strmodheisencomm} represents
the phase space version of the noncommutative quantum uncertainties that were
derived in the previous subsection. We note that, even for a single D-particle,
at higher orders in $\bar g_s$ the phase space uncertainty relations here are
different from the ones derived in \cite{ven} in that the modifications depend
on the recoil velocities and not only on the uncertainties in the momenta. In
fact, the present approach gives a formal prescription for evaluating the
higher-order stringy corrections to the Heisenberg uncertainty relations in
string perturbation theory, in principle to arbitrary order in the (weak)
string coupling constant.

\newsubsection{Space--time Uncertainty Principles}

Upon summation over worldsheet genera the physical target space time coordinate
\eqn{phystimeexpl} becomes a quantum operator $\widehat{T}$ \cite{a-c}, unlike
the situation in conventional quantum mechanics. Within the present
Born-Oppenheimer approximation, we can expand the function \eqn{calFdef} as a
power series in $\|\bar U_{ab}\|/\|u_c\|\ll1$, $a\neq b$, using the identity
\beq
\tr\,\bar U_i\bar U^i=\frac{E_{\rm tot}}{|\bar g_s|^2}\left(1+|\bar
g_s|^2\sum_{a\neq b}\frac{\bar U_i^{ab}\bar U_{ba}^i}{E_{\rm tot}}\right)
\label{kinenid}\eeq
where $E_{\rm tot}=|\bar g_s|^2\sum_{a=1}^N\|\bar U_{aa}\|^2$ is the total
kinetic energy (per unit string length) of the individual D-particles. We
substitute \eqn{kinenid} into \eqn{phystimeexpl}, expand the square root to
lowest order in the off-diagonal velocities, and average over the worldsheet
renormalization group time parameter $\log\Lambda$. Identifying the velocity
operators with $\widehat{\Pi}^j_{ab}$ as described in the previous subsection
and using the Heisenberg commutation relations \eqn{heisencomms} we arrive at
the space--time quantum commutation relations
\beq
\left[\!\left[\widehat{Y}_i^{ab},\widehat{T}\right]\!\right]\simeq\frac{i
\alpha'\hbar_{\cal M}}{2|\bar
g_s|}\left(\delta^{ab}+\left(1-\delta^{ab}\right)\frac{\sqrt{\alpha'}}{4|\bar
g_s|}\,\frac{\widehat{\Pi}_i^{ab}}{\sqrt{E_{\rm tot}}}\right)
\label{na}\eeq
to leading order in $\bar g_s$ (or equivalently in the off-diagonal velocity
expansion).

{}From \eqn{hbarM} and \eqn{na} we infer the space--time uncertainty relation
\be
\Delta\bar Y_i^{aa}\,\Delta T\geq|\bar g_s|^{\chi/2}\alpha'
\label{timespace}\ee
for the individual D-particle coordinates. For $\chi=0$, \eqn{timespace} yields
the standard lower bound \eqn{tyuncert} which is independent of the string
coupling, as argued in \cite{yoneya}--\cite{liyoneya} from basic string
ideas. But then the minimal distance \eqn{minimum} doesn't
probe scales down to the 11-dimensional Planck length. This fact can be
understood by noting that the physical target space (Liouville) time coordinate
$T$ is not the same as the longitudinal worldline coordinate of a D-particle,
as is assumed in the arguments leading to the hypothesis \eqn{tyuncert}, but is
rather a collective time coordinate of the D-particle system which is induced
by all of the string interactions among the particles. However, we can adjust
the uncertainty relations to match the dynamical properties of 11-dimensional
supergravity by multiplying the definition (\ref{phystimeexpl}) by an overall
factor of $|\bar g_s|^{-\chi/2}$. This redefinition will be assumed below, and
it implies that with weak string interactions the target space propagation time
for the D-particles is very long.

To see the effects of the string interactions between D-particles in this
space--time framework, we again use the canonical (minimal) smearing
\eqn{heisenuncerts} between $\widehat{Y}^{ab}_i$ and $\widehat{\Pi}_{ab}^i$ for
$a\ne b$ in \eqn{na} to arrive at a triple uncertainty relation
\be
\left(\Delta\bar Y_i^{ab}\right)^2\Delta
T\geq\frac{|\bar g_s|^{\chi/2}\alpha'^{3/2}}{2\sqrt{E_{\rm
tot}}}~~~~~~,~~~~~~a\neq b
\label{triple}\ee
The uncertainty principle \eqn{triple} depends on the total kinetic energy of
the constituent D-branes. It implies that the system of D-particles, through
their open string interactions, can probe distances much smaller than the
characteristic distance scale in \eqn{triple}, which for $\chi=\frac23$ is
$\ell_{\rm P}^{(11)}\ell_s^2$, provided that their kinetic energies are large
enough. In the fully relativistic case the existence of a limiting speed
$\|u_a\|<1$ implies a lower bound on \eqn{triple}. With the minimum spatial
extensions obtained in subsection 6.3, this bound yields, for $\chi=\frac23$,
the characteristic temporal length
\be
\Delta T\geq|\bar g_s|^{-1/3}\,\sqrt{\alpha'}
\label{temlength}\ee
for D-particles \cite{liyoneya} (see \eqn{minyt}). Triple uncertainty relations
involving only the 11 dimensional Planck length have been suggested in
\cite{liyoneya} based on the holographic principle of M-theory.

Again the present approach formally gives a prescription for evaluating
higher-order contributions to the space--time quantum commutator \eqn{na} in
the
string coupling $\bar g_s$ (or in the velocity expansion). A characteristic
feature of the uncertainty relations we have derived in this section, which
distinguishes D-particle dynamics from ordinary quantum mechanics, is their
dependences on the recoil momenta. The dependence of quantum uncertainties in
the measurement of certain quantities on the magnitude of the quantities
themselves (here the kinetic energies of the D-branes) is characteristic of
decoherence effects which are induced by quantum gravity \cite{a-cqg}. It was
argued in \cite{diffusion} that the quantum recoil degrees of freedom are
responsible for inducing decoherence in low-energy systems. In the case of a
single D-particle, the analysis of \cite{ekmnw} demonstrates explicitly the
induced decoherence by exhibiting particle creation in the direction of the
recoiling velocities for the scattering of a spectator light mode in the
presence of a D-particle due to the scattering of another closed string state
off the defect. The analysis of this section thus shows that multiple
D-particle field theory in flat target spaces naturally incorporates quantum
gravity
effects into the sub-Planckian spacetime structure. It therefore illuminates
the manner in which D-particle interactions probe very short distances where
the effects of quantum gravity are significant.

\newsection{Conclusions}

In this paper we have employed a worldsheet approach to the study of the
collective dynamics of $N$ parallel D-branes, interacting through the
exchange of open (or closed) strings, which are scattered off them. This is the
simplest model of multi-brane dynamics, where the branes do not intersect.
Working with Neumann boundary conditions, in which the coupling constants of
the pertinent $\sigma$-model are $U(N)$ gauge potentials, we have developed a
formalism for describing recoil of the multi-brane system after scattering with
low-energy string states. This formalism utilizes generic properties of
logarithmic conformal field theories on the worldsheet. In this way we have
shown that the recoil deformations define a system of collective coordinates
and momenta which are consistent with the corresponding ones derived from a
(symmetrized) non-abelian Born-Infeld effective action. We have argued
that worldsheet genus expansion produces quantum fluctuations (in target space)
of these $\sigma$-model couplings. For a specific choice of consistent gauge
field backgrounds, therefore, a quantum phase space arises, which however
involves noncommutativity among all coordinate directions as a result of the
interactions of the branes. We also derived new coordinate uncertainty
relations, among different components of the coordinate matrices of the
interacting D-branes, consistent with generic expectations from noncommutative
geometry analyses. These relations justify properly the association of Lie
algebraic noncommutativity with quantum mechanical noncommutativity, and as we
have discussed this is a non-trivial fact. We have also discussed the
definition of target time in the context of the Liouville approach, and shown
that it becomes an operator in this formalism, which exhibits unconventional
uncertainty relations with the collective coordinates.

There are many aspects of the approach of this paper that still require
examination. The most glaring one is the arbitrariness of the exponent $\chi$
in \eqn{deltaepcorr}. In the present approach, which considers only string
interactions, we have not found any way to fix its value, but it may be fixed
upon considering brane exchanges between the system of D-branes. Another aspect
that needs to be worked out is the explicit calculation of the perturbation
expansion to some higher-orders which will begin to involve not just the
velocities of the D-particles, but also their collective coordinates. The
resulting moduli space geometry, which as we have shown naturally describes the
structure of spacetime at sub-Planckian scales, will then contain information
not only about the kinematics of the D-particles, but also of their dynamics
which are governed by terms such as the Yang-Mills potential \eqn{D0potential}.
This would then lead to spacetime noncommutativity from the quantum phase space
structure itself, and presumably new forms of spacetime uncertainty relations.
Of course, the present results only strictly apply to the simplest physical
system whose motion is governed by fat brane dynamics. It would be interesting
to consider more complicated matrix D-brane couplings involving, for example,
higher-rank Jordan blocks in the spectrum of the underlying logarithmic
conformal field theory. Such generalizations may probe deeper into the nature
of the string interactions among the branes, and hence into the small-scale
structure of spacetime. Another generalization involves the incorporation of
intersecting D-branes in this formalism. It would be interesting to see whether
there exists an appropriate generalization of logarithmic operators that
describes quantum fluctuations of such systems. Such constructions are crucial
to the understanding of the stringy quantum spacetime at sub-Planckian scales.
They may also shed further light on the short-distance structure, fundamental
degrees of freedom and dynamics of M-theory within the geometrical framework of
moduli space dynamics.

It would be interesting to see if the present worldsheet approach, which
exhibits unconventional properties of string spacetimes, is amenable in some
way to experimental verification. The presence of multi D-brane domain wall
structures, like the ones considered in this paper, may act as traps of
low-energy string states, thereby resulting in a decoherent medium nature of
quantum gravity spacetime foam. In the present case the quantum coordinate
fluctuations, due to the open string excitations between the D-particles, can
lead to quantum decoherence for a low-energy observer who cannot detect such
recoil fluctuations in the sub-Planckian spacetime structure. These foamy
properties of the noncommutative structure of the D-particle spacetime might
require a reformulation of the phenomenological analyses of length measurements
as probes of quantum gravity. If one accepts the generic $\ell_{\rm P}$ maximal
suppression effects by the gravitational (Planck) mass scales, then, as
described in some recent literature, there may be sensitive probes such as
neutral kaon sytems \cite{dafne} or cosmological gamma-ray burst spectroscopy
\cite{grb}. However, such approaches do not incorporate length measurements
in the transverse directions to the probe, so that it is unclear how to
incorporate noncommutative uncertainty relations such as \eqn{x1x2} into these
analyses.

\bigskip

\noindent
{\bf Acknowledgements:} We are grateful to J. Ellis, A. Kempf, F. Lizzi, J.
Wheater and E. Winstanley for helpful discussions. A preliminary version of
this paper was presented by {\sc r.j.s.} at the {\it SUSY '98} conference in
Oxford, England in July 1998. We thank the organisors and participants of the
conference for having provided a stimulating environment.

\setcounter{section}{0}
\setcounter{subsection}{0}
\setcounter{equation}{0}
\renewcommand{\thesection}{Appendix \Alph{section}}
\renewcommand{\theequation}{\Alph{section}.\arabic{equation}}

\newsection{Correlation Functions of Logarithmic Operators}

In this appendix we will describe some properties and compute the first few
correlation functions of the $C$ and $D$ logarithmic operators that were
introduced in section 3. They are calculated using fundamental string averages
which are evaluated with the propagator
\beq
\left\langle x^\mu(z_1,\bar z_1)x^\nu(z_2,\bar
z_2)\right\rangle_0=2\alpha'\,\eta^{\mu\nu}\,\log|z_1-z_2|
\label{xprop}\eeq
associated with the action \eqn{freesigma}. The coincidence limit of the
two-point function \eqn{xprop} is defined using the short-distance cutoff
$\Lambda$ as
\beq
\left\langle x^\mu(z,\bar z)x^\nu(z,\bar
z)\right\rangle_0=2\alpha'\,\eta^{\mu\nu}\,\log\Lambda
\label{2ptLambda}\eeq

The correlators of the logarithmic operators \eqn{Depsilonop} and
\eqn{Cepsilonop} can now be evaluated using the regulated step function
\eqn{Thetafnreg}. Note that upon integrating by parts the $D$ operator can be
written as
\beq
D(x^0;\epsilon)=-\frac1{2\pi}\int_{-\infty}^\infty\frac{dq}{(q-i\epsilon)^2}~
\e^{iqx^0}=-\frac1{\epsilon}\frac\partial{\partial\epsilon}\,C(x^0;\epsilon)
\label{DCderiv}\eeq
The second equality in \eqn{DCderiv} also follows from the general property
$D=\alpha'\partial C/\partial\Delta_\epsilon$ of logarithmic conformal field
theories \cite{W}. This property enables one to deduce expressions for many of
the correlators once the expectation values of the $C$ operator are known
\cite{r-tak}. Using these identities one can compute explicitly the one-point
correlation functions in the correlated limit $\epsilon\to0^+$ with the
relation \eqn{scalerel} to get
\cite{kmw}
\beq
\left\langle C(x^0;\epsilon)\right\rangle_0={\cal
O}(\epsilon)~~~~~~,~~~~~~\left\langle D(x^0;\epsilon)\right\rangle_0=a/\epsilon
\label{1ptfns}\eeq
where here and in the following $a,b,\dots$ denote (in principle arbitrary)
dimensionless constants.

The higher-point correlators can be computed using the Koba-Nielsen formula
\beq
\left\langle\prod_{i=1}^n\e^{iq_ix^0(z_i,\bar
z_i)}\right\rangle_0=\prod_{i,j}\e^{-q_iq_j\langle x^0(z_i,\bar
z_i)x^0(z_j,\bar z_j)\rangle_0/2}
\label{knformula}\eeq
For the two-point functions one finds, always in the correlated limit
$\epsilon\to0^+$, the expressions \cite{kmw}
\bea
\left\langle C(z,\bar z;\epsilon)C(w,\bar w;\epsilon)\right\rangle_0&=&{\cal
O}(\epsilon^2)\label{2ptCC}\\\left\langle C(z,\bar z;\epsilon)D(w,\bar
w;\epsilon)\right\rangle_0&=&\frac
b{|z-w|^{2\Delta_\epsilon}}\label{2ptCD}\\\left\langle D(z,\bar
z;\epsilon)D(w,\bar w;\epsilon)\right\rangle_0&=&\frac1{\epsilon^2}\left\langle
C(z,\bar z;\epsilon)D(w,\bar w;\epsilon)\right\rangle_0\nn\\&=&-\frac
{b\alpha'}{|z-w|^{2\Delta_\epsilon}}\left(\frac1{2\Delta_\epsilon}+\log
\left|\frac{z-w}\Lambda\right|^2\right)
\label{2ptDD}\eea
{}From \eqn{scalerel} it follows that \eqn{2ptCC}--\eqn{2ptDD} have the
canonical form of the two-point correlation functions of a generic logarithmic
conformal field theory. The constant in \eqn{2ptDD} which depends on the
anomalous dimension $\Delta_\epsilon$ can be made arbitrary by shifting the $D$
operator according to \eqn{DCscaletransfs} (i.e. by a worldsheet scale
transformation), whereas the coefficient $b$ is fixed by the leading
logarithmic terms in the conformal blocks. Note that the correlators in
\eqn{1ptfns} and \eqn{2ptCC}--\eqn{2ptDD} involving solely the $C$ field vanish
while those involving only the $D$ field diverge as $\epsilon\to0^+$.

The three-point functions of the logarithmic pair can be calculated using the
canonical forms derived for general logarithmic conformal field theories in
\cite{r-tak}. As in \eqn{2ptCC}--\eqn{2ptDD}, these correlators involve some
arbitrary (integration) constants, while the coefficients of the
logarithmically divergent terms are fixed by the leading logarithmic behaviours
of the conformal blocks. We can therefore apply the results of \cite{r-tak} to
the present case (using the behaviours of \eqn{2ptCC}--\eqn{2ptDD}) provided we
know the leading behaviours of the three-point functions as $\epsilon\to0^+$.
For example, consider the three-point function of the $C$ fields, which using
\eqn{knformula}, \eqn{xprop} and \eqn{2ptLambda} is given by
\beq
\left\langle C(z_1,\bar z_1;\epsilon)C(z_2,\bar z_2;\epsilon)C(z_3,\bar
z_3;\epsilon)\right\rangle_0=\frac{\epsilon^3}{(2\pi
i)^3}\int_{-\infty}^\infty\prod_{k=1}^3\frac{dq_k}{q_k-i\epsilon}~
\e^{-\alpha'q_k^2\log\Lambda}\prod_{k<j}\e^{-2\alpha'q_kq_j\log|z_{kj}|}
\label{3ptfnCdef}\eeq
Using \eqn{scalerel} and rescaling the integration variables in \eqn{3ptfnCdef}
as $q_k=\epsilon\tilde q_k$, we have
\beq
\left\langle C(z_1,\bar z_1;\epsilon)C(z_2,\bar z_2;\epsilon)C(z_3,\bar
z_3;\epsilon)\right\rangle_0=\frac{\epsilon^3}{(2\pi
i)^3}\int_{-\infty}^\infty\prod_{k=1}^3\frac{d\tilde q_k}{\tilde
q_k-i}~\e^{-\tilde q_k^2/2}\prod_{k<j}\e^{-2\alpha'\epsilon^2\tilde q_k\tilde
q_j\log|z_{kj}|}
\label{3ptfnCscaled}\eeq
The last product in \eqn{3ptfnCscaled} has the form
$\prod_{k<j}\e^{-2\alpha'\epsilon^2\tilde q_k\tilde q_j\log|z_{kj}|}\sim1+{\cal
O}(\epsilon^2)$, so that the three-point function has leading constant term
which vanishes as $\epsilon^3$, while the remaining $z$-dependent terms coming
from the final product in \eqn{3ptfnCscaled} are sub-leading in $\epsilon$.
Thus $\langle CCC\rangle_0\sim\epsilon^3$. Using exactly the same method one
shows that $\langle CCD\rangle_0\sim\epsilon$, $\langle
CDD\rangle_0\sim1/\epsilon$ and $\langle DDD\rangle_0\sim1/\epsilon^3$.

{}From these leading behaviours in $\epsilon$ we can now read off from
\cite{r-tak} the three-point correlation functions,\footnote{In the
perturbative calculations of sections 4 and 5 we neglected throughout the parts
of the correlators which involve exponents of the scaling dimension
$\Delta_\epsilon$, as these terms do not contribute to the leading divergences
as $\epsilon\to0^+$.}
\bea
& &\left\langle C(z_1,\bar z_1;\epsilon)C(z_2,\bar z_2;\epsilon)C(z_3,\bar
z_3;\epsilon)\right\rangle_0~=~\frac{c\epsilon^3}{|z_{12}z_{23}z_{31}|^{2
\Delta_\epsilon}}\label{3ptCCC}\\& &\left\langle D(z_1,\bar
z_1;\epsilon)C(z_2,\bar z_2;\epsilon)C(z_3,\bar
z_3;\epsilon)\right\rangle_0~=~\frac1{|z_{12}z_{23}z_{31}|^{2
\Delta_\epsilon}}\left(d\epsilon+\frac
c2\epsilon^3\alpha'\log\left|\frac{z_{23}\Lambda}{z_{12}z_{31}}\right|^2\right)
\nn\\& &~~~~\label{3ptCCD}\\& &\left\langle D(z_1,\bar z_1;\epsilon)D(z_2,\bar
z_2;\epsilon)C(z_3,\bar
z_3;\epsilon)\right\rangle_0~=~\frac1{|z_{12}z_{23}z_{31}|^{2
\Delta_\epsilon}}\left(\frac
e\epsilon-d\epsilon\alpha'\log\left|\frac{z_{12}}\Lambda\right|^2\right.
\nn\\& &~~~~~~~~~~~~~~~~~~~~~~~~~~~~~~~~~~~~~~~~~~~~~
\left.+\,c\epsilon^3\alpha'^2\left[\left(\frac{\log\left|
\frac{z_{23}}\Lambda\right|^2}{\log\left|\frac{z_{31}}\Lambda\right|^2}
\right)^2+\frac14\left(\log\left|\frac{z_{12}}\Lambda\right|^2
\right)^2\right]\right)\nn\\& &~~~~\label{3ptDDC}\\& &\left\langle D(z_1,\bar
z_1;\epsilon)D(z_2,\bar z_2;\epsilon)D(z_3,\bar
z_3;\epsilon)\right\rangle_0~=~\frac1{|z_{12}z_{23}z_{31}|^{2
\Delta_\epsilon}}\left\{\frac f{\epsilon^3}-\frac
e{2\epsilon}\alpha'\log\left|\frac{z_{12}z_{23}z_{31}}{\Lambda^3}\right|^2
\right.\nn\\& &
{}~~~~~~~~~~+\,d\epsilon\alpha'^2\left[\log\left|\frac{z_{12}}
\Lambda\right|^2\log\left|\frac{z_{23}}\Lambda\right|^2+\log\left|\frac{z_{12}}
\Lambda\right|^2\log\left|\frac{z_{31}}\Lambda\right|^2+\log\left|\frac{z_{23}}
\Lambda\right|^2\log\left|\frac{z_{31}}\Lambda\right|^2\right]
\nn\\& &~~~~~~~~~~
-\frac{d\epsilon}4\alpha'^2\left(\log\left|\frac{z_{12}z_{23}z_{31}}
{\Lambda^3}\right|^2\right)^2+c\epsilon^3\alpha'^3\log\left|\frac{z_{12}}
\Lambda\right|^2\log\left|\frac{z_{23}}\Lambda\right|^2\log\left|\frac{z_{31}}
\Lambda\right|^2\nn\\& &~~~~~~~~~~-\frac
c2\epsilon^3\alpha'^3\log\left|\frac{z_{12}z_{23}z_{31}}{\Lambda^3}\right|^2
\left[\log\left|\frac{z_{12}}\Lambda\right|^2\log\left|\frac{z_{23}}\Lambda
\right|^2+\log\left|\frac{z_{12}}\Lambda\right|^2\log\left|\frac{z_{31}}
\Lambda\right|^2\right.\nn\\& &~~~~~~~~~~
\left.\left.+\log\left|\frac{z_{23}}
\Lambda\right|^2\log\left|\frac{z_{31}}\Lambda\right|^2\right]+\frac
c8\epsilon^3\alpha'^3\left(\log\left|\frac{z_{12}z_{23}z_{31}}{\Lambda^3}
\right|^2\right)^3\right\}
\label{3ptDDD}\eea
Note that on the boundary of the worldsheet $\Sigma$ where $z_i=\e^{2\pi
is_i}$, $s_i\in[0,1]$, the propagator \eqn{xprop} becomes
\beq
\left\langle
x^\mu(s_1)x^\nu(s_2)\right\rangle_0=\alpha'
\,\eta^{\mu\nu}\,\log\left[2-2\cos2\pi(s_1-s_2)\right]
\label{xpropboundary}\eeq
This can be used to express all correlators above in terms of the boundary
variables. Comparing \eqn{xpropboundary} with \eqn{2ptLambda} we see that the
short-distance cutoff on the boundary variables is
\beq
s_\Lambda=\mbox{$\frac1{2\pi}$}\arccos\left(\mbox{$1-\frac{\Lambda^2}2$}
\right)=\mbox{$\frac{\Lambda^2}{2\pi}+\frac{\Lambda^4}{48\pi}$}\,+{\cal
O}(\Lambda^6)
\label{bdrycutoff}\eeq
Furthermore, differentiating \eqn{xpropboundary} we arrive at the correlator
\beq
\left\langle\frac d{ds_1}x^i(s_1)\,\frac
d{ds_2}x^j(s_2)\right\rangle_0=\frac{4\pi^2\alpha'\,\eta^{ij}}
{1-\cos2\pi(s_1-s_2)}
\label{derivprop}\eeq

The calculation of $n$-point functions with $n\geq4$ is quite cumbersome. As
described in \cite{r-tak}, they can be evaluated in principle by noting that
the $C$ operators are primary fields and hence have standard conformal field
theoretical correlation functions, from which all other correlators of the
logarithmic pair may be found via differentiation using the identity
\eqn{DCderiv}. Their behaviours as $\epsilon\to0^+$ can be deduced rather
directly using relations analogous to \eqn{3ptcoeffsgen} between the
three-point functions and the operator product expansion coefficients, which
remain valid in
the presence of logarithmic deformations \cite{gflcft}. The logarithmic pair
$C,D$ form a complete set of states in the $2\times2$ Jordan cell of the
Virasoro generator $L_0$. From \eqn{2ptCC}--\eqn{2ptDD} it follows that the
Zamolodchikov metric in the $C,D$ basis behaves as \cite{lm,gflcft}
\beq
G_{CC}\sim\epsilon^2~~~~~~,~~~~~~G_{DD}\sim\epsilon^{-2}~~~~~~,~~~~~~
G_{CD}=G_{DC}\sim{\rm const.}
\label{ZCDscale}\eeq
Then \eqn{123rel} yields, for example, the scaling behaviour
\beq
\langle CC\rangle_0\sim G^{CC}\langle CCC\rangle_0\langle
C\rangle_0+G^{DD}\langle CCD\rangle_0\langle D\rangle_0+G^{CD}\Bigl(\langle
CCC\rangle_0\langle D\rangle_0+\langle CCD\rangle_0\langle C\rangle_0\Bigr)
\label{CC123}\eeq
{}From \eqn{2ptCC} we see that the left-hand side of \eqn{CC123} is ${\cal
O}(\epsilon^2)$. Then using \eqn{1ptfns} and \eqn{ZCDscale} we can immediately
deduce the anticipated small $\epsilon$ behaviours of $\langle CCC\rangle_0$
and $\langle CCD\rangle_0$. The general result is
\beq
\left\langle\prod_{i=1}^nC(z_i,\bar z_i;\epsilon)\prod_{j=1}^mD(w_j,\bar
w_j;\epsilon)\right\rangle_0\sim{\cal O}\left(\epsilon^{n-m}\right)
\label{CDgenscale}\eeq
This relation does not, however, yield any information about the logarithmic
scaling violations present in the correlation functions, i.e. their dependences
on the worldsheet renormalization group scale $\log\Lambda$.

\newsection{Renormalization of the Canonical Momentum}

In this appendix we shall derive the expression \eqn{Pirenordern} for the
renormalized canonical momentum. From \eqn{etaavgs} it follows that the
momentum contribution \eqn{Piordern} can be written as a sum over permutations
$P\in S_{n+2}$. This sum can be decomposed into a sum over permutations $P\in
S_n\times S_2$ which permute only contractions among the
$\prod_{k=1}^n\bar\xi_{a_k}(s_k-\epsilon)\xi_{b_k}(s_k)$ part
of the auxilliary field expectation value in \eqn{Piordern} among themselves,
and the remaining $(abc)$ part of this correlator among themselves, plus a sum
over the remaining ones $P\in S_{n+2}-(S_n\times S_2)$. Let us first
introduce some short-hand notation. For each positive integer $m$ we define an
$m$-dimensional integration measure $d\mu_m$ on $[0,1]^m$ by
\bea
\int_{[0,1]^m}d\mu_m(s_1,\dots,s_m)&\equiv&\int_0^1\prod_{k=1}^mds_k~
\left(\prod_{l=1}^{m-1}\Theta\left(s_{l+1}-s_l\right)\right)\Theta
\left(s_1-s_m\right)\nn\\&=&\int_0^1ds_m~\int_0^1\prod_{k=2}^{2[\frac
m2]-2}ds_k~\int_{\alpha(m)}^{s_m}ds_{m-1}\nn\\&
&~~~~~~~~~~~~~~~\times\left(\prod_{k=2[\frac
m2]-3}^2\int_{s_k}^{s_{k+2}}ds_{k+1}\right)~\int_{s_m}^{s_2}ds_1
\label{mmeas}\eea
where $[\frac m2]$ is the integer part of $\frac m2$, and $\alpha(m)=s_{m-2}$
for $m$ even and $\alpha(m)=0$ for $m$ odd. We define the initial value $\int
d\mu_{m=0}\equiv1$. We also define an $N\times N$ Hermitian matrix ${\cal T}_m$
by
\beq
{\cal
T}_m[Y,x;s_1,\dots,s_m]_{ab}\equiv\left[\prod_{k=1}^mY_{j_k}\left(x^0(s_k)
\right)\,\frac d{ds_k}x^{j_k}(s_k)\right]_{ab}
\label{calTdef}\eeq
with the initial value $[{\cal T}_{m=0}]_{ab}\equiv\delta_{ab}$.

We begin by evaluating the contribution to \eqn{Piordern} from $P\in S_n\times
S_2$, which give
\bea
& &{\cal P}_{ab}^{(n)j}[Y;s]\Bigm|_{S_n\times
S_2}=\lim_{\epsilon\to0^+}\sum_{c=1}^N~\sum_{b_1,\dots,b_n}~\sum_{P\in
S_n}~\int_0^1\prod_{k=1}^nds_k~\Theta\left(s_{P(k)}-s_k\right)\nn\\&
&~~~~~~\times\left(\Theta(\epsilon)\delta_{ab}
\delta_{cc}+\delta_{ac}\delta_{cb}\right)\left\langle\frac
d{ds}x^j(s)\prod_{k=1}^nY_{i_k}^{b_{P(k)},b_k}\left(x^0(s_k)\right)\,\frac
d{ds_k}x^{i_k}(s_k)\right\rangle_0
\label{PiSn}\eea
where we have explicitly summed over the $S_2$ part. To express \eqn{PiSn} in a
more succinct form, we decompose each permutation $P\in S_n$ into a product of
disjoint cycles $C_i(P)$,
\beq
P=\prod_{i=1}^nC_i(P)
\label{Permdecomp}\eeq
and let $L_i(P)\geq0$ denote the length of the cycle $C_i(P)$, so that the set
of integers $\{L_i(P)\}$ form a partition of $n$,
\beq
\sum_{i=1}^nL_i(P)=n
\label{LiPn}\eeq
It is then possible to express the boundary measure and $Y$-matrix products in
\eqn{PiSn} in a more explicit form by writing products $\prod_{k=1}^n$ in terms
of this cyclic decomposition as $\prod_{i=1}^n\prod_{m=1}^{L_i(P)}$ for each
$P\in S_n$. We can explicitly combine the products in the correlation functions
in \eqn{PiSn} into matrix products, using the cyclicity of each $C_i(P)$ and
summing over the $b_i$'s. We can also label the boundary integrations
$s_k\equiv s_{C_i(P)_{(k)}}$ in terms of the components of the cycles in
\eqn{Permdecomp}, giving the integration measure \eqn{mmeas} for each
$i=1,\dots,n$. In this way the sum over permutations in \eqn{PiSn} can be
written as a sum over partitions \eqn{LiPn}, and the result after some algebra
is
\bea
& &{\cal P}_{ab}^{(n)j}[Y;s]\Bigm|_{S_n\times
S_2}=\lim_{\epsilon\to0^+}\left(N\Theta(\epsilon)+1\right)\,\delta_{ab}~
\sum_{{\buildrel{0\leq L_1,\dots,L_n\leq
n}\over{\sum_iL_i=n}}}~\Theta(\epsilon)^{\sum_i\delta_{L_i,1}}\nn\\&
&~~~~~~~~~~
\times\prod_{i=1}^n\int_{[0,1]^{L_i}}d\mu_{L_i}\left(s_1^{(i)},
\dots,s_{L_i}^{(i)}\right)~\left\langle\frac d{ds}x^j(s)\,\prod_{i=1}^n
{}~\tr\left({\cal T}_{L_i}\left[Y,x;s_1^{(i)},\dots,s_{L_i}^{(i)}\right]
\right)\right\rangle_0
\nn\\& &~~~~\label{PiSnfinal}\eea

The contributions from the remaining permutations $P\in S_{n+2}-(S_n\times
S_2)$ are somewhat more involved. We decompose this sum into three disjoint
sums of permutations. In the first class, whose contributions we denote by
${\cal P}_{ab}^{(n)j}[Y;s]^{[1]}$, for each $P$ there is a unique integer
$k_0\in\{1,\dots,n\}$ for which $P(k_0)\notin\{1,\dots,n\}$ with $P(k_0)=n+1$,
while the second class of permutations, whose contributions we denote by ${\cal
P}_{ab}^{(n)j}[Y;s]^{[2]}$, are those for which there is a unique integer
$k_0\in\{1,\dots,n\}$ with $P(k_0)\notin\{1,\dots,n\}$ and $P(k_0)=n+2$. The
final contributions ${\cal P}_{ab}^{(n)j}[Y;s]^{[3]}$ come from permutations
for which there are two integers $k_1,k_2\in\{1,\dots,n\}$ with
$P(k_1),P(k_2)\notin\{1,\dots,n\}$ and $P(k_1)=n+1,P(k_2)=n+2$.

We have
\bea
{\cal
P}_{ab}^{(n)j}[Y;s]^{[1]}&=&\sum_{c=1}^N~\sum_{b_1,\dots,b_n}~\sum_{k_0=1}^n~
\sum_{{\buildrel{P\in S_{n+2}-(S_n\times
S_2)}\over{P(k_0)=n+1}}}~\int_0^1\prod_{{\buildrel{k=1}\over{k\neq
k_0}}}^nds_k~\Theta\left(s_{P(k)}-s_k\right)\nn\\&
&~~\times\int_0^1ds_{k_0}~\Theta
\left(s-s_{k_0}\right)~\left(\delta_{c,b_{P(n+2)}}\delta_{ac}+\Theta(s_{
P(n+1)}-s)\delta_{cc}\delta_{a,b_{P(n+1)}}
\right)\nn\\& &~~\times\left\langle\frac
d{ds}x^j(s)\,\prod_{{\buildrel{k=1}\over{k\neq
k_0}}}^nY_{i_k}^{b_{P(k)},b_k}\left(x^0(s_k)\right)\,\frac
d{ds_k}x^{i_k}(s_k)\right.\nn\\& &~~~~~~~~~~~~~~~~~~~~\left.\times
\,Y_{i_{k_0}}^{b,b_{k_0}}\left(x^0(s_{k_0})\right)\,\frac
d{ds_{k_0}}x^{i_{k_0}}(s_{k_0})\right\rangle_0
\label{P[1]def}\eea
In \eqn{P[1]def} the terms with $\delta_{c,b_{P(n+2)}}$ correspond to
permutations with $P(n+1)=n+2,P(n+2)\in\{1,\dots,n\}$, while the $\delta_{cc}$
terms come from those with $P(n+2)=n+2,P(n+1)\in\{1,\dots,n\}$. In the
former terms, we consider the orbit of the integer $n+2$ under a given
permutation $P$, and let $\ell(P)+2\geq3$ be the order of the orbit of $k_0$
under $P$, i.e. $P^{\ell(P)}(n+2)=P^{\ell(P)+2}(k_0)=k_0$.  The sum over
$b_{k_0},b_{P(k_0)},\dots,b_{P^{\ell(P)+2}(k_0)=P^{\ell(P)-1}(n+2)}$ then
yields ${\cal T}_{\ell(P)}[Y,x;s_1,\dots,s_{\ell(P)}]_{ba}$ for the
corresponding $Y$-matrix products in \eqn{P[1]def}. The corresponding boundary
integration measure is $d\mu_{\ell(P)}(s_1,\dots,s_{\ell(P)})$, as before
(after appropriate relabellings of the $s$-indices), with additional step
function restrictions as given in \eqn{P[1]def} which must be carefully
incorporated into the integration measure \eqn{mmeas}. The remaining part of
$P$ that does not act on this particular orbit is an element of
$S_{n-\ell(P)}$, so that the remaining sums and products can be decomposed into
cycles exactly as in \eqn{PiSnfinal}. For each $\ell\geq1$ there are
$\frac{(n-1)!}{(n-\ell)!}$ permutations $P$ under which $k_0$ has an orbit of
order $\ell$.

For the latter $\delta_{cc}$ terms in \eqn{P[1]def}, the integer
$\ell(P)+1\geq2$ is the order of the orbit of $k_0$ under $P$, i.e.
$P^{\ell(P)}(n+1)=P^{\ell(P)+1}(k_0)=k_0$. The sums over $b_i$'s and all
products in \eqn{P[1]def} give the same contribution as for the former
$c$-dependent terms. It follows that the sum over permutations in \eqn{P[1]def}
can be written as a sum over the orbit integers $\ell(P)$ and, for each such
integer, a sum over partitions of $n-\ell(P)$. After some algebra the result is
finally
\bea
{\cal
P}_{ab}^{(n)j}[Y;s]^{[1]}&=&\lim_{\epsilon\to0^+}
\sum_{\ell=1}^n\frac{n!}{(n-\ell)!}
{}~\sum_{{\buildrel{0\leq L_1,\dots,L_{n-\ell}\leq
n-\ell}\over{\sum_iL_i=n-\ell}}}~\Theta(\epsilon)^{\sum_i\delta_{L_i,1}}
\,\prod_{i=1}^{n-\ell}
\int_{[0,1]^{L_i}}d\mu_{L_i}\left(s_1^{(i)},\dots,s_{L_i}^{(i)}\right)\nn\\&
&~~\times\int_{[0,1]^\ell}d\mu_\ell(r_1,\dots,r_\ell)~\Theta(s-r_1)
\left(1+N\Theta(r_\ell-s)\right)\nn\\& &~~\times\left\langle\frac
d{ds}x^j(s)\,\prod_{i=1}^{n-\ell}~\tr\left({\cal
T}_{L_i}\left[Y,x;s_1^{(i)},\dots,s_{L_i}^{(i)}\right]\right)\,{\cal
T}_\ell[Y,x;r_1,\dots,r_\ell]_{ba}\right\rangle_0
\label{P[1]final}\eea

The resummation of ${\cal P}_{ab}^{(n)j}[Y;s]^{[2]}$ carries through in an
identical fashion, with the roles of the integers $n+1$ and $n+2$ interchanged.
The result is identical to \eqn{P[1]final}, except for some changes in the
combinatorics of the indices. We find
\bea
{\cal
P}_{ab}^{(n)j}[Y;s]^{[2]}~&=&~\lim_{\epsilon\to0^+}
\sum_{c=1}^N~\sum_{b_1,\dots,b_n}~
\sum_{k_0=1}^n~\sum_{{\buildrel{P\in S_{n+2}-(S_n\times
S_2)}\over{P(k_0)=n+2}}}~\int_0^1\prod_{{\buildrel{k=1}\over{k\neq
k_0}}}^nds_k~\Theta\left(s_{P(k)}-s_k\right)\nn\\&
&~~\times\int_0^1ds_{k_0}~\left(\Theta(s_{P(n+1)}-s)\delta_{cb}
\delta_{a,b_{P(n+1)}}+\Theta(\epsilon)\delta_{c,b_{P(n+2)}}\delta_{ab}
\right)\nn\\& &~~\times\left\langle\frac
d{ds}x^j(s)\,\prod_{{\buildrel{k=1}\over{k\neq
k_0}}}^nY_{i_k}^{b_{P(k)},b_k}\left(x^0(s_k)\right)\,\frac
d{ds_k}x^{i_k}(s_k)\right.\nn\\& &~~~~~~~~~~~~~~~~~~~~\left.\times
\,Y_{i_{k_0}}^{c,b_{k_0}}\left(x^0(s_{k_0})\right)\,\frac
d{ds_{k_0}}x^{i_{k_0}}(s_{k_0})\right\rangle_0
\nn\\& &~~~~~\nn\\~&=&~\lim_{\epsilon\to0^+}
\sum_{\ell=1}^n\frac{n!}{(n-\ell)!}~\sum_{{\buildrel{0\leq
L_1,\dots,L_{n-\ell}\leq
n-\ell}\over{\sum_iL_i=n-\ell}}}~\Theta(\epsilon)^{\sum_i\delta_{L_i,1}}
\nn\\& &~~\times\prod_{i=1}^{n-\ell}
\int_{[0,1]^{L_i}}d\mu_{L_i}\left(s_1^{(i)},\dots,s_{L_i}^{(i)}\right)~
\int_{[0,1]^\ell}d\mu_\ell(r_1,\dots,r_\ell)~\Theta(s-r_1)\nn\\&
&~~\times\biggm\langle\frac
d{ds}x^j(s)\,\prod_{i=1}^{n-\ell}~\tr\left({\cal
T}_{L_i}\left[Y,x;s_1^{(i)},\dots,s_{L_i}^{(i)}\right]\right)
\nn\\& &~~~~~~~~~~\times\Bigm\{\Theta(r_\ell-s)\,{\cal
T}_\ell[Y,x;r_1,\dots,r_\ell]_{ba}\nn\\&
&~~~~~~~~~~~~~~~+\,\Theta(\epsilon)~\tr\Bigl({\cal
T}_\ell[Y,x;r_1,\dots,r_\ell]\Bigr)\,\delta_{ab}\Bigm\}\biggm\rangle_0
\label{P[2]final}\eea

Finally, the combinatorics of the resummation of ${\cal
P}_{ab}^{(n)j}[Y;s]^{[3]}$ now involve tracing the orbits of both
$P(n+1),P(n+2)\in\{1,\dots,n\}$, i.e. we introduce two integers $\ell_1(P)$ and
$\ell_2(P)$ representing the orders of the orbits of $k_1$ and $k_2$,
respectively, under a given permutation $P$. The evaluation is then identical
to that above with these two orbits taken into account, and we find
\bea
{\cal
P}_{ab}^{(n)j}[Y;s]^{[3]}~&=&~\sum_{c=1}^N~\sum_{b_1,\dots,b_n}~\sum_{1\leq
k_1\neq k_2\leq n}~\sum_{{
\buildrel{P\in S_{n+2}-(S_n\times S_2)}\over{P(k_1)=n+1,P(k_2)=n+2}}}
{}~\int_0^1\prod_{{\buildrel{k=1}\over{k\neq
k_1,k_2}}}^nds_k~\Theta\left(s_{P(k)}-s_k\right)\nn\\&
&~~\times\int_0^1ds_{k_1}~\Theta\left(s-s_{k_1}\right)~\int_0^1ds_{k_2}~
\Theta\left(s_{P(n+1)}-s\right)\delta_{c,b_{P(n+2)}}\delta_{a,b_{P(n+1)}}\nn\\&
&~~\times\left\langle\frac d{ds}x^j(s)\,\prod_{{\buildrel{k=1}\over{k\neq
k_1,k_2}}}^nY_{i_k}^{b_{P(k)},b_k}\left(x^0(s_k)\right)\,\frac
d{ds_k}x^{i_k}(s_k)\right.\nn\\& &~~~~~~\left.\times\,Y_{i_{k_1}}
^{b,b_{k_1}}\left(x^0(s_{k_1})\right)\,\frac d{ds_{k_1}}x^{i_{k_1}}(s_{k_1})~
Y_{i_{k_2}}^{c,b_{k_2}}\left(x^0(s_{k_2})\right)\,\frac
d{ds_{k_2}}x^{i_{k_2}}(s_{k_2})\right\rangle_0
\nn\\& &~~~~~\nn\\~&=&~\lim_{\epsilon\to0^+}
\sum_{\ell_1=1}^{n-1}~\sum_{\ell_2=1}^{n-\ell_1}\frac{n!}
{(n-\ell_1)(n-\ell_1-\ell_2)!}~
\sum_{{\buildrel{0\leq L_1,\dots,L_{n-\ell_1-\ell_2}\leq
n-\ell_1-\ell_2}\over{\sum_iL_i=n-\ell_1-\ell_2}}}~
\Theta(\epsilon)^{\sum_i\delta_{L_i,1}}\nn\\&
&~~\times\prod_{i=1}^{n-\ell_1-\ell_2}
\int_{[0,1]^{L_i}}d\mu_{L_i}\left(s_1^{(i)},\dots,s_{L_i}^{(i)}\right)
{}~\int_{[0,1]^{\ell_2}}d\mu_{\ell_2}(t_1,\dots,t_{\ell_2})\nn\\& &~~\times
\int_{[0,1]^{\ell_1}}d\mu_{\ell_1}(r_1,\dots,r_{\ell_1})~
\Theta(r_{\ell_1}-s)\Theta(s-r_1)\nn\\& &~~\times\biggm\langle\frac
d{ds}x^j(s)\,\prod_{i=1}^{n-\ell_1-\ell_2}~\tr\left({\cal
T}_{L_i}\left[Y,x;s_1^{(i)},\dots,s_{L_i}^{(i)}\right]\right)
\nn\\& &~~~~~~~~~~~~~~~~~~~~\times\Bigl({\cal
T}_{\ell_1}[Y,x;r_1,\dots,r_{\ell_1}]\cdot{\cal
T}_{\ell_2}[Y,x;t_1,\dots,t_{\ell_2}]\Bigr)_{ba}\biggm\rangle_0
\label{P[3]final}\eea

The total order $n$ contribution \eqn{Piordern} is now the sum of
\eqn{PiSnfinal} and \eqn{P[1]final}--\eqn{P[3]final}, and after some algebra we
arrive at the final expression for the terms in the momentum
expansion \eqn{Pipowerseries},
\bea
{\cal P}_{ab}^{(n)j}[Y;s]&=&\lim_{\epsilon\to0^+}
\sum_{\ell_1=0}^{n-1}~\sum_{\ell=\ell_1+1-
\delta_{\ell_1,0}}^n\frac{n!}{(n-\ell)!}~\sum_{{\buildrel{0\leq
L_1,\dots,L_{n-\ell}\leq
n-\ell}\over{\sum_iL_i=n-\ell}}}~\Theta(\epsilon)^{\sum_i\delta_{L_i,1}}
\nn\\& &~~\times\prod_{i=1}^{n-\ell}
\int_{[0,1]^{L_i}}d\mu_{L_i}\left(s_1^{(i)},\dots,s_{L_i}^{(i)}\right)~
\int_{[0,1]^{\ell-\ell_1}}d\mu_{\ell-\ell_1}(r_1,\dots,r_{\ell-\ell_1})
\nn\\& &~~\times\int_{[0,1]^{\ell_1}}d\mu_{\ell_1}(t_1,\dots,t_{\ell_1})
{}~\Theta(t_{\ell_1}-s)\Theta(s-t_1)\nn\\& &~~\times\Biggm\langle\frac
d{ds}x^j(s)\,\prod_{i=1}^{n-\ell}~\tr\left({\cal
T}_{L_i}\left[Y,x;s_1^{(i)},\dots,s_{L_i}^{(i)}\right]\right)\nn\\& &~~
\times\biggm\{\biggm(\mbox{$\frac1{n-\ell_1}$}
+\delta_{\ell_1,0}\Bigm[\Theta(s-r_1)\Bigm(1+(N+1)
\Theta(r_\ell-s)\Bigm)-\mbox{$\frac1n$}\Bigm]\biggm)\nn\\&
&~~~~~~~~~~~~~~~~~~~~\times\Bigl({\cal
T}_{\ell_1}[Y,x;t_1,\dots,t_{\ell_1}]\cdot{\cal
T}_{\ell-\ell_1}[Y,x;r_1,\dots,r_{\ell-\ell_1}]\Bigr)_{ba}\nn\\&
&~~~~~~~~~~+\,\delta_{\ell_1,0}\,\Theta(\epsilon)~\tr\Bigl({\cal
T}_\ell[Y,x;r_1,\dots,r_\ell]\Bigr)\,\delta_{ab}\biggm\}\Biggm\rangle_0
\label{Pinfinal}\eea
This expression contains ambiguous factors of $\Theta(\epsilon),\epsilon\to0^+$
and products such as $\Theta(s)\Theta(-s)$ which depend on the particular
choice of regularization of the step function. The auxilliary quantum field
theory contains most of the information about the non-abelian dynamics, and, to
obtain an expression which is explicitly independent of such regularizations,
we need to choose an appropriate renormalization scheme for it.\footnote{Note
that with the regularization \eqn{Thetafnreg} we have
$\Theta(s)\Theta(-s)=-\Theta(\epsilon)^2-\Theta(\epsilon)$, so that such a
renormalization scheme can be understood as removing all powers of the
ambiguous term $\Theta(\epsilon)$.} The removal of these ambiguous factors is
also required in order that \eqn{wilsonloop} be a proper representation of the
Wilson loop operator. This renormalization has been discussed in \cite{dorn}.
In terms of the Feynman diagram representation of the $\bar\xi\xi$ field
averages in \eqn{Piordern}, we keep only those graphs corresponding to Wick
contractions in which there is a single continuous line connecting the (same)
boundary points $s=0$ and $s=1$, i.e. we restrict to connected Feynman graphs.
This will also ensure that the final result is independent of the
$s$-dependence of the auxilliary field representation, as it should be. From
\eqn{etaavgs} this means that we restrict the sum over permutations $P\in
S_{n+2}$ to those whose cyclic decomposition contains only a single cycle
$C_{n+2}(P)$ of length $L_{n+2}(P)=n+2$. This is achieved essentially by
normalizing the functional integration measure $D\bar\xi\,D\xi$ in
\eqn{auxprop} so that $\langle\!\langle1\rangle\!\rangle=1$.

The renormalized canonical momentum is thus calculated by restricting to cyclic
permutations of length $n+2$. By definition, this immediately eliminates the
contributions ${\cal P}_{ab}^{(n)j}[Y;s]|_{S_n\times S_2}$ and ${\cal
P}_{ab}^{(n)j}[Y;s]^{[3]}$ above. This scheme removes the $\delta_{cc}$ terms
in \eqn{P[1]def} and the $\delta_{c,b_{P(n+2)}}$ terms in the first equality in
\eqn{P[2]final}. Then, we keep only the orbits of length $\ell=n$ in
\eqn{P[1]final} and in the second equality of \eqn{P[2]final}. The sum ${\cal
P}_{ab}^{(n)j}[Y;s]^{\rm ren}$ of these two terms contains no ambiguities from
the step functions involved. Furthermore, after some careful algebra one can
rewrite the resulting integration measure from this sum as
$\int_0^1\prod_{k=1}^nds_k$, and the corresponding integrand with the
appropriate relabelling of indices is readily seen to form a symmetrized matrix
product. The result is finally
\beq
{\cal P}_{ab}^{(n)j}[Y;s]^{\rm
ren}=\int_0^1\prod_{k=1}^nds_k~\left\langle\frac{dx^j(s)}{ds}
\,~{\rm Sym}~{\cal T}_n[Y,x;s_1,\dots,s_n]_{ba}\right\rangle_0
\label{Pirenorderns}\eeq
which yields the expression \eqn{Pirenordern}.

\newsection{Boundary Correlation Functions}

In this appendix we will present the results of the boundary integrations which
are used in the perturbative calculations of sections 4 and 5. In general, the
integrals are divergent, and difficult to do analytically. However, we need
only determine their most divergent parts as $\Lambda\to0$, dropping
sub-divergent pieces which vanish upon taking the limit $\epsilon\to0^+$ with
the correlation \eqn{scalerel}. To see how these calculations proceed, let us
consider as an example the boundary integral
\beq
I_c^{(1)}\equiv\int_0^1ds_1~ds_2~ds_3~\frac{\log[2-2\cos2\pi(s_1-s_2)]}
{[1-\cos2\pi s_1][1-\cos2\pi(s_2-s_3)]}
\label{ic1}\eeq
which arises in the evaluation of the $Y^2U$ contributions to the canonical
momentum of subsection 4.3. The integral over $s_3$ can be done as in
\eqn{order1def} to give
\beq
I_c^{(1)}=\frac1\pi\int_0^1ds_1~ds_2~\frac1{\tan\pi s_2}\frac{\log(2|\sin\pi
s_1\cos\pi s_2-\sin\pi s_2\cos\pi s_1|)}{\sin^2\pi s_1}
\label{ic1s3}\eeq
The divergent contributions to the integral over $s_1$ come from the short
distance boundary behaviour $s_1-s_2\sim s_\Lambda$, i.e. $\sin\pi
s_1\sim\sin\pi s_2$. Expanding the integrand of \eqn{ic1s3} about this point
gives the most divergent contribution leading to
\beq
I_c^{(1)}\simeq\frac1{\pi^2}\log(2\sin\pi s_\Lambda)\int_0^1ds_2~\frac{\cos\pi
s_2}{\sin^3\pi s_2}
\label{ic1s1}\eeq
where here and in the following $\simeq$ denotes the most divergent
contribution as $\Lambda\to0$. Using the boundary cutoff \eqn{bdrycutoff} and
evaluating the final integration over $s_2$ using this cutoff we arrive finally
at
\beq
I_c^{(1)}\simeq-\frac2{\pi^3}\frac{\log\Lambda}{\tan^2\pi s_\Lambda}
\label{ic1final}\eeq

All other boundary integrations are evaluated using similar sorts of asymptotic
approximation techniques. Below we list their leading divergent behaviours as
$\Lambda\to0$. For the $Y^2U$ terms of the canonical momentum calculation of
subsection 4.3, which come from the correlation function \eqn{3ptCCD}, in
addition to (\ref{ic1}, \ref{ic1final}) we used the integrals
\bea
I_0&\equiv&\int_0^1ds_1~ds_2~ds_3~\frac1{[1-\cos2\pi s_1][1-\cos2\pi(s_2-s_3)]}
\nonumber\\&\simeq&\frac4{\pi^3}\frac{\log\Lambda}{\tan\pi
s_\Lambda}\label{i0}\\I_u^{(1)}&\equiv&\int_0^1ds_1~ds_2~ds_3~\frac{
\log[2-2\cos2\pi(s_2-s_3)]}{[1-\cos2\pi s_1][1-\cos2\pi(s_2-s_3)]}\nonumber\\
&\simeq&-\frac8{\pi^3}\frac{(\log\Lambda)^2}{\tan\pi s_\Lambda}\label{iu1}
\eea

The additional integrals involved in the calculation of the $YU^2$ part in
subsection 4.3, which come from the correlators \eqn{3ptDDC}, are
\bea
I_u^{(2)}&\equiv&\int_0^1ds_1~ds_2~ds_3~\frac{(\log[2-2\cos2\pi(s_2-s_3)])^2}
{[1-\cos2\pi s_1][1-\cos2\pi(s_2-s_3)]}\nonumber\\&\simeq&
-\frac{16}{3\pi^3}\frac{(\log\Lambda)^3}{\tan\pi
s_\Lambda}\label{iu2}\\I_c^{(2)}&\equiv&
\int_0^1ds_1~ds_2~ds_3~\frac{(\log[2-2\cos2\pi(s_1-s_3)])^2}{[1-\cos2\pi
s_1][1-\cos2\pi(s_2-s_3)]}\nonumber\\&\simeq&
-\frac4{3\pi^3}\frac{(\log\Lambda)^2}{\tan^2\pi
s_\Lambda}\label{ic2}\\I_q^{(1)}&\equiv&
\int_0^1ds_1~ds_2~ds_3~\frac1{[1-\cos2\pi s_1][1-\cos2\pi(s_2-s_3)]}
\nonumber\\& &~~~~~~~~~~~~~~~~~~~~\times
\left(\frac{\log\frac1{\Lambda^2}[2-2\cos2\pi(s_1-s_3)]}{\log\frac1{\Lambda^2}
[2-2\cos2\pi(s_1-s_2)]}\right)^2\nonumber\\&\simeq&-\frac2{\pi^3}
\frac{(\log\Lambda)^2}{\tan\pi
s_\Lambda}\log\log\Lambda\label{iq1}\\I_q^{(2)}&\equiv&
\int_0^1ds_1~ds_2~ds_3~\frac1{[1-\cos2\pi s_1][1-\cos2\pi(s_2-s_3)]}
\nonumber\\& &~~~~~~~~~~~~~~~~~~~~\times
\left(\frac{\log\frac1{\Lambda^2}[2-2\cos2\pi(s_2-s_3)]}{\log\frac1{\Lambda^2}
[2-2\cos2\pi(s_1-s_2)]}\right)^2\nonumber\\&\simeq&\frac1{8\pi^3}
\frac1{\tan^2\pi s_\Lambda}\label{iq2}\\I_q^{(3)}&\equiv&
\int_0^1ds_1~ds_2~ds_3~\frac1{[1-\cos2\pi s_1][1-\cos2\pi(s_2-s_3)]}
\nonumber\\& &~~~~~~~~~~~~~~~~~~~~\times
\left(\frac{\log\frac1{\Lambda^2}[2-2\cos2\pi(s_1-s_3)]}{\log\frac1{\Lambda^2}
[2-2\cos2\pi(s_2-s_3)]}\right)^2\nonumber\\&\simeq&-\frac2{\pi^3}
\frac{(\log\Lambda)^2}{\tan\pi s_\Lambda}\log\log\Lambda\label{iq3}
\eea

For the $U^3$ terms of subsection 4.3, which come from the correlation function
\eqn{3ptDDD}, we use the integrals
\bea
I_m^{(1)}&\equiv&\int_0^1ds_1~ds_2~ds_3~\frac{\log[2-2\cos2\pi(s_1-s_2)]
\log[2-2\cos2\pi(s_2-s_3)]}{[1-\cos2\pi s_1][1-\cos2\pi(s_2-s_3)]}\nonumber\\
&\simeq&\frac8{\pi^3}\frac{(\log\Lambda)^2}{\tan^2\pi s_\Lambda}\label{im1}\\
I_m^{(2)}&\equiv&\int_0^1ds_1~ds_2~ds_3~\frac{\log[2-2\cos2\pi(s_1-s_2)]
\log[2-2\cos2\pi(s_1-s_3)]}{[1-\cos2\pi s_1][1-\cos2\pi(s_2-s_3)]}\nonumber\\
&\simeq&\frac{16}{\pi^3}\frac{(\log\Lambda)^2}{\tan^2\pi s_\Lambda}\label{im2}
\\I_t^{(1)}&\equiv&\int_0^1ds_1~ds_2~ds_3~\frac{\log[2-2\cos2\pi(s_1-s_2)]
\log[2-2\cos2\pi(s_2-s_3)]}{[1-\cos2\pi s_1][1-\cos2\pi(s_2-s_3)]}\nonumber\\
& &~~~~~~~~~~~~~~~~~~~~\times\,\log[2-2\cos2\pi(s_1-s_3)]\nonumber\\
&\simeq&\frac{64}{\pi^3}\frac{(\log\Lambda)^3}{\tan^2\pi s_\Lambda}\label{it1}
\\I_t^{(2)}&\equiv&\int_0^1ds_1~ds_2~ds_3~\frac{(\log[2-2\cos2\pi(s_1-s_2)])^2
\log[2-2\cos2\pi(s_2-s_3)]}{[1-\cos2\pi s_1][1-\cos2\pi(s_2-s_3)]}\nonumber\\
&\simeq&\frac{32}{\pi^3}\frac{(\log\Lambda)^3}{\tan^2\pi s_\Lambda}\label{it2}
\\I_t^{(3)}&\equiv&\int_0^1ds_1~ds_2~ds_3~\frac{(\log[2-2\cos2\pi(s_1-s_2)])^2
\log[2-2\cos2\pi(s_1-s_3)]}{[1-\cos2\pi s_1][1-\cos2\pi(s_2-s_3)]}\nonumber\\
&\simeq&\frac{24}{\pi^3}\frac{(\log\Lambda)^3}{\tan^2\pi s_\Lambda}\label{it3}
\\I_t^{(4)}&\equiv&\int_0^1ds_1~ds_2~ds_3~\frac{\log[2-2\cos2\pi(s_1-s_2)]
(\log[2-2\cos2\pi(s_2-s_3)])^2}{[1-\cos2\pi
s_1][1-\cos2\pi(s_2-s_3)]}\nonumber\\
&\simeq&\frac{16}{\pi^3}\frac{(\log\Lambda)^3}{\tan^2\pi s_\Lambda}\label{it4}
\\I_t^{(5)}&\equiv&\int_0^1ds_1~ds_2~ds_3~\frac{(\log[2-2\cos2\pi(s_1-s_2)])^3
}{[1-\cos2\pi s_1][1-\cos2\pi(s_2-s_3)]}\nonumber\\
&\simeq&-\frac{16}{\pi^3}\frac{(\log\Lambda)^3}{\tan^2\pi s_\Lambda}\label{it5}
\\I_t^{(6)}&\equiv&\int_0^1ds_1~ds_2~ds_3~\frac{(\log[2-2\cos2\pi(s_2-s_3)])^3
}{[1-\cos2\pi s_1][1-\cos2\pi(s_2-s_3)]}\nonumber\\
&\simeq&-\frac{32}{\pi^3}\frac{(\log\Lambda)^4}{\tan\pi s_\Lambda}\label{it6}
\eea

Finally, the boundary integrals arising in the evaluation of the Zamolodchikov
metric of section 5, which come from the two-point correlation functions
\eqn{2ptCD} and \eqn{2ptDD} of the logarithmic operators, are
\bea
I_g^{(1)}&\equiv&\int_0^1ds_1~ds_2~\frac1{1-\cos2\pi(s_1-s_2)}\nonumber\\
&\simeq&\frac4{\pi^2}\log\Lambda\label{ig1}\\I_g^{(2)}&\equiv&\int_0^1ds_1~
ds_2~\frac1{(1-\cos2\pi s_1)(1-\cos2\pi
s_2)}\nonumber\\&\simeq&\frac2{\pi^2\Lambda^2}\frac1{\tan\pi
s_\Lambda}\label{ig2}\\I_g^{(3)}&\equiv&\int_0^1ds_1~ds_2~\frac
{\log[2-2\cos2\pi(s_1-s_2)]}{1-\cos2\pi(s_1-s_2)}\nonumber\\
&\simeq&-\frac8{\pi^2}(\log\Lambda)^2\label{ig3}\\I_g^{(4)}&\equiv&
\int_0^1ds_1~ds_2~\frac{\log[2-2\cos2\pi(s_1-s_2)]}{(1-\cos2\pi s_1)(1-\cos2\pi
s_2)}\nonumber\\&\simeq&-\frac2{3\pi^2\Lambda^2}\frac{\log\Lambda}{\tan^2\pi
s_\Lambda}\label{ig4}
\eea

\newsection{Ward Identities and Leading Divergences in the Genus Expansion}

In this appendix we shall show how the leading $(\log\delta)^2$ modular
divergences which appear in \eqn{mixing} can be removed by invoking an
appropriate Ward identity for the fundamental string fields of the matrix
$\sigma$-model. As we shall show, this is equivalent to imposing momentum
conservation for scattering processes in the matrix D-brane background. This
has been demonstrated explicitly for the single D-particle case in \cite{lm}.
Within the framework of the auxilliary field representation of the Wilson loop
operator, the effective abelianization of the matrix $\sigma$-model leads to a
relatively straightforward generalization of this proof, as we now demonstrate.

The pertinent bilocal term induced by (\ref{mixing}), which exponentiates upon
summing over pinched topologies, can be written as a local worldsheet effective
action using the wormhole parameters $[\rho_{C,D}]_i^{ab}$
to give
\bea
\e^{\Delta S^{CD}}&=&\lim_{\epsilon\to0^+}\int d\rho_C~d\rho_D~
\exp\left[\sum_{a,b=1}^N\left(-\frac1{2g_s^2(\log\delta)^2}\,G^{LM}
\sum_{c,d=1}^NG_{ab;cd}^{ij}\,[\rho_L]_i^{ab}[\rho_M]_j^{cd}\right.\right.
\nn\\& &~~~~~~~~~~~~~~~+\frac{ig_s[\rho_C]_i^{ab}}{2\pi\alpha'}
\int_0^1ds~C(x^0(s);\epsilon)\,
\bar\xi_a(s-\epsilon)\xi_b(s)\,\frac d{ds}x^i(s)\nn\\& &~~~~~~~~~~~~~~~
\left.\left.+\frac{ig_s[\rho_D]_i^{ab}}{2\pi\alpha'}\int_0^1ds~D(x^0(s);
\epsilon)\,\bar\xi_a(s-\epsilon)\xi_b(s)\,\frac d{ds}x^i(s)\right)\right]
\label{wlocal2}\eea
Here we have for simplicity considered only the zero frequency modes of the
fields involved with respect to the Fourier transformations defined at the
beginning of subsection 6.1. They will be sufficient to describe the relevant
cancellations. In \eqn{wlocal2} the (dimensionless) moduli space metric
$G^{LM}$ (where $L,M=C,D$) is an appropriate off-diagonal $2\times2$ matrix
with respect to the decomposition \eqn{modulidecomp} (see \eqn{ZCDscale}) which
is required to reproduce the initial bilocal operator with the $CD$-mixing of
the logarithmic operators. This off-diagonal metric includes all the
appropriate normalization factors ${\cal N}_L$ for the zero mode states. These
factors are essentially the inverse of the $CD$ two-point function \eqn{2ptCD}
which is finite.

We consider the propagation of two (closed string) matter tachyon states
$T_{1,2}=\e^{i(k_{1,2})_ix^i}$ in the background of (\ref{wlocal2}) at tree
level. In what follows the effects of the $C$ operator are sub-leading and can
be ignored. Then, we are interested in the amplitude
\bea
{\cal A}_{CD}&\equiv&\left\langle~\left\langle\!\!\left\langle\sum_{c'=1}^N
\bar\xi_{c'}(0)\,T_1T_2~\e^{\Delta
S^{CD}}\,\xi_{c'}(1)\right\rangle\!\!\right\rangle~\right\rangle_0\nn\\
&=&\lim_{\epsilon\to0^+}\sum_{c'=1}^N\int d\rho_C~d\rho_D~\int
Dx~D\bar\xi~D\xi~\bar\xi_{c'}(0)\nn\\&
&\times\exp\left(-N^2S_0[x]-\sum_{c=1}^N\int_0^1ds~
\bar\xi_c(s-\epsilon)\frac d{ds}\xi_c(s)\right)\nn\\&
&\times\,T_1[x]\,T_2[x]~\exp\left[\sum_{a,b=1}^N
\left(-\frac1{2g_s^2(\log\delta)^2}\,G^{LM}
\sum_{c,d=1}^NG_{ab;cd}^{ij}\,[\rho_L]_i^{ab}[\rho_M]_j^{cd}\right.\right.
\nn\\& &\left.\left.+\frac{ig_s[\rho_D]_i^{ab}}{2\pi\alpha'}
\int_0^1ds~D(x^0(s);\epsilon)\,
\bar\xi_a(s-\epsilon)\xi_b(s)\,\frac d{ds}x^i(s)\right)\right]\xi_{c'}(1)+\dots
\label{dsv1v2}\eea
where $\dots$ represent sub-leading terms. The scaling property
\eqn{DCscaletransfs} of the logarithmic operators must be taken into account.
Under a scale transformation \eqn{Lambdarescale} on the worldsheet the $C$
operator emerges from $D$ due to mixing with a scale-dependent coefficient
$\sqrt{\alpha'}\,t$. This will contribute to the scaling infinities we are
considering here.

The composite $D$ operator insertion in (\ref{dsv1v2}) needs to be
normal-ordered on the disc. Normal ordering in the present case amounts to
subtracting scaling infinities originating from divergent contributions of
$D(x^0(s);\epsilon)$ as $\epsilon \rightarrow 0^+$. To determine these
infinities, we first note that the one-point function of the composite $D$
operators, computed with respect to the free $\sigma$-model and auxilliary
field actions, can be written as
\bea
& &\left\langle~\left\langle\!\!\left\langle\sum_{c'=1}^N\bar\xi_{c'}(0)
\exp\left(\sum_{a,b=1}^N\frac{ig_s[\rho_D]_i^{ab}}{2\pi\alpha'}
\int_0^1ds~D(x^0(s);\epsilon)\,\bar\xi_a(s-\epsilon)\xi_b(s)\,\frac
d{ds}x^i(s)\right)\xi_{c'}(1)\right\rangle\!\!\right\rangle~
\right\rangle_0\nn\\&
&=~\left\langle\!\!\left\langle\sum_{c'=1}^N\bar\xi_{c'}(0)\exp\left(-
\sum_{a,b,c,d}\frac{g_s^2[\rho_D]_i^{ab}[\rho_D]_j^{cd}}{2(2\pi\alpha')^2}
\int_0^1ds~ds'~\left\langle
D(x^0(s);\epsilon)\,D(x^0(s');\epsilon)\right\rangle_0\right.\right.
\right.\nn\\& &~~~~~~~~~~~~~~~~~~~~~~~~~~~~~~
\Biggl.\Biggl.\Biggl.\times\,\bar\xi_a(s-\epsilon)
\xi_b(s)\bar\xi_c(s'-\epsilon)\xi_d(s')\left\langle\mbox{$\frac
d{ds}$}\,x^i(s)\,\mbox{$\frac
d{ds'}$}\,x^j(s')\right\rangle_0\Biggr)\xi_{c'}(1)\Biggr\rangle\!\!\Biggr
\rangle\nn\\&
&=~\exp\left(-\sum_{a,b=1}^N\frac{g_s^2[\rho_D]_i^{ab}[\rho_D]_j^{ba}}
{2(2\pi\alpha')^2}\int_0^1ds~ds'~\left\langle
D(x^0(s);\epsilon)\,D(x^0(s');\epsilon)\right\rangle_0\left\langle\mbox{$\frac
d{ds}$}\,x^i(s)\mbox{$\frac d{ds'}$}\,x^j(s')\right\rangle_0\right)\nn\\& &~~~~
\label{two-point}\eea
where we have used Wick's theorem. The second equality in \eqn{two-point}
follows after removing ambiguous $\Theta(\epsilon)$ type terms from the Wick
expansion in the auxilliary fields using the renormalization scheme described
in appendix B. One finds that this procedure has the overall effect of
replacing the product of auxilliary fields in the first equality in
\eqn{two-point} by the delta-functions $\delta_{ad}\delta_{bc}$.

In what follows we shall ignore, for simplicity, the basic divergences that
come from the fundamental string propagator in \eqn{two-point}. Such
divergences will appear globally in all correlators below and will not affect
the final result. As a consequence of the logarithmic algebra \eqn{2ptCD} and
the scale transformation \eqn{Lambdarescale},\eqn{DCscaletransfs}, there are
leading (scaling) divergences in \eqn{two-point} for
$\epsilon\rightarrow 0^+$ which behave as
\beq
g_s^2b\alpha'^{-1/2}t~\tr\,[\rho_D]_i[\rho_D]^i
\label{counterterm}\eeq
Thus, normal ordering of the $D$ operator amounts to adding a term of opposite
sign to \eqn{counterterm} into the argument of the exponential in
(\ref{dsv1v2}) in order to cancel such divergences.

Let us now introduce a complete set of states $|{\cal E}_I\rangle$ into the
two-point function of string matter fields on the disc,
\be
\langle T_1T_2\rangle_0=\sum_I|{\cal N}_I|^2\,\langle T_1|{\cal
E}_I\rangle_0\,\langle{\cal E}_I|T_2\rangle_0
\label{completeset}\ee
where ${\cal N}_I$ is a normalization factor for the fundamental string states
(determined by the Zamolodchikov metric). Taking into account the effects of
the $C$ operator included in $D$ under the scaling (\ref{Lambdarescale}), we
see that the leading divergent contributions to (\ref{completeset}) are of the
form
\be
\langle T_1T_2\rangle_0\simeq-\sqrt{\alpha'}\,t\,\langle
T_1|C\rangle_0\,\langle
C|T_2\rangle_0+\dots
\label{vc}\ee
where we have used \eqn{ZCDscale} and \eqn{scalerel}. We now notice that the
$C$ deformation vertex operator plays the role of the Goldstone mode for the
translation symmetry of the fundamental string coordinates $x^i$, and as such
we can apply the corresponding Ward identity in the matrix $\sigma$-model path
integral to represent the action of the $C$ deformation on physical states by
$-i\delta/\delta x^i$ \cite{kogmav,paban}. The leading contribution to
\eqn{completeset} can thus be exponentiated to yield
\bea
\langle T_1T_2\rangle_0&\simeq&\lim_{\epsilon\to0^+}\sum_{c'=1}^N\int
Dx~D\bar\xi~D\xi~\bar\xi_{c'}(0)~\exp\left(-N^2S_0[x]-\sum_{c=1}^N\int_0^1ds~
\bar\xi_c(s-\epsilon)\frac d{ds}\xi_c(s)\right)\nn\\&
&~~~~~~\times\,T_1[x]\exp\left(-\frac{g_s^2\sqrt{\alpha'}\,t}2
\sum_{a,b=1}^N\int_0^1ds~ds'~\bar
\xi_a(s-\epsilon)\xi_b(s)\bar\xi_b(s'-\epsilon)\xi_a(s')\right.\nn\\&
&~~~~~~\left.\times\,\frac{\buildrel\leftarrow\over\delta}{\delta
x_i(s)}\frac{\buildrel\rightarrow\over\delta}{\delta
x^i(s')}\right)T_2[x]~\xi_{c'}(1)
\label{expon}\eea
where we have used the on-shell condition $T_j(\frac\delta{\delta
x_i}\frac\delta{\delta x^i})T_k=0$ for the tachyon fields. \eqn{expon}
expresses the non-abelian version of the Ward identity in the presence of
logarithmic deformations.

Using \eqn{counterterm}, \eqn{expon} and normalizing the parameters of the
logarithmic conformal algebra appropriately, it follows that (\ref{dsv1v2}) can
be written as
\bea
{\cal A}_{CD}&\simeq&\lim_{\epsilon\to0^+}\sum_{c'=1}^N\int
d\rho_C~d\rho_D~\int Dx~D\bar\xi~D\xi~\bar\xi_{c'}(0)\nn\\&
&\times\exp\left(-N^2S_0[x]-\sum_{c=1}^N\int_0^1ds~
\bar\xi_c(s-\epsilon)\frac d{ds}\xi_c(s)\right)\nn\\&
&\times\,T_1[x]\exp\left[\sum_{a,b=1}^N
\left(-\frac1{2g_s^2(\log\delta)^2}\,G^{LM}
\sum_{c,d=1}^NG_{ab;cd}^{ij}\,[\rho_L]_i^{ab}[\rho_M]_j^{cd}\right.\right.
\nn\\& &-\frac{g_s^2\alpha'^{-1/2}t}2\,\eta^{ij}[\rho_D]_i^{ab}[\rho_D]_j^{ba}
+ig_st\,[\rho_D]_i^{ab}\int_0^1ds~
\bar\xi_a(s-\epsilon)\xi_b(s)\,\frac{\buildrel\leftrightarrow\over\delta}
{\delta x_i(s)}\nn\\&
&\left.\left.-\frac{g_s^2\sqrt{\alpha'}\,t}2\int_0^1ds~ds'~\bar
\xi_a(s-\epsilon)\xi_b(s)\bar\xi_b(s'-\epsilon)\xi_a(s')
\,\frac{\buildrel\leftarrow\over\delta}{\delta
x_i(s)}\frac{\buildrel\rightarrow\over\delta}{\delta
x^i(s')}\right)\right]T_2[x]~\xi_{c'}(1)+\dots\nn\\&
&~~~~\nn\\&=&\lim_{\epsilon\to0^+}
\sum_{c'=1}^N\int d\rho_C~d\rho_D~\int Dx~D\bar\xi~D\xi~\bar\xi_{c'}(0)\nn\\&
&\times\exp\left(-N^2S_0[x]-\sum_{c=1}^N\int_0^1ds~
\bar\xi_c(s-\epsilon)\frac d{ds}\xi_c(s)\right)\nn\\&
&\times\,T_1[x]\exp\left[\sum_{a,b=1}^N
\left\{-\frac1{2g_s^2(\log\delta)^2}\,G^{LM}
\sum_{c,d=1}^NG_{ab;cd}^{ij}\,[\rho_L]_i^{ab}[\rho_M]_j^{cd}\right.\right.
\nn\\& &-\frac{g_s^2\alpha'^{-1/2}t}2\,\eta^{ij}\left([\rho_D]_i^{ab}-\frac{i
\sqrt{\alpha'}}{g_s}\int_0^1ds~\bar\xi_a(s-\epsilon)\xi_b(s)
\,\frac{\buildrel\leftrightarrow\over\delta}{\delta x_i(s)}\right)\nn\\&
&~~~~~~\left.\left.\times\left([\rho_D]_j^{ba}-\frac{i\sqrt{\alpha'}}{g_s}
\int_0^1ds~\bar\xi_b(s-
\epsilon)\xi_a(s)\,\frac{\buildrel\leftrightarrow\over\delta}{\delta
x_j(s)}\right)\right\}\right]T_2[x]~\xi_{c'}(1)+\dots
\label{cdsq2}\eea
{}From (\ref{cdsq2}) it follows that the limit $t\to\infty$ localizes the
worldsheet wormhole parameter integrations with delta-function support
\beq
\prod_{a,b=1}^N\,\prod_{i=1}^9\delta\left(\mbox{$[\rho_D]_i^{ab}-\mbox{$
\frac{\sqrt{\alpha'}}{g_s}$}\,
\,(k_1+k_2)_i\,\int_0^1ds~\bar\xi_a(s-\epsilon)\xi_b(s)$}\right)
\label{dfunc}\eeq
where $(k_{1,2})_{i}$ are the momenta of the closed string matter states. This
result shows that the leading modular divergences in the genus expansion are
cancelled by the scattering of (closed) string states off the matrix D-brane
background. Upon rescaling $\rho_D$ by $g_s^2$, averaging over the auxilliary
boundary fields, and incorporating \eqn{dfunc} as an effective shift in the
velocity recoil operator (see \eqn{quantumcoords}), we can identify this
renormalization as fixing the velocity matrix
\beq
U_i^{ab}=-\sqrt{\alpha'}\,g_s\,(k_1+k_2)_i\,\delta^{ab}
\label{velocitydiag}\eeq
of the fat brane background. Thus momentum conservation for the D-brane
dynamics guarantees conformal invariance of the matrix $\sigma$-model as far as
leading divergences are concerned.

\end{document}